\documentclass[%
 preprint,
 amsmath,amssymb,
 aps,
]{revtex4-1}

\usepackage{graphicx}% Include figure files
\usepackage{dcolumn}% Align table columns on decimal point
\usepackage{bm}
\usepackage{mathtools}
\usepackage{breqn}
\usepackage{physics}
\usepackage{caption}
\usepackage{subcaption}
\usepackage{subcaption}
\usepackage{listings}
\usepackage[section]{placeins}
\usepackage{mathrsfs}

\begin{document}

%\preprint{APS/123-QED}

\title{Towards Uncovering Generic Effects Of Matter Sources In Anisotropic Quantum Cosmologies Via Taub Models}% Force line breaks with \\

\author{Daniel Berkowitz}
 \altaffiliation{Physics Department, Yale University.\\ daniel.berkowitz@yale.edu \\ This work is in memory of my parents, Susan Orchan Berkowitz, and Jonathan Mark Berkowitz}%Lines break automatically or can be forced with \\

\date{\today}% It is always \today, today,

\begin{abstract}
We solve the Wheeler DeWitt equation for the Taub models in closed form when both a cosmological constant and an electromagnetic field are present. In doing so, we examine the 'excited states' of the quantum Taub models with the aforementioned matter sources, and prove the existence of a countably infinite number of 'excited' states. Additionally, we prove the existence of multiple asymptotic solutions to the Lorentzian signature Wheeler DeWitt equation using the Taub analogues of the 'wormhole' \cite{moncrief1991amplitude}, and 'no boundary' \cite{graham1993supersymmetric} solutions of its Euclidean-signature Hamilton Jacobi equation; we also study their 'excited states. In the end we investigate qualitatively how our matter sources affect the behavior of our wave functions and argue how these effects within the context of quantum cosmology point to how a primordial electromagnetic field could have played a profound role in the early universe. 
\end{abstract}

\pacs{Valid PACS appear here}% PACS, the Physics and Astronomy
\maketitle

%\tableofcontents

\section{Introduction}
The Bianchi IX models possess a rich history and have been studied in a plethora of contexts. Investigations into them began with the works of Misner\cite{misner1969mixmaster}, Ryan\cite{ryan1972oscillatory}, and Belinskii, Khalatnikov and Lifshitz \cite{belinskii1970oscillatory}\cite{belinskii1978asymptotically}. What partly made these models so appealing was that they shared the 3-sphere spatial topology of the k=1 FLRW models, which were widely believed to be a good approximation for our physical universe until more precise cosmological observations \cite{de2002multiple} showed otherwise. In addition the equations which govern the dynamics of the Bianchi IX mini-superspace variables, which we will take to be the Misner variables $(\alpha,\beta_+,\beta_-)$ \cite{misner1969mixmaster}\cite{misner1969quantum} admit chaotic solutions \cite{barrow1982chaotic} \cite{chernoff1983chaos} \cite{cornish1997mixmaster}. It was originally thought that the chaos present near its singularity, which took the form of an erratic sequence of Kasner contractions and expansions could explain why we observe the universe to be mostly homogeneous, hence the origin of the name Mixmaster\cite{misner1969mixmaster}. 

Even though on large scales our universe is incredibly isotropic and homogeneous it is possible that our early universe possessed a considerable amount of anisotropy. Thus it is useful to study anistropic classical/quantum cosmologies so we can better understand what our universe could have been like when it was extremely young. To accomplish this we will study the quantum Taub models with matter sources in order to determine what possible effects an electromagnetic field and cosmological constant could have induced in our early universe. 

The quantum diagonalized Bianchi IX models were first studied by Misner \cite{misner1969quantum}, and later by Moncrief-Ryan \cite{moncrief1991amplitude}; among others in a plethora of different contexts, such as supersymmetric quantum cosmology\cite{graham1992supersymmetric}\cite{macias1993supersymmetric}. They were later revisited in the 2010s by Bae and Moncrief\cite{bae2015mixmaster}\cite{moncrief2014euclidean}, when they applied the Euclidean-signature semi classical method\cite{marini2019euclidean}\cite{moncrief2014euclidean} to prove for the wormhole case that a smooth and globally defined asymptotic solution exists for arbitrary Hartle-Hawking\cite{hartle1983wave} ordering parameter. They also studied the 'excited' states of the Bianchi IX models. 

Our focus will be on the Taub models \cite{taub1951empty}, otherwise known as the LRS Bianchi IX models, when a cosmological constant and a primordial electromagnetic field are present. In comparison to scalar fields and other matter sources, very little attention has been given to the study\cite{kobayashi2019early,jimenez2009cosmological,louko1988quantum,karagiorgos2018quantum} of primordial electromagnetic fields within the context of quantum cosmology. Considering that new evidence\cite{neronov2010evidence,tavecchio2010intergalactic} for the existence of a femto Gauss strength intergalactic magnetic field has been uncovered by observing gamma rays there is now an additional incentive to study what effects electromagnetic fields have on cosmological evolution. Through studying the effects of electromagnetic fields on quantum universes we can better understand nucleogenesis and how seeds of anisotropy developed in our early universe. In this work we will examine how a primordial electromagnetic field alters the behavior of our wave functions of the universe and extrapolate physical implications from them.

The classical Taub models and its extension to the Taub-NUT models have a rich history of their own, which includes being used to model the space-time around a black hole \cite{newman1963empty,kerner2006tunnelling}. Recently work has been conducted in finding new solutions to the symmetry reduced Wheeler DeWitt equation of the LRS Bianchi IX models \cite{karagiorgos2019quantum} using Killing vectors. Their quantum cosmology has also been investigated within the context of the generalized uncertainty principle \cite{battisti2008quantum} and using the WKB approximation\cite{de2020dynamics}. We will further expand upon what was previously done by finding a variety of asymptotic and closed form solutions to the Taub Wheeler DeWitt equation for both the vacuum case and when matter sources are present. In addition we will study in depth the 'excited' states of the quantum Taub models as was done for the Bianchi IX models \cite{bae2015mixmaster}. 

Classically the Taub models are defined via the following metric

\begin{equation}
\begin{aligned}
&d s^{2}=-N^{2}d t^{2}+\frac{L^{2}}{6\pi}e^{2 \alpha(t)}\left(e^{2 \beta(t)}\right)_{a b} \omega^{a} \omega^{b}\\& \left(e^{2 \beta(t)}\right)_{a b}=\operatorname{diag}\left(e^{2 \beta_{+}\left(t\right)}, e^{2 \beta_{+}\left(t\right)}, e^{-4 \beta_{+}\left(t\right)}\right).
\end{aligned}
\end{equation}
The $\omega^{i}$ terms are one forms defined on the spatial hypersurface of each Bianchi cosmology and obey $ d \omega^{i}=\frac{1}{2} C_{j k}^{i} \omega^{j} \wedge \omega^{k}$ where $C_{j k}^{i}$ are the structure
constants of the invariance Lie group associated with each particular class of Bianchi models. For the Taub models the one forms are 

\begin{equation}
\begin{aligned}
&\omega^1=\cos{\psi}d\theta+\sin{\psi}\sin{\theta}d\phi \\
&\omega^2=\sin{\psi}d\theta-\cos{\psi}\sin{\theta}d\phi \\
&\omega^3=d\psi+\cos{\theta}d\phi.
\end{aligned}
\end{equation}
Furthermore $L$ has units of length and sets a scale for the spatial size of our cosmology.

In Misner variables, for a particular class of operator ordering, the Wheeler DeWitt equation for the Taub models is the ordinary Bianchi IX Wheeler Dewitt equation when $\beta_{-}$ = 0, and the $\beta_{-}$ degree of freedom in the kinetic term is removed. Using the methodology presented in \cite{waller1984bianchi} \cite{uggla1995classifying} for writing out cosmological potentials for Hamiltonians of Bianchi A models with certain matter sources such as a cosmological constant and a primordial electromagnetic field, we write down the following symmetry reduced Wheeler DeWitt equation for the Taub models 

\begin{equation}
\frac{\partial^2 \Psi}{\partial \alpha^2}-B\frac{\partial \Psi}{\partial \alpha}-\frac{\partial^2 \Psi}{\partial \beta_+^{2}}+\left( \frac{e^{4 \alpha-8 \beta_+}}{3} \left(1-4 e^{6 \beta_+} \right)+\frac{2e^{6\alpha} \Lambda}{9\pi}+2b^{2}e^{2\alpha-4\beta_{+}}\right)\Psi=0,
\end{equation}
where B is the Hartle-Hawking \cite{hartle1983wave} ordering parameter, $b^{2}$ is the strength of the electromagnetic field; and $\Lambda$ is normalized differently than in \cite{uggla1995classifying}, and is dimensionless. The above Wheeler DeWitt equation can be obtained by applying canonical methods to the following action

\begin{equation}
\mathcal{S}=\frac{c^{3}}{16 \pi G} \int_{\Omega} \sqrt{-\operatorname{det}^{(4)} g} \left(^{4}R\left(^{(4)} g\right)-2\Lambda\right) d^{4} x +\mathcal{S}_{matter},
\end{equation}
resulting in a Hamiltonian which can be quantized.
The $\mathcal{S}_{matter}$ term is the classical electromagnetic action and is given in (24).

The Wheeler DeWitt (3) equation is the analogue to the time dependent Schr$\text{\" o}$dinger equation for Taub quantum cosmology. Viewing the Wheeler DeWitt(WDW) equation as $\hat{\mathcal{H}}_{\perp} \Psi=0$, and trying to relate it to the conventional Schr$\text{\" o}$dinger equation results in the problem of time manifesting itself as
\begin{equation}
i \hbar \frac{\partial \Psi}{\partial t}=N \hat{\mathcal{H}}_{\perp} \Psi=0,
\end{equation}
where $\frac{\partial \Psi}{\partial t}=0$. Due to the absence of the time derivative term of the Schr$\text{\" o}$dinger equation in the WDW equation, the construction of a unitary time evolution operator is not trivial, thus leading to the potential breakdown of a simple probabilistic interpretation of the wave function of the universe.

A Klein-Gordon current 
 \begin{equation}
\mathcal{J}=\frac{i}{2}\left(\Psi^{*} \nabla \Psi-\Psi \nabla \Psi^{*}\right)
\end{equation}
can be defined \cite{vilenkin1989interpretation} \cite{mostafazadeh2004quantum} which could be used to construct a probability density. It however, possesses unattractive features such as it vanishing when the wave function used to construct the current is purely real or imaginary and not always being positive definite. With our closed form solutions which we will present shortly, complex linear combinations can be constructed which yield non trivial Klein-Gordon currents which can be used to extrapolate physics from these wave functions.

Besides the issue of constructing a probability density function, it appears the quantized Hamiltonian constraint admits only zero's as eigenvalues. This may lead one to the conclusion that all of the states which satisfy the WDW equation possess vanishing energy. This on the surface makes it impossible to distinguish between ground and excited states because all states seemingly have the same energy. This apparent obstacle to delineate 'ground' and 'excited' states can be overcome by examining the nuanced nature of the ADM formalism \cite{arnowitt1959dynamical}. When cast in the ADM formalism general relativity is a constrained theory with four Lagrange multipliers, the lapse and the three components of the shift. The constraint associated with the lapse is due to general relativity being invariant under reparameterization of the evolution parameter. Likewise the constraint associated with the shift is due to diffeomorphism invariance and is called the diffeomorphism constraint. The diffeomorphism constraint is due to the configuration space $h_ {ab} $ being too large to the point of it being physically redundant. To remedy this one can define a superspace \cite{fischer1970theory} \cite{giulini2009superspace} where an equivalence class for $h_ {ab} $ is constructed such that two $h_ {ab} $ are in the same class if they can be carried into one another by a diffeomorphism. This shrinks the configuration space, allowing the diffeomorphism constraint to be satisfied. The same cannot be done for the reparameterization constraint \cite{wald1984general}. This explains why it wouldn't even make sense for a time derivative to be present because there is no unique "time" to use and partially explains the origins of the "problem of time". 

To get a better feel for what is going on, one can examine the vanishing Hamiltonian of a fully constrained system. One can formulate the Lagrangian of a free particle moving in one dimension , and introduce another configuration variable by defining the function t(T) where T is some arbitrary evolution parameter. If one were to treat both X(t(T)) and t(T) as configuration variables and formulate the system's Hamiltonian, they would notice that the Hamiltonian vanishes. Obviously the energy of a free one dimensional particle moving at a particular velocity cannot be zero. This is resolved by realizing that the dynamics of the system are now encoded in how X(t(T)) evolves with respect to t(T) where both are configuration variables. For an explicit demonstration of the above vanishing Hamiltonian construction, we refer the reader to \cite{rovelli2014covariant}. In other words, for these types of constrained systems the Hamiltonian no longer corresponds to the total energy. Thus the Hamiltonian constraint we quantized does not represent the total energy of a space-time in general relativity, and its vanishing eigenvalues do not mean that only states which possess zero energy are physically allowed. This allows leeway in defining 'ground' and 'excited' states in which features of ordinary quantum mechanics manifest as will be demonstrated in the next section.

A more in depth discussion in regards to how the Euclidean-signature semi classical method can be used to define 'ground' and 'excited' states despite them both being annihilated by the quantized Hamiltonian constraint can be found in \cite{moncrief2014euclidean}. To deal with the problem of time we will choose one of the Misner variables to be our "time" \cite{dewitt1967quantum}. A good time parameter increases monotonically, and out of the variables we can choose from $\alpha$ is the best candidate for "time", which roughly measures the size of our Taub universe as can be seen from (1).

This paper will be organized as follows. In the next section we will describe the Euclidean-signature semi classical method that we will use to solve (3)  for a plethora of cases. While discussing our modified semi-classical method we will define within the contexts of it 'ground' and 'excited' states for quantized homogeneous cosmological models. Then we will derive the electromagnetic potential term present in (3). Afterwards we will apply our formalism to the quantum Taub models when both a cosmological constant and a primordial electromagnetic field are present to obtain a closed form 'ground' state solution. In addition we will find two closed form 'excited' state solutions when $\Lambda=0$ and $b\ne 0$ which can aid us in understanding how electromagnetic fields could have impacted the evolution of the early universe. We will then prove that a sequence composed of a countably infinite number of  'excited' states exist as well and express them in closed form. Afterwards we will prove the existence of 'ground' state asymptotic solutions for the 'wormhole' \cite{moncrief1991amplitude} and 'no boundary'  \cite{graham1993supersymmetric} cases and analyze their 'excited' states. Finally we will qualitatively discuss the plethora of wave functions computed in this work, while giving special attention to the physical implications of a primordial electromagnetic field within the context of our results.

\section{\label{sec:level1}The Euclidean-signature semi classical method} 
Our outline of this method will follow closely \cite{moncrief2014euclidean} and will be tailored for the quantum Taub models. For an outline of the method which applies to the general Bianchi A models we refer the reader to the aforementioned original source. 
 
The first step we will take in solving the WDW equation is to introduce the ansatz
\begin{equation}
\stackrel{(0)}{\Psi}_{\hbar}=e^{-S_{\hbar} / \hbar}
\end{equation}
where $S_{\hbar}$ is a function of $\left(\alpha,\beta_+\right)$. We will rescale $S_{\hbar}$ in the following way  
\begin{equation}
\mathcal{S}_{\hbar} :=\frac{G}{c^{3} L^{2}} S_{\hbar}
\end{equation}
where $\mathcal{S}_{\hbar}$ is dimensionless and admits the following power series in terms of this dimensionless parameter
\begin{equation}
X :=\frac{L_{\text { Planck }}^{2}}{L^{2}}=\frac{G \hbar}{c^{3} L^{2}}.
\end{equation}
The series is given by 
\begin{equation}
\mathcal{S}_{\hbar}=\mathcal{S}_{(0)}+X \mathcal{S}_{(1)}+\frac{X^{2}}{2 !} \mathcal{S}_{(2)}+\cdots+\frac{X^{k}}{k !} \mathcal{S}_{(k)}+\cdots,
\end{equation}
and as a result our initial ansatz now takes the following form 
\begin{equation}
\stackrel{(0)}{\Psi}_{\hbar}=e^{-\frac{1}{X} \mathcal{S}_{(0)}-\mathcal{S}_{(1)}-\frac{X}{2 !} \mathcal{S}_{(2)}-\cdots}
\end{equation}.

Substituting this ansatz into the WDW equation and
requiring satisfaction order-by-order in powers of X leads immediately to the sequence of equations

\begin{equation}
\begin{aligned}
&{\left(\frac{\partial \mathcal{S}_{(0)}}{\partial \alpha}\right)^{2}-\left(\frac{\partial \mathcal{S}_{(0)}}{\partial \beta_{+}}\right)^{2}}+\frac{e^{4 \alpha-8 \beta_+}}{3} \left(1-4 e^{6 \beta_+} \right)+\frac{2e^{6\alpha} \Lambda}{9\pi}+2b^{2}e^{2\alpha-4\beta_{+}}=0,
\end{aligned}
\end{equation}
\begin{equation}
\begin{aligned}
& 2\left[\frac{\partial \mathcal{S}_{(0)}}{\partial \alpha} \frac{\partial \mathcal{S}_{(1)}}{\partial \alpha}-\frac{\partial \mathcal{S}_{(0)}}{\partial \beta_{+}} \frac{\partial \mathcal{S}_{(1)}}{\partial \beta_{+}}\right]  +B \frac{\partial \mathcal{S}_{(0)}}{\partial \alpha}-\frac{\partial^{2} \mathcal{S}_{(0)}}{\partial \alpha^{2}}+\frac{\partial^{2} \mathcal{S}_{(0)}}{\partial \beta_{+}^{2}}=0,
\end{aligned}
\end{equation},
\begin{equation}
\begin{aligned}
& 2\left[\frac{\partial \mathcal{S}_{(0)}}{\partial \alpha} \frac{\partial \mathcal{S}_{(k)}}{\partial \alpha}-\frac{\partial \mathcal{S}_{(0)}}{\partial \beta_{+}} \frac{\partial \mathcal{S}_{(k)}}{\partial \beta_{+}}\right]  {+k\left[B \frac{\partial \mathcal{S}_{(k-1)}}{\partial \alpha}-\frac{\partial^{2} \mathcal{S}_{(k-1)}}{\partial \alpha^{2}}+\frac{\partial^{2} \mathcal{S}_{(k-1)}}{\partial \beta_{+}^{2}}\right]} \\ & + \sum_{\ell=1}^{k-1} \frac{k !}{\ell !(k-\ell) !}\Biggr(\frac{\partial \mathcal{S}_{(\ell)}}{\partial \alpha} \frac{\partial \mathcal{S}_{(k-\ell)}}{\partial \alpha}-\frac{\partial \mathcal{S}_{(\ell)}}{\partial \beta_{+}} \frac{\partial \mathcal{S}_{(k-\ell)}}{\partial \beta_{+}}\Biggl) =0
\end{aligned}
\end{equation}
We will refer to $\mathcal{S}_{(0)}$ in our WDW wave functions as the leading order term, which can be used to construct a semi-classical approximate solution to the Lorentzian signature WDW equation, and call $\mathcal{S}_{(1)}$ the first order term. The $\mathcal{S}_{(1)}$ term can also be viewed as our first quantum correction, with the other $\mathcal{S}_{(k)}$ terms being additional higher order quantum corrections.  

If one can find a solution to the $\mathcal{S}_{(1)}$ equation which allows the $\mathcal{S}_{(2)}$ equation to be satisfied by zero, then one can write down the following as an exact solution to the WDW equation for either a particular value of the Hartle-Hawking ordering parameter, or for an arbitrary ordering parameter depending on the properties of the $\mathcal{S}_{(1)}$ which is found.   

\begin{equation}
\stackrel{(0)}{\Psi}_{\hbar}=e^{-\frac{1}{X} \mathcal{S}_{(0)}-\mathcal{S}_{(1)}}
\end{equation}.
This can be easily shown. Lets take $\mathcal{S}_{(0)}$ and $\mathcal{S}_{(1)}$ as arbitrary known functions which allow the $\mathcal{S}_{(2)}$ transport equation to be satisfied by zero then the $k=3$ transport equation can be expressed as 
\begin{equation}
{2\left[\frac{\partial \mathcal{S}_{(0)}}{\partial \alpha} \frac{\partial \mathcal{S}_{(3)}}{\partial \alpha}-\frac{\partial \mathcal{S}_{(0)}}{\partial \beta_{+}} \frac{\partial \mathcal{S}_{(3)}}{\partial \beta_{+}}\right]}=0
\end{equation}
which is clearly satisfied by $\mathcal{S}_{(3)}$=0. The $\mathcal{S}_{(4)}$ equation can be written in the same form and one of its solution is 0 as well, thus resulting in the $\mathcal{S}_{(5)}$ equation possessing the same form as (16). One can easily convince oneself that this pattern continues for all of the $k\geq 3$ $\mathcal{S}_{(k)}$ transport equations as long as the solution of the $\mathcal{S}_{(k-1)}$ transport equation is chosen to be 0. Thus in some situations an $\mathcal{S}_{(1)}$ exists which allows one to set the solutions to all of the higher order transport equations to zero, which results in the infinite sequence of transport equations generated by our ansatz truncating to a finite sequence of equations whose solutions allow us to construct a closed form wave function satisfying the WDW equation.

Not all solutions to the $\mathcal{S}_{(1)}$ transport equation will allow the $\mathcal{S}_{(2)}$ transport equation to be satisfied by zero; however for the case when a cosmological constant and an electromagnetic field are present, we were able to find an $\mathcal{S}_{(1)}$ which causes the $\mathcal{S}_{(2)}$ transport equation to be satisfied by zero when $B=0$, thus allowing one to set all of the solutions to the higher order transport equations to zero as shown above. This will enable us to construct a closed form 'ground' state solution to the Lorentzian signature Taub WDW equation for the case when our aforementioned matter sources are present. Furthermore this will allow us to prove that a sequence composed of a countably infinite number of closed form 'excited' state solutions exist for arbitrary values of the cosmological constant $\Lambda$. 

To calculate 'excited' states we introduce the following ansatz. 
\begin{equation}
{\Psi}_{\hbar}={\phi}_{\hbar} e^{-S_{\hbar} / \hbar}
\end{equation}
where $$
S_{\hbar}=\frac{c^{3} L^{2}}{G} \mathcal{S}_{\hbar}=\frac{c^{3} L^{2}}{G}\left(\mathcal{S}_{(0)}+X \mathcal{S}_{(1)}+\frac{X^{2}}{2 !} \mathcal{S}_{(2)}+\cdots\right)
$$
is the same series expansion as before and ${\phi}_{\hbar}$ can be expressed as the following series 
\begin{equation}
{\phi_{\hbar}=\phi_{(0)}+X \phi_{(1)}+\frac{X^{2}}{2 !} \phi_{(2)}+\cdots+\frac{X^{k(*)}}{k !} \phi_{(k)}+\cdots}
\end{equation}
with X being the same dimensionless quantity as before. 
Inserting $\left(17\right)$ with the expansions given by $\left(10\right)$ and $\left(18\right)$ into the WDW equation and by matching equations in powers of X leads to the following sequence of equations. 
\begin{equation}
-\frac{\partial \phi_{(0)}}{\partial \alpha} \frac{\partial \mathcal{S}_{(0)}}{\partial \alpha}+\frac{\partial \phi_{(0)}}{\partial \beta_{+}} \frac{\partial \mathcal{S}_{(0)}}{\partial \beta_{+}}=0,
\end{equation}
\begin{equation}
\begin{aligned}
&{-\frac{\partial \phi_{(1)}}{\partial \alpha} \frac{\partial \mathcal{S}_{(0)}}{\partial \alpha}+\frac{\partial \phi_{(1)}}{\partial \beta_{+}} \frac{\partial \mathcal{S}_{(0)}}{\partial \beta_{+}}} {+\left(-\frac{\partial \phi_{(0)}}{\partial \alpha} \frac{\partial \mathcal{S}_{(1)}}{\partial \alpha}+\frac{\partial \phi_{(0)}}{\partial \beta_{+}} \frac{\partial \mathcal{S}_{(1)}}{\partial \beta_{+}}\right)} \\ & {+\frac{1}{2}\left(-B \frac{\partial \phi_{(0)}}{\partial \alpha}+\frac{\partial^{2} \phi_{(0)}}{\partial \alpha^{2}}-\frac{\partial^{2} \phi_{(0)}}{\partial \beta_{+}^{2}}\right)=0},
\end{aligned}
\end{equation}
\begin{equation}
\begin{aligned}
& -\frac{\partial \phi_{(k)}}{\partial \alpha} \frac{\partial \mathcal{S}_{(0)}}{\partial \alpha}+\frac{\partial \phi_{(k)}}{\partial \beta_{+}} \frac{\partial \mathcal{S}_{(0)}}{\partial \beta_{+}}
+k\Biggr(-\frac{\partial \phi_{(k-1)}}{\partial \alpha} \frac{\partial \mathcal{S}_{(1)}}{\partial \alpha}+\frac{\partial \mathcal{S}_{(1)}}{\partial \beta_{+}} \frac{\partial \mathcal{S}_{(1)}}{\partial \beta_{+}}\Biggr) \\ & 
+\frac{k}{2}\Biggr(-B \frac{\partial \phi_{(k-1)}}{\partial \alpha}+\frac{\partial^{2} \phi_{(k-1)}}{\partial \alpha^{2}}-\frac{\partial^{2} \phi_{(k-1)}}{\partial \beta_{+}^{2}}-\frac{\partial^{2} \phi_{(k-1)}}{\partial \beta_{-}^{2}}\Biggr)\\ & -
\sum_{\ell=2}^{k} \frac{k !}{\ell !(k-\ell) !}\Biggr( \frac{\partial \phi_{(k-\ell)}}{\partial \alpha} \frac{\partial \mathcal{S}_{(\ell)}}{\partial \alpha}-\frac{\partial \phi_{(k-\ell)}}{\partial \beta_{+}} \frac{\partial \mathcal{S}_{(\ell)}}{\partial \beta_{+}}\Biggr) =0.
\end{aligned}
\end{equation}

It can be seen from computing $\frac{d\phi_{(0)}\left(\alpha,\beta_+\right)}{dt}=\dot{\alpha}\frac{\partial \phi_{(0)}}{\partial \alpha}+\dot{\beta_+}\frac{\partial \phi_{(0)}}{\partial \beta_+}$, and inserting $\left(4.9, \hspace{1 mm} 4.18-4.20\right)$ from\cite{moncrief2014euclidean} that $\phi_{(0)}$ is a conserved quantity under the flow of $S_{0}$. This means that any function $F\left(\phi_{(0)}\right)$ is also a solution of equation $\left(19\right)$. Wave functions constructed from these functions of $\phi_{0}$ are only physical if they are smooth and globally defined. Beyond the semi-classical limit, if smooth globally defined solutions can be proven to exist for the higher order $\phi$ transport equations then one can construct asymptotic or closed form 'excited' state solutions for the quantum Taub models.

If our 'excited' states ${\Psi}_{\hbar}={\phi}_{\hbar} e^{-S_{\hbar} / \hbar}$ behave as bound states then they qualitatively in the sense of their mathematical structure possess the same form as excited states for the quantum harmonic oscillator $ \psi_{n}(x)=H_{n}\left(\sqrt{\frac{m \omega}{\hbar}} x\right)\frac{1}{\sqrt{2^{n} n !}}\left(\frac{m \omega}{\pi \hbar}\right)^{1 / 4}  e^{-\frac{m \omega x^{2}}{2 \hbar}}$, where $H_n$ are the Hermite polynomials in which n is a positive integer which specifies its form. Because the solutions of the $\phi_{(0)}$ equation are quantities conserved along the flow generated by $\mathcal{S}_{(0)}$, any multiple $\phi^{n}_{(0)}$ also satisfies equation $\left(19\right)$. On purely physical grounds the amount of numbers required to specify an 'excited' state equals the number of excitable degrees of freedom present. For the Taub models with non dynamical matter sources that number is one which corresponds to the one anistropic degree of freedom. As a result our $\phi_{(0)}$ which distinguishes 'excited' states from 'ground' states has the following form $f\left(\alpha,\beta_+\right)^{m}$; where $f\left(\alpha,\beta_+\right)$ is a conserved quantity satisfying equation $\left(19\right)$ and $m$ is a quantity which can plausibly be interpreted as a graviton excitation number\cite{bae2014quantizing}. If $f\left(\alpha,\beta_+\right)$ vanishes at some point or points in minisuperspace then to ensure that our wave function is smooth and globally defined we must restrict $m$ to be a positive integer which results in our 'excited' states being 'bound' states just like the quantum harmonic oscillator. This discretization of the quantity that denotes our 'excited' states is the mathematical manifestation of quantization one would expect excited states to possess in quantum dynamics. If our conserved quantity $f\left(\alpha,\beta_+\right)$ does not vanish in minisuperspace then our 'excited' states are 'scattering' states akin to the quantum free particle and $m$ can be any real number. Additional information for why we can call the above 'excited' states despite them being solutions to an equation which does not have the same form as the Schr$\text{\" o}$dinger equation can be found in \cite{moncrief2014euclidean}. In what follows we will set $X=1$.  

\section{\label{sec:level1}Electromagnetic Potentials For The Taub Models}

In this section we will compare two methods for obtaining the WDW equation (3). The first method will be based on directly quantizing the class of classical Hamilitonians for Bianchi A models that was developed in \cite{waller1984bianchi} and repurposing it for the Taub models. This will lead to a semi-classical treatment of our electromagnetic degree of freedom and will be what we use in the following sections to analyze how matter sources affect our wave functions. However we will also do a full quantum treatment of the electromagnetic degree of freedom and compare the two approaches. We will assume all of our electric and magnetic fields are parallel to each other as is justified in \cite{jacobs1970homogeneous}.

With this in mind our first task is to obtain solutions for Maxwell's equations in the space-time (1) in terms of the Misner variables. In our calculations we will set $L=\sqrt{6\pi} \ell$ where $\ell$ is a quantity which equals 1 and has units of length. Starting from 
\begin{equation}
\boldsymbol{A}=A_{0}dt+A_{1}\omega^{1}+A_{2}\omega^{3}+A_{3}\omega^{3}
\end{equation}
and using the fact that $  d\omega^{i}=\frac{1}{2} C_{j k}^{i} \omega^{j} \wedge \omega^{k}$
to aide us in computing $\boldsymbol{F}=d\boldsymbol{A}= \frac{1}{2}F_{\mu v}\omega^{\mu} \wedge \omega^{v}$ results in the following expression for $F_{\mu v}$
\begin{equation}
 F_{\mu v}=A_{v,\mu}-A_{ \mu,v}+A_{\alpha} C_{\mu v}^{\alpha}.
\end{equation} 
In (23) differentiation is done through a vector dual acting on our one forms $\omega^{\mu}$ which we denote as $X_{\mu}$. Thus $A_{v, \mu}=X_{\mu}A_{v}$. The action for the electromagnetic contribution (4) is the following
\begin{equation}
\mathcal{S}_{matter}=\int dt dx^{3}N\sqrt{h}\left(-\frac{1}{16 \pi} F_{\mu \nu} F^{\mu \nu}\right),
\end{equation}
where 
\begin{equation}
\begin{aligned}
h_{ab}=e^{2\alpha(t)}\operatorname{diag}\left(e^{2 \beta\left(t\right)_{+}}, e^{2 \beta\left(t\right)_{+}}, e^{-4 \beta\left(t\right)_{+}}\right).
\end{aligned}
\end{equation}
Writing the action (24) in terms of its vector potential $A$ and our structure constants results in the Lagrangian density which is derived in \cite{waller1984bianchi}
\begin{equation}
\begin{aligned}
&\mathscr{L}=\Pi^{s} A_{0, s}-NH \\&
\mathscr{L}=\Pi^{s} A_{0, s}-\Pi^{s} A_{s, 0}-N\frac{2 \pi}{\sqrt{h}} \Pi^{s} \Pi^{p} h_{s p} \\& - \frac{N\sqrt{h}}{16 \pi} h^{i k} h^{s l}\left(2A_{[i,s]}+A_{m} C^{m}_{i s}\right)\left(2A_{[k,l]}+A_{m} C^{m}_{k l}\right),
\end{aligned}
\end{equation}
where
\begin{equation}
\Pi^{s}=\frac{\partial \mathscr{L} }{\partial\left(X_{0} A_{s}\right)}=\frac{ h^{s j}\sqrt{h}}{4 N\pi} \left(-A_{0, j}+A_{j, 0}+A_{\alpha} C^{\alpha}_{0 j}\right).
\end{equation}
We allow the shift $N^{k}$ to vanish. If we invoke the homogeneity of (1) then we can say that $A_{i, j}=0$, and $ A_{0, j}=0$ which results in (26) simplifying to 

\begin{equation}
\mathscr{L}=\Pi^{s} A_{s, 0}-N\left[\frac{2 \pi}{\sqrt{h}} \Pi^{s} \Pi^{p} h_{p s}+\frac{\sqrt{h}}{16 \pi} h^{i k} h^{s l} C^{m}_{k l}C^{n}_{i s} A_{m} A_{n}\right].
\end{equation}
The structure constants for the Bianchi IX and the Taub models are 
\begin{equation}
\begin{aligned}
C_{j k}^{i}=\epsilon_{ijk}
\end{aligned}
\end{equation}
We will now set $A_{2}$, $A_{3}$, $\Pi^{2}$, and $\Pi^{3}$ to zero as is justified in \cite{waller1984bianchi}  and only consider the electromagnetic field produced by $A_{1}$ and $\Pi^{1}$; doing so results in the following Lagrangian density
\begin{equation}
\mathscr{L}=\Pi^{1} A_{1,0}-N\left[\frac{2 \pi}{\sqrt{h}} \Pi^{1} \Pi^{1} h_{1 1}+\frac{\sqrt{h}}{16 \pi} h^{i k} h^{s l} C^{1}_{k l}C^{1}_{i s} A_{1} A_{1}\right].
\end{equation}
As the reader can easily verify $ h^{i k} h^{s l}\epsilon_{1kl}\epsilon_{1is}=\frac{2h_{11}}{h}$, where $h$ is the determinant of (25). This allows us to obtain the following set of Maxwell's equations when $A_{1}$ and $\Pi^{1}$ are varied in (30)
\begin{equation}
\dot{A}_{1}-4 \pi \frac{1}{\sqrt{h}} \Pi^{1}h_{11}=0
\end{equation}
and 
\begin{equation}
\dot{\Pi}^{1}+\frac{1}{4 \pi} \frac{1}{\sqrt{h}} h_{1 1} A_{1}=0.
\end{equation}
For the last equation we applied an integration by parts to the term $\Pi^{1} A_{0, 1}$ and dropped the total derivative which vanishes at the spatial  boundary. The solutions for (31) and (32) are 
\begin{equation}
A_{1}=\sqrt{2}B_{0}cos(\theta(t))
\end{equation}
\begin{equation}
\Pi^{1}=\frac{1}{2\sqrt{2}\pi}B_{0}sin(\theta(t))
\end{equation}
where $\theta(t)$ is an integral which is immaterial for our purposes and $B_{0}$ is an integration constant. Inserting (33) and (34) back into (30) results in 
\begin{equation}
\begin{aligned}
\mathscr{L}=\Pi^{1} A_{1,0}-N\frac{B_{0}^{2}}{4\pi\sqrt{h}}h_{11}\left(sin(\theta(t))^{2}+cos(\theta(t))^{2}\right)  =\Pi^{1} A_{1,0}-N\frac{B_{0}^{2}}{4\pi}e^{-\alpha(t)+2\beta(t)_{+}},
\end{aligned}
\end{equation}
From (26) we can easily identify the electromagnetic Hamiltonian as 
\begin{equation}
\begin{aligned}
H_{em}=\frac{B_{0}^{2}}{4\pi}e^{-\alpha(t)+2\beta(t)_{+}},
\end{aligned}
\end{equation}
which can be added to the gravitational Hamiltonian constraint resulting from the action we had earlier (4)
\begin{equation}
\begin{aligned}
&e^{-3 \alpha(t)}\left(-p_{\alpha}^{2}+p_{+}^{2}+p_{-}^{2}\right)+U_{g}+\frac{B_{0}^{2}}{4\pi}e^{-\alpha(t)+2\beta(t)_{+}+2\sqrt{3}\beta(t)_{-}}\\&=0.
\end{aligned}
\end{equation}
Quantizing (37) using the factor ordering we chose before, multiplying each side by $e^{3\alpha(t)}$, and rescaling $B_{0}$ results in a WDW equation with the electromagnetic potential
\begin{equation}
\begin{aligned}
U_{1 \hspace{1mm} em }=2b^{2}e^{2\alpha+2\beta_{+}}.
\end{aligned}
\end{equation}
As the reader can verify for the cases when $A_{1}=0$, $A_{3}=0$, $\Pi^{1}=0$, and $\Pi^{3}=0$ the resulting electromagnetic term is 
\begin{equation}
\begin{aligned}
U_{2 \hspace{1mm} em }=2b^{2}e^{2\alpha+2\beta_{+}},
\end{aligned}
\end{equation}
and the case when $A_{1}=0$, $A_{2}=0$, $\Pi^{1}=0$, and $\Pi^{2}=0$ results in the term reported in \cite{waller1984bianchi}
\begin{equation}
\begin{aligned}
U_{3 \hspace{1mm} em }=2b^{2}e^{2\alpha-4\beta_{+}},
\end{aligned}
\end{equation}
which is the potential term which appears in (3). 

If we start with (30) and directly quantize our component of the total Hamiltonian constraint which is proportional to the lapse N we obtain a similar, but slightly different contribution to the potential. Simplifying the term in brackets of (30) results in the following contribution to the Hamiltonian constraint derived with (4)

\begin{equation}
\begin{aligned}
H_{em}=N\left[\frac{e^{-\alpha+2\beta_{+}}\left(16 \pi ^2 \Pi^{1}\Pi^{1}+A^{2}_{1}\right)}{8 \pi }\right].
\end{aligned}
\end{equation}
The term $\left(16 \pi ^2 \Pi^{1}\Pi^{1}+A^{2}_{1}\right)$ commutes with our total Hamiltonian constraint $H_{gravity} +H_{em}$. Thus we can solve the following WDW equation constructed by directly quantizing (41) with the rest of our constraint 

\begin{equation}
\frac{\partial^2 \Psi}{\partial \alpha^2}-\frac{\partial^2 \Psi}{\partial \beta_+^{2}}-e^{2\alpha+2\beta_{+}}\left(2\pi \frac{\partial^2 \Psi}{\partial A^{2}_{1}}+\frac{1}{8\pi}A^{2}_{1}\Psi\right)+\left( \frac{e^{4 \alpha-8 \beta_+}}{3} \left(1-4 e^{6 \beta_+} \right)+\frac{2e^{6\alpha} \Lambda}{9\pi}\right)\Psi=0
\end{equation}
by first solving this eigenvalue problem 
\begin{equation}
-2\pi \frac{\partial^2 \Psi}{\partial A^{2}_{1}}+\frac{1}{8\pi}A^{2}_{1}\Psi=b_{n}\Psi,
\end{equation}
where $\Psi$ is a function of $\alpha$, $\beta_{+}$, and $A^{2}_{1}$. This is simply the Schr$\text{\" o}$dinger equation for a harmonic oscillator whose solutions are well known
\begin{equation}
\begin{aligned}
&\Psi=\psi\left(\alpha,\beta_{+}\right)e^{-\frac{A^{2}_{1}}{8 \pi }} H_{b_{n}-\frac{1}{2}}\left(\frac{A_{1}}{2 \sqrt{\pi }}\right)\\& b_{n}=\frac{1}{2}\left(1+2n\right).
\end{aligned}
\end{equation}
Inserting our $\Psi$ from (44) into (42) yields 
\begin{equation}
\frac{\partial^2 \psi}{\partial \alpha^2}-B\frac{\partial \psi}{\partial \alpha}-\frac{\partial^2 \psi}{\partial \beta_+^{2}}+ b_{n}e^{2\alpha+2\beta_{+}}\psi+\left( \frac{e^{4 \alpha-8 \beta_+}}{3} \left(1-4 e^{6 \beta_+} \right)+\frac{2e^{6\alpha} \Lambda}{9\pi}\right)\psi=0
\end{equation}
where the ordering parameter B has been reinstated. This WDW equation is similar to what we had before except for the fact that the strength $b_{n}$ of the electromagnetic field is now quantized thanks to (44). By first solving the classical $A_{i}$ equations (33) in terms of the Misner variables we eliminate the electromagnetic field degree of freedom. By keeping it we can in theory study much more general quantum cosmologies for anisotropic models that involve electromagnetic fields. For now though working with just (37) is sufficient for what will follow.

\section{\label{sec:level1}Closed Form 'Ground' And 'Excited' State Solutions For The Quantum Taub Models With A Cosmological Constant And An Electromagnetic Field}  

The author found the following $\mathcal{S}_{(0)}$ associated with (3)

\begin{equation}
\begin{aligned}
\mathcal{S}_{(0)} := \text{b}^2 (-\alpha-\beta_{+})-\frac{\Lambda e^{4 (\alpha+\beta_{+})}}{36 \pi }+\frac{1}{6} \left(2 e^{6 \beta_{+}}+1\right) e^{2\alpha-4 \beta_{+}},
\end{aligned}
\end{equation}
which is the standard 'wormhole' solution with two additional terms for our matter sources. Inserting (46) into our $\mathcal{S}_{(1)}$ equation results in

\begin{equation}
\begin{aligned}
&-9\pi \text{b}^2 (2  \frac{\partial \mathcal{S}_{(1)}}{\partial \alpha}-2  \frac{\partial \mathcal{S}_{(1)}}{\partial \beta_+}+\text{B})-\Lambda e^{4 (\alpha+\beta_{+})} (2  \frac{\partial \mathcal{S}_{(1)}}{\partial \alpha}-2  \frac{\partial \mathcal{S}_{(1)}}{\partial \beta_+}+\text{B})\\&+6\pi e^{2 (\alpha+\beta_{+})} (2  \frac{\partial \mathcal{S}_{(1)}}{\partial \alpha}-2  \frac{\partial \mathcal{S}_{(1)}}{\partial \beta_+}+\text{B})+3\pi e^{2 \alpha-4 \beta_{+}} (2  \frac{\partial \mathcal{S}_{(1)}}{\partial \alpha}+4  \frac{\partial \mathcal{S}_{(1)}}{\partial \beta_+}+\text{B}+6)=0,
\end{aligned}
\end{equation}
which is satisfied by the very simple expression 

\begin{equation}
\begin{aligned}
\mathcal{S}_{(1)} := \left(-\frac{\text{B}}{2}-1\right)\alpha -\beta_+,
\end{aligned}
\end{equation}
that surprisingly is independent of $\Lambda$ and $b$.

There are infinitely many solutions to equation (47); and it is possible that some of them do a better job of aiding one in finding solutions to the WDW equation than (48). Keeping that in mind we will proceed using (48) and insert it into equation (14), yielding

\begin{equation}
\begin{aligned}
&-9 \pi  \left(4 \text{b}^2 ( \frac{\partial \mathcal{S}_{(2)}}{\partial \alpha}- \frac{\partial \mathcal{S}_{(2)}}{\partial \beta_+})+\text{B}^2\right)-4 \Lambda e^{4 (\alpha+\beta_{+})} ( \frac{\partial \mathcal{S}_{(2)}}{\partial \alpha}- \frac{\partial \mathcal{S}_{(2)}}{\partial \beta_+})\\&+12
   \pi  e^{2 \alpha-4 \beta_{+}} \left(2 e^{6 \beta_{+}} ( \frac{\partial \mathcal{S}_{(2)}}{\partial \alpha}- \frac{\partial \mathcal{S}_{(2)}}{\partial \beta_+})+ \frac{\partial \mathcal{S}_{(2)}}{\partial \alpha}+2  \frac{\partial \mathcal{S}_{(2)}}{\partial \beta_+}\right)=0.
\end{aligned}
\end{equation}

When $B=0$ this equation is satisfied by $\mathcal{S}_{(2)}=0$ which as was previously shown allows us to construct the following closed form solution to the Taub WDW equation.

\begin{equation}
\begin{aligned}
\Psi_{matter}=e^{ \left(\left(\text{b}^2+1\right) (\alpha+\beta_{+})+\frac{\Lambda e^{4 (\alpha+\beta_{+})}}{36 \pi }-\frac{1}{6} \left(2 e^{6
   \beta_{+}}+1\right) e^{2 \alpha-4 \beta_{+}}\right)}.
\end{aligned}
\end{equation}
Comparing (50) to our two ans$\text{\" a}$tze (17) and (7) indicates that our closed form solution is a 'ground' state. Figures 1(a) to 1(d) further support this. Because our $\mathcal{S}_{(0)}$ is related to the 'wormhole' solution, by setting $\Lambda=0$ and $b=0$ we recover the Taub wave function found earlier by Moncrief and Ryan\cite{moncrief1991amplitude}. In our discussion section will analyze the physical implications of the behavior of these wave functions(figs 1a-1d) as our matter sources $\Lambda$ and $b^{2}$ are varied.

\begin{figure}
\centering
\begin{subfigure}{.4\textwidth}
  \centering
  \includegraphics[scale=.12]{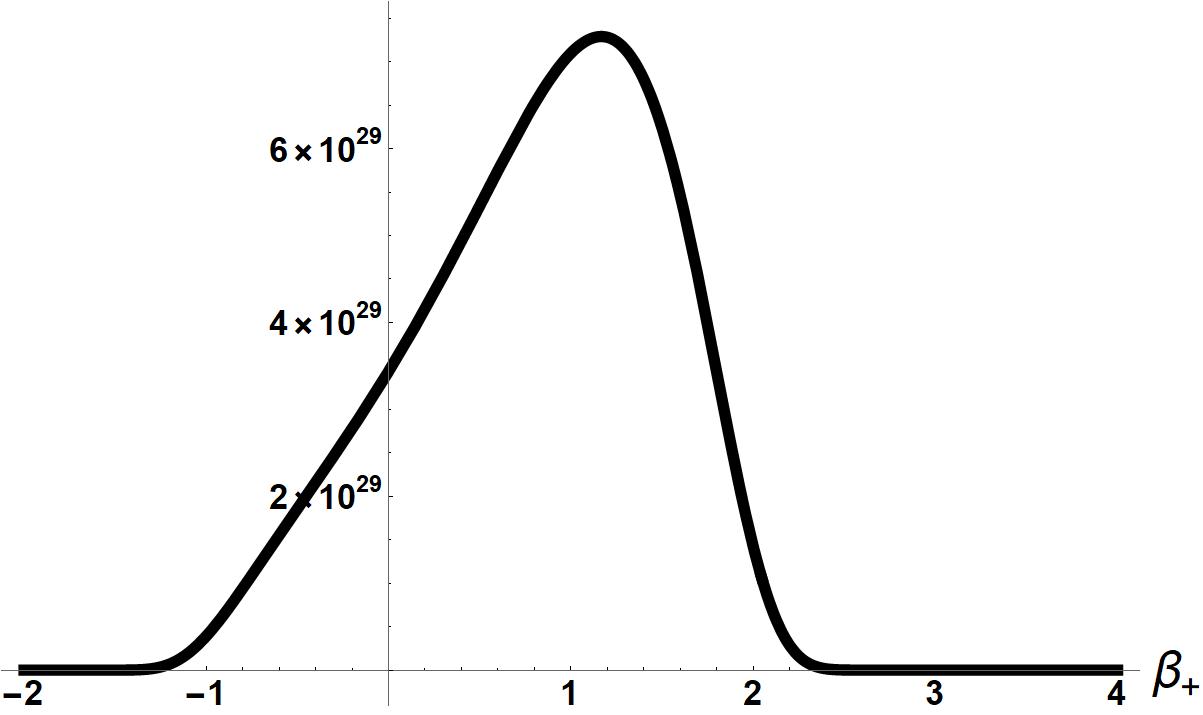}
  \caption{ $\alpha=-1$ \hspace{1mm} $\Lambda=-1$ \hspace{1mm} $b=0$}
  \label{fig:sub1}
\end{subfigure}%
\begin{subfigure}{.4\textwidth}
  \centering
  \includegraphics[scale=.12]{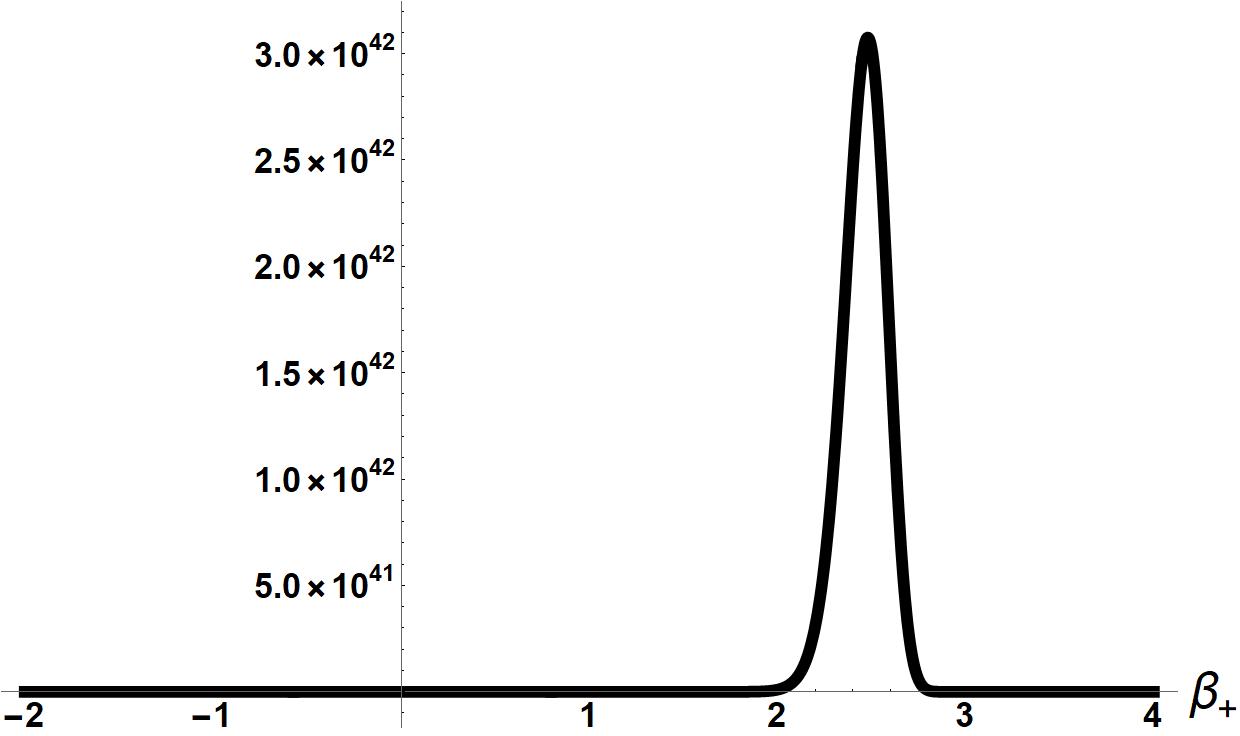}
  \caption{ $\alpha=-1$ \hspace{1mm} $\Lambda=-1$ \hspace{1mm} $b=5$}
  \label{fig:sub2}
\end{subfigure}
\begin{subfigure}{.4\textwidth}
  \centering
  \includegraphics[scale=.17]{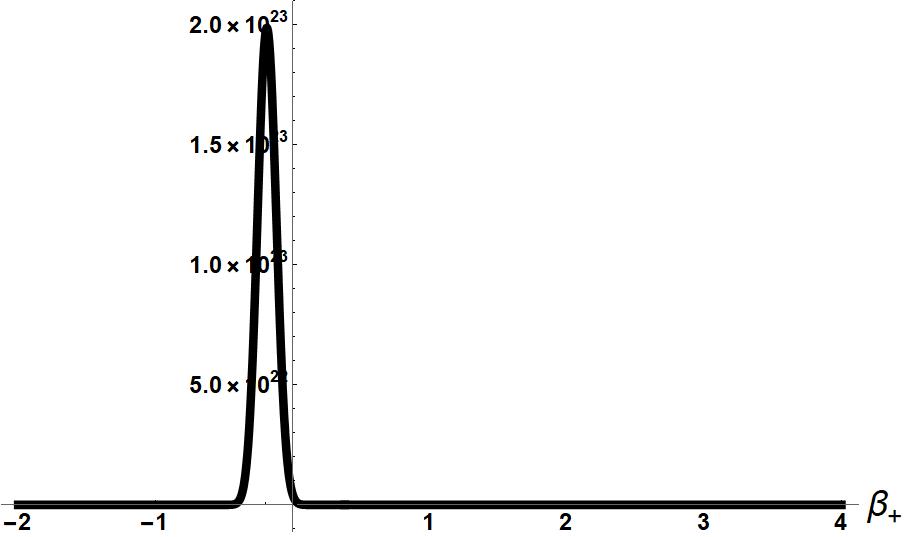}
  \caption{ $\alpha=1.5$ \hspace{1mm} $\Lambda=-3$ \hspace{1mm} $b=0$}
  \label{fig:sub1}
\end{subfigure}%
\begin{subfigure}{.4\textwidth}
  \centering
  \includegraphics[scale=.17]{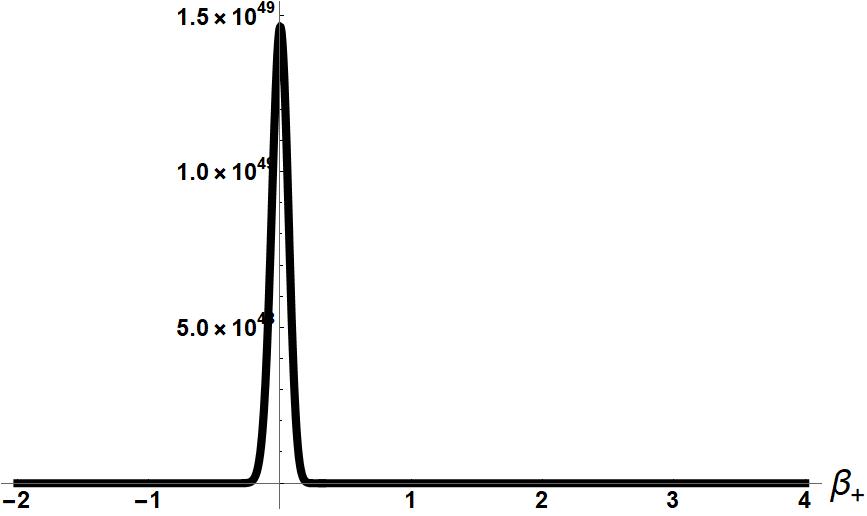}
  \caption{ $\alpha=1.5$ \hspace{1mm} $\Lambda=-3$ \hspace{1mm} $b=6.5$}
  \label{fig:sub1}
\end{subfigure}%
\caption{Plots of (50) when our matter sources and $\alpha$ are varied}
\label{fig:test}
\end{figure}

As of the writing of this paper the author found two 'excited' state solutions for the quantum Taub models when an electromagnetic field is present. Using (46) when $\left(\Lambda=0\right)$, the author found the following family of conserved quantities which satisfies (19) 

\begin{equation}
\begin{aligned}
\phi_{(0)}:=\left(-3 \text{b}^2 e^{2 (\alpha+\beta_{+})}-e^{4 \alpha-2 \beta_{+}}+e^{4 (\alpha+\beta_{+})}\right)^{n}.
\end{aligned}
\end{equation}
Because (51) vanishes for finite values of the Misner variables our 'excited' states constructed using it are bound states. In this paper we will find and analyze the first  two closed form 'excited' states. Later though in a future work the author will prove that a countably infinite number of 'excited' states exist when both an electromagnetic field and cosmological constant are present.

For the first 'excite' state when $n=1$ the author found the following solution to the the $\phi_{1}$ equation which causes the rest of the $\phi_{k}$ equations to be satisfied by zero as will be explained in more detail in the next section

\begin{equation}
\begin{aligned}
\phi_{(1)}:=9 \text{b}^2-6 e^{2 (\alpha+\beta_{+})}.
\end{aligned}
\end{equation}
For the case when $n=2$ the author had to solve up to the $\phi_{3}$ equation to find a closed form solution to (3). The sequence of $\phi_{k}$s needed to find the second 'excited' state are given below
 
\begin{equation}
\begin{aligned}
\phi_{(1)}:=63 \text{b}^2 e^{4 (\alpha+\beta_{+})}-24 e^{6 (\alpha+\beta_{+})}+18 e^{6 \alpha},
\end{aligned}
\end{equation}

\begin{equation}
\begin{aligned}
\phi_{(2)}:=108 \left(6 \text{b}^2 e^{2 (\alpha+\beta_{+})}+2 e^{4 \alpha-2 \beta_{+}}-9 \text{b}^4\right),
\end{aligned}
\end{equation}

\begin{equation}
\begin{aligned}
\phi_{(3)}:=972 \left(4 e^{2 (\alpha+\beta_{+})}-6 \text{b}^2\right).
\end{aligned}
\end{equation}

We will plot these 'excited' states below in figures(2a-2f) and discuss the physical consequences of the electromagnetic field $b^{2}$ by examining the qualitative behavior of our wave functions towards the end of this paper.

\begin{figure}
\centering
\begin{subfigure}{.4\textwidth}
  \centering
  \includegraphics[scale=.12]{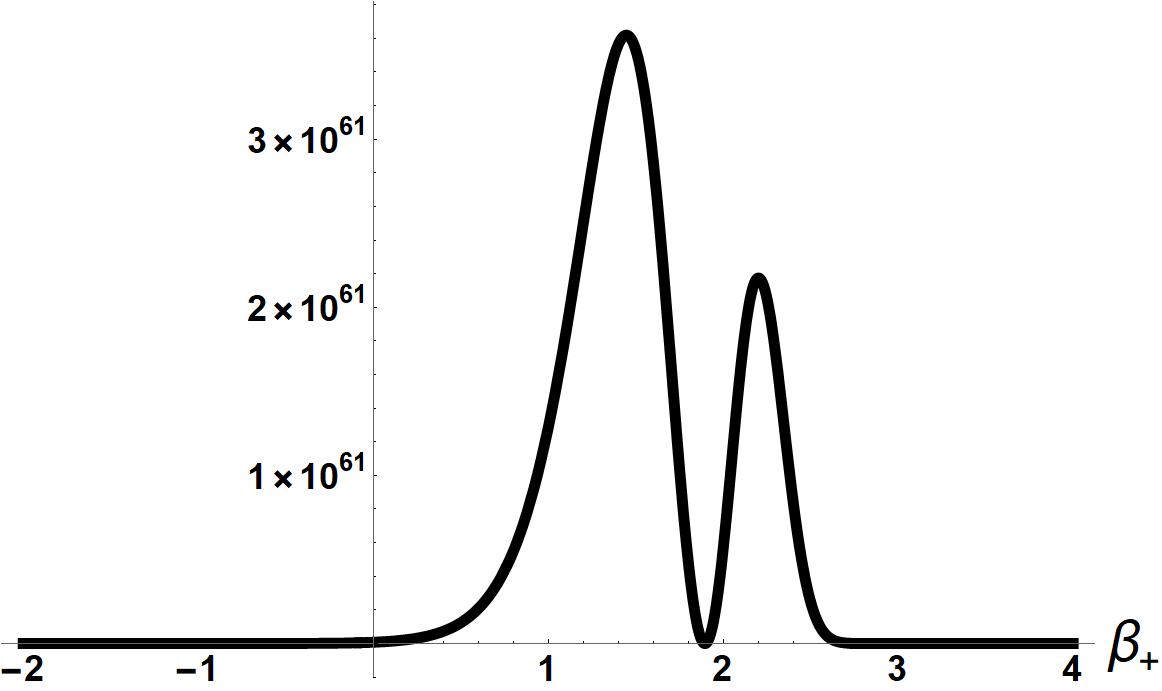}
  \caption{$n=1$ \hspace{1mm}  $\alpha=-1$ \hspace{1mm} $b=0$}
  \label{fig:sub1}
\end{subfigure}%
\begin{subfigure}{.4\textwidth}
  \centering
  \includegraphics[scale=.12]{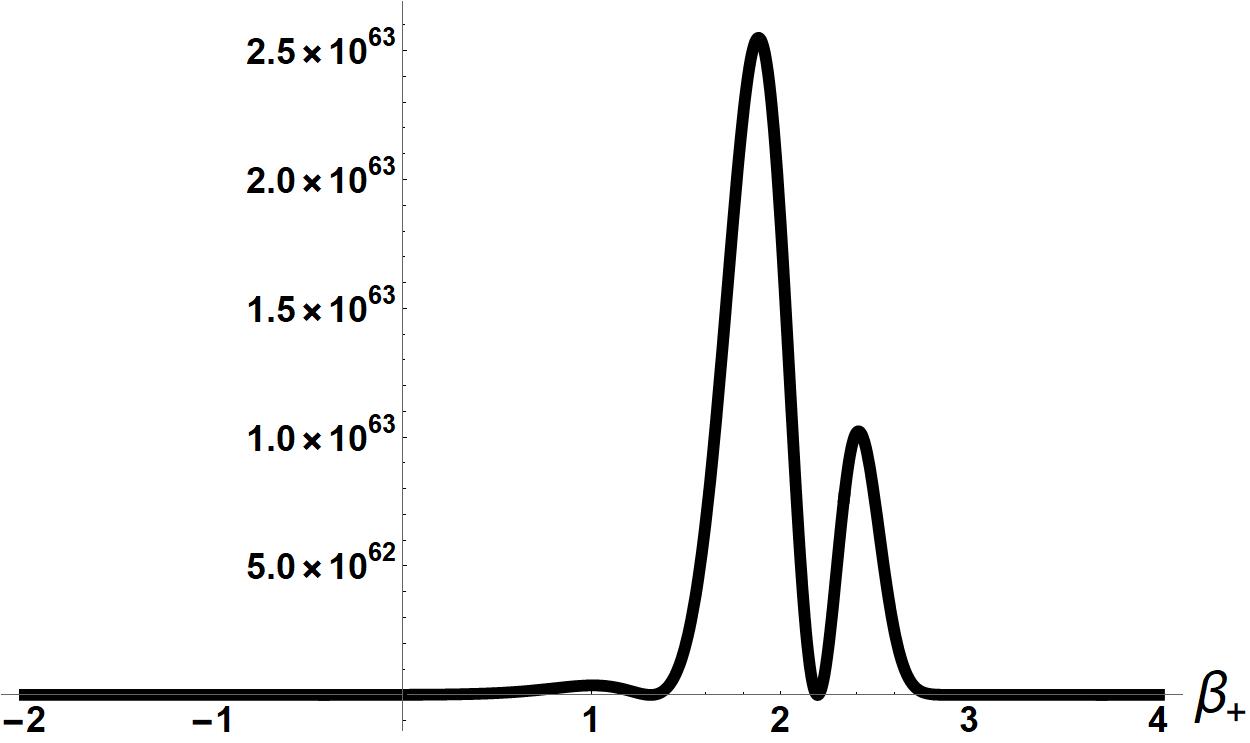}
  \caption{ $n=1$ \hspace{1mm} $\alpha=-1$  \hspace{1mm} $b=1.5$}
  \label{fig:sub2}
\end{subfigure}
\begin{subfigure}{.4\textwidth}
  \centering
  \includegraphics[scale=.12]{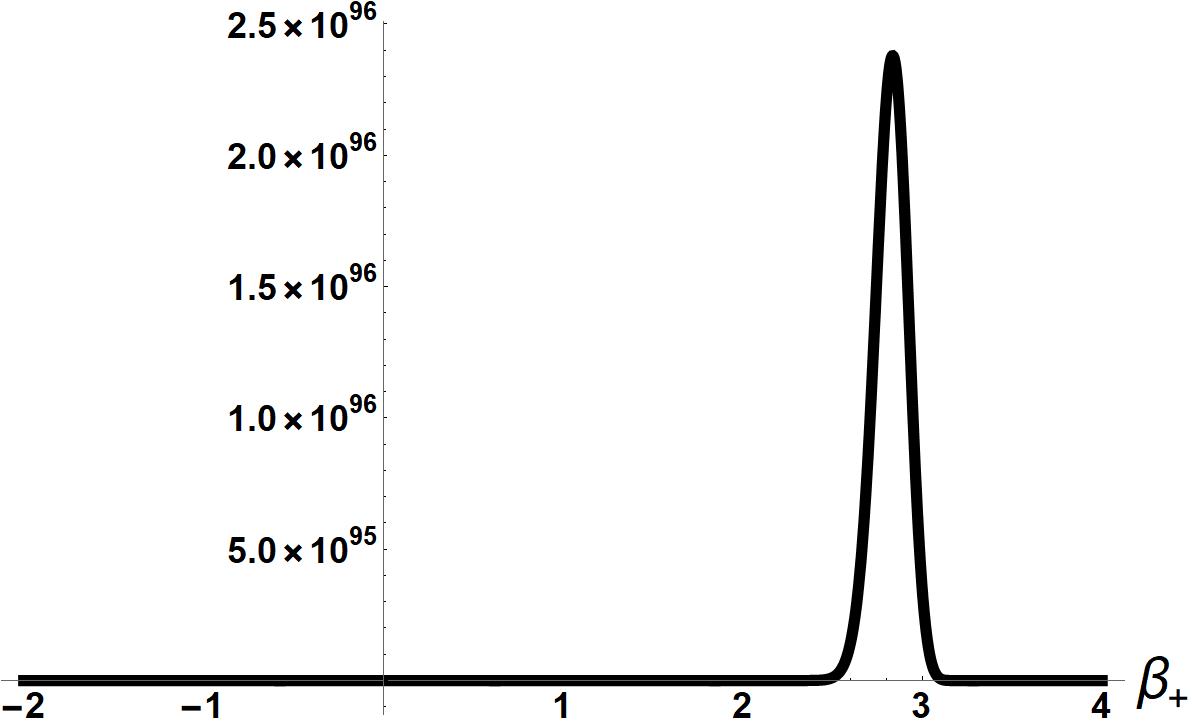}
  \caption{ $n=1$ \hspace{1mm} $\alpha=-1$ \hspace{1mm} $b=5$}
  \label{fig:sub1}
\end{subfigure}
\begin{subfigure}{.4\textwidth}
  \centering
  \includegraphics[scale=.12]{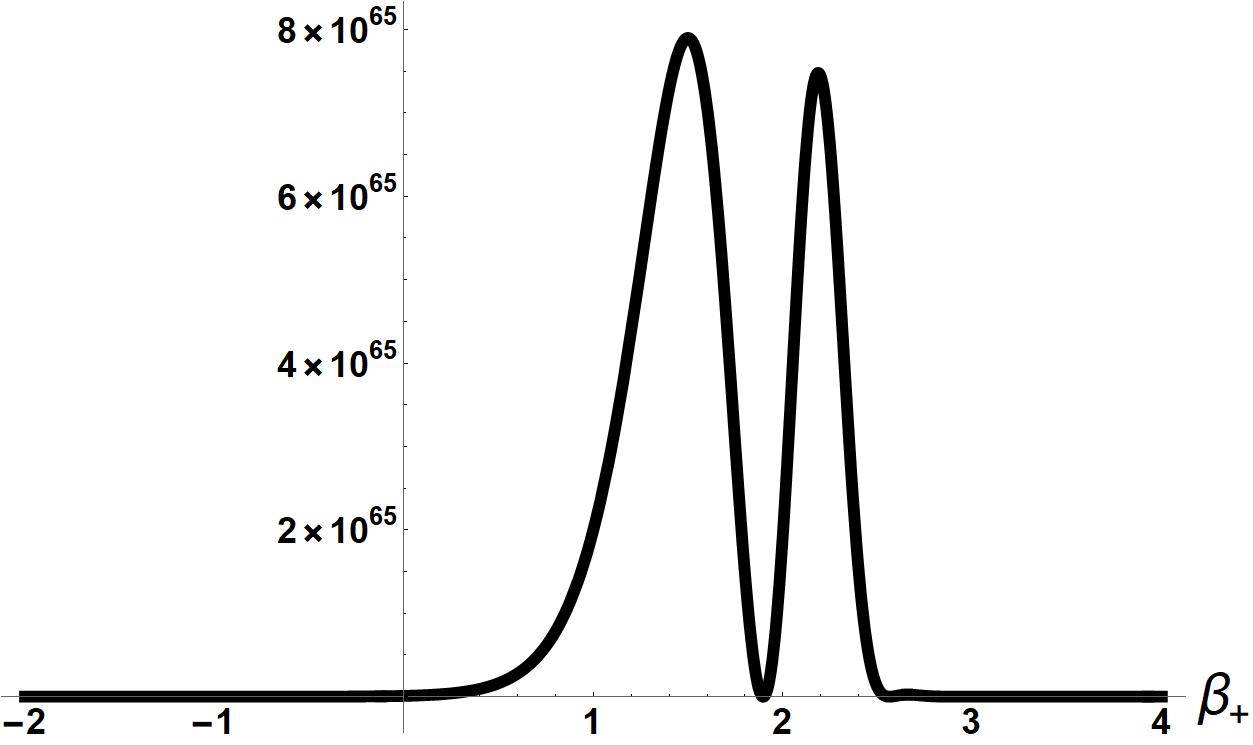}
  \caption{ $n=2$ \hspace{1mm} $\alpha=-1$ \hspace{1mm} $b=0$}
  \label{fig:sub1}
\end{subfigure}
\begin{subfigure}{.4\textwidth}
  \centering
  \includegraphics[scale=.12]{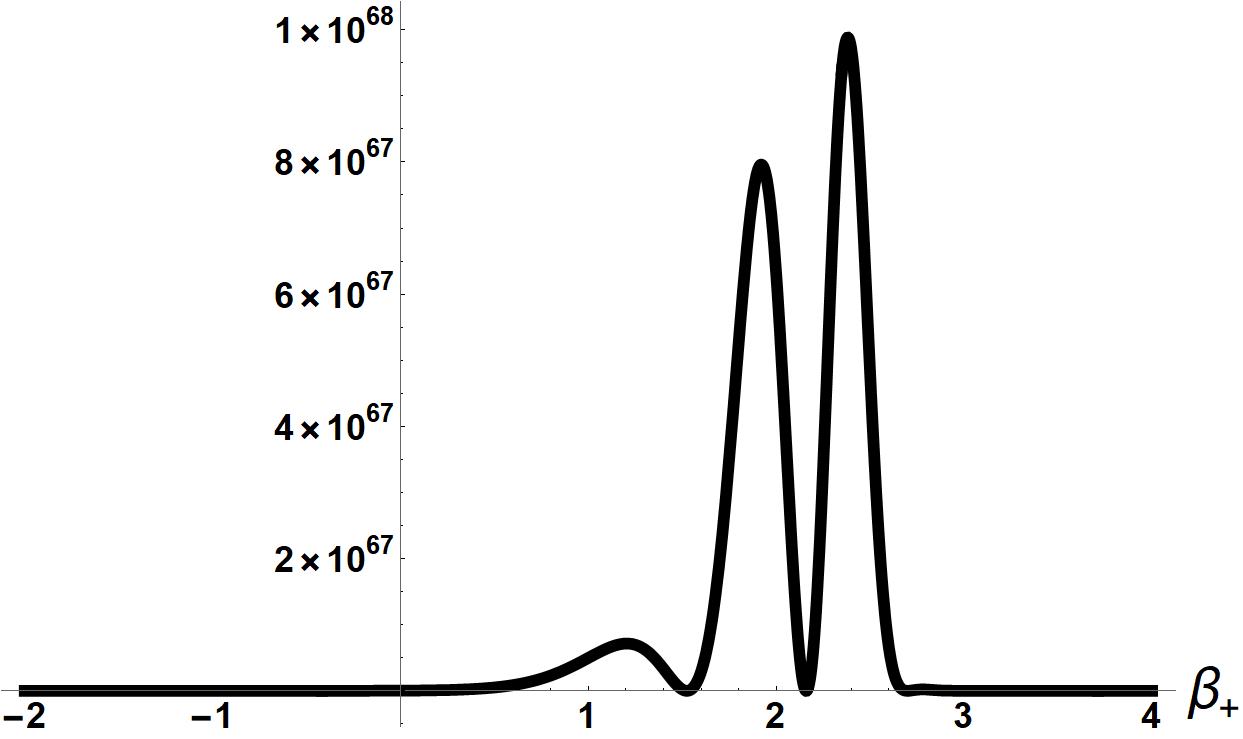}
  \caption{ $n=2$ \hspace{1mm} $\alpha=-1$ \hspace{1mm} $b=1.5$}
  \label{fig:sub1}
\end{subfigure}
\begin{subfigure}{.4\textwidth}
  \centering
  \includegraphics[scale=.12]{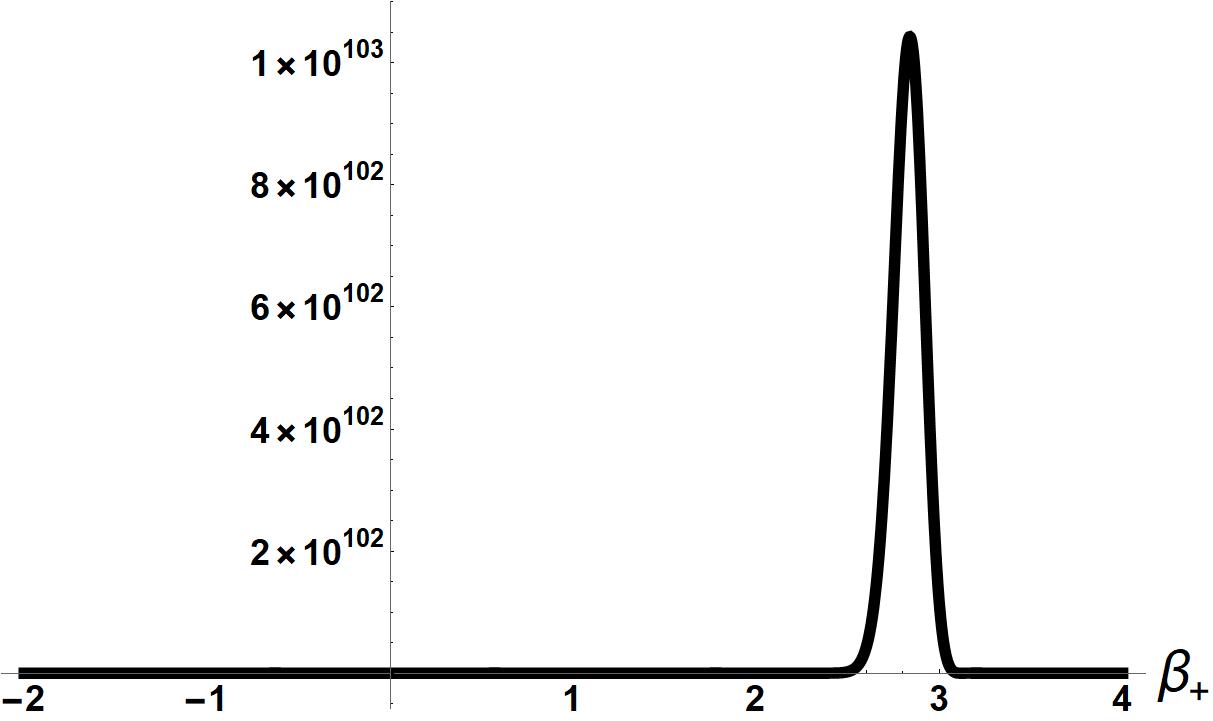}
  \caption{ $n=2$ \hspace{1mm} $\alpha=-1$ \hspace{1mm} $b=5$}
  \label{fig:sub1}
\end{subfigure}%
\caption{Figures (a)-(c) are of the first closed form 'excited' state when an electromagnetic field is present while figures (d)-(f) are of the second 'excited' state when an electromagnetic field is present. For both sets of figures we varied the values of the electromagnetic field.}
\label{fig:test}
\end{figure}

When $ B \ne 0$ (49) becomes more challenging to solve. To proceed, we will set $b=0$ and find a suitable coordinate change $u(\alpha,\beta_+)$, $v(\alpha,\beta_+)$ which simplifies our transport equation

\begin{equation}
\begin{aligned}
\frac{\partial \mathcal{S}_{(2)}}{\partial \alpha}=\frac{\partial \mathcal{S}_{(2)}}{\partial u} \frac{\partial u}{\partial \alpha}+\frac{\partial \mathcal{S}_{(2)}}{\partial v} \frac{\partial v}{\partial \alpha} \\\frac{\partial \mathcal{S}_{(2)}}{\partial \beta_+}=\frac{\partial \mathcal{S}_{(2)}}{\partial u} \frac{\partial u}{\partial \beta_+}+\frac{\partial \mathcal{S}_{(2)}}{\partial v} \frac{\partial v}{\partial \beta_+}.
\end{aligned}
\end{equation}
Inserting (56) into (49) and choosing $u=e^{\alpha+\beta_+}$, $v=e^{-\alpha+\beta_+}$ results in the following simplification

\begin{equation}
\begin{aligned}
\left(8 u^5 v^4 \Lambda+12 \pi  v \left(1-4 u^3 v^3\right)\right) \frac{\partial \mathcal{S}_{(2)}}{\partial v}-9 \pi  u
   \left(\text{B}^2 v^3-4 \frac{\partial \mathcal{S}_{(2)}}{\partial u}\right)=0
\end{aligned}
\end{equation}
whose solutions can be found using Mathematica's DSolve. Once a family of solutions for (57) is found one can insert $u=e^{\alpha+\beta_+}$ and  $v=e^{-\alpha+\beta_+}$ back into it to obtain 

\begin{equation}
\begin{aligned}
\\&
f=\frac{3}{8} \pi  \text{B}^2 \\&
\mathcal{S}_{(2)}:=f \text{RootSum}\left[\text{x}^3 \left(-e^{2 \beta _+}\right) \Lambda+9 \pi  \text{x}^2 e^{2 \beta _+}-9 \pi  e^{4 \alpha
   +6 \beta _+}+9 \pi  e^{4 \alpha }+\Lambda e^{6 \alpha +8 \beta _+}\&,\frac{\log \left(e^{2 \alpha +2 \beta _+}-\text{x}\right)}{6 \pi 
   \text{x}-\text{x}^2 \Lambda}\&\right] \\&+c_1\left(\frac{1}{9} e^{4 \alpha -2 \beta _+} \left(-9 e^{6 \beta _+}+\frac{\Lambda e^{2
   \alpha +8 \beta _+}}{\pi }+9\right)\right).
\end{aligned}
\end{equation}
RootSum in (58) means summing $\frac{\log \left(e^{2 \alpha +2 \beta _+}-\text{x}\right)}{6 \pi 
   \text{x}-\text{x}^2 \Lambda}$ over the roots of $\text{x}^3 \left(-e^{2 \beta _+}\right) \Lambda+9 \pi  \text{x}^2 e^{2 \beta _+}-9 \pi  e^{4 \alpha
   +6 \beta _+}+9 \pi  e^{4 \alpha }+\Lambda e^{6 \alpha +8 \beta _+}=0$.

\section{\label{sec:level1} Proof Of Infinite Sequence Of ’Excited’ State Solutions For Quantum Taub
Models} 

To construct our 'excited' states we need a family of smooth and globally defined solutions to equation (19). The author was able to obtain the following family of such solutions to equation (19) when both a cosmological constant and an electromagnetic field are present

\begin{equation}
\begin{aligned}
\phi_{(0)}:=\left(27 \pi  \text{b}^2 e^{2 (\alpha+\beta_{+})}+\Lambda e^{6 (\alpha+\beta_{+})}-9 \pi  \left(e^{6
   \beta_{+}}-1\right) e^{4 \alpha-2 \beta_{+}}\right)^n.
\end{aligned}
\end{equation}
The author in a future work hopes to prove that an infinite number of 'excited' states exist for the quantum Taub models using the above $\phi_{0}$ which includes both of our matter sources. For the time being though, in this paper we will study the semi-classical 'excited' states that this $\phi_{0}$ (59) produces. Before that though we will prove that an infinite number of smooth and globally defined 'excited' states do exist for the quantum Taub models when only $\Lambda$ is present. 

To begin we will use this $\phi_{0}$

\begin{equation}
\begin{aligned}
\phi_{(0)}:=\left(\Lambda e^{6 (\alpha+\beta_{+})}-9 \pi  \left(e^{6
   \beta_{+}}-1\right) e^{4 \alpha-2 \beta_{+}}\right)^n.
\end{aligned}
\end{equation}
Because our $\phi_{(0)}$(60) clearly vanishes for some finite values of the Misner variables, in order to ensure that our wave functions are smooth and globally defined we must restrict n to be either zero or a positive integer. As was mentioned before, this manifestation of quantization is one of the reasons why we refer to these states as 'excited' states.

We will now proceed to argue that there are a countably infinite number of closed form solutions to the Taub WDW equation, each corresponding to an 'n' for the case when $\phi_{0}$ is (60) and 

\begin{equation}
\begin{aligned}
&\mathcal{S}_{(0)} := -\frac{\Lambda e^{4 (\alpha+\beta_{+})}}{36 \pi }+\frac{1}{6} \left(2 e^{6 \beta_{+}}+1\right) e^{2\alpha-4 \beta_{+}},\\&\mathcal{S}_{(1)} := -\alpha -\beta_+.
\end{aligned}
\end{equation}

We will start by observing that the source term for our 'excited' state transport equations (21) for $k \geq 1$ reduces to 

\begin{equation}
\begin{aligned}
{k\left(\frac{\partial \phi_{(k-1)}}{\partial \alpha} -\frac{\partial \phi_{(k-1)}}{\partial \beta_{+}}\right)}  {+\frac{k}{2}\left(\frac{\partial^{2} \phi_{(k-1)}}{\partial \alpha^{2}}-\frac{\partial^{2} \phi_{(k-1)}}{\partial \beta_{+}^{2}}\right)},
\end{aligned}
\end{equation}
because when B=0 the non-trivial solutions to the 'ground' state transport equations, $\mathcal{S}_{(0)}$ and $\mathcal{S}_{(1)}$, allow all of the higher order $\mathcal{S}_{(k \geq 2)}$ transport equations to be satisfied by zero, which results in the $\mathcal{S}_{(\ell)}$ terms in (21) to vanish. Thus our 'excited' state transport equation for any kth quantum correction to the nth 'excited' state is  
\begin{equation}
\begin{aligned}
-\frac{\partial \phi_{(k)}}{\partial \alpha} \frac{\partial \mathcal{S}_{(0)}}{\partial \alpha}+\frac{\partial \phi_{(k)}}{\partial \beta_{+}} \frac{\partial \mathcal{S}_{(0)}}{\partial \beta_{+}}+{k\left(\frac{\partial \phi_{(k-1)}}{\partial \alpha} -\frac{\partial \phi_{(k-1)}}{\partial \beta_{+}}\right)}  {+\frac{k}{2}\left(\frac{\partial^{2} \phi_{(k-1)}}{\partial \alpha^{2}}-\frac{\partial^{2} \phi_{(k-1)}}{\partial \beta_{+}^{2}}\right)}=0.
\end{aligned}
\end{equation}

Because of the linearity and signs of the derivatives present in (62) all expressions of the form $x_{1}e^{x_{2}\alpha+x_{2}\beta_+}$ which are contained, and expressed within $\phi_{k-1}$ as a sum $\Sigma_{i}x_{i}e^{x_{i}\alpha+x_{i}\beta_+}+\Sigma_{i} \cdots$ vanish from the source term of the $\phi_{k}$ transport equations. If we insert (60) into (62) when k=1 it is evident that we obtain a sum of exponentials which is composed of $\frac{1}{2} n (1 + n)$ terms, as can be checked by examining the cumbersome resultant expression. In addition the homogeneous portion of the transport equations (63) contains a sum of exponentials thanks to (61). Thus it is natural to try as an ansatz for $\phi_{1}$ a sum of exponentials composed of $\frac{1}{2} n (1 + n)$ terms, $\phi_{1}:=\sum _{i=1}^{\frac{1}{2} n (n+1)} x_i e^{y_i\alpha + z_i\beta_+}$. The author for n=1, 2, and 3 inserted the aforementioned ansatz into (63) when k=1 and was able to find unique solutions for the three, nine, and eighteen coefficients $\left(x_{i},y_{i},z_{i}\right)$ required to respectively express in closed form the first quantum correction to the first, second and third $\Lambda \ne 0$ 'excited' quantum Taub states. For the k=2 transport equations the source term for the n=1 case vanished because its respective $\phi_{1}$ was $54 \pi  e^{2 \alpha+2 \beta_+}$; thus allowing the sequence of 'excited' state transport equations to truncate, resulting in a closed form 'excited' state solution to (3). On the other hand for n=2 and n=3 states the author modified the aforementioned ansatz $\sum _{i=1}^{\frac{1}{2} (-k+n+1) (-k+n+2)} x_i e^{y_{i}\alpha+z_{i}\beta_{+}}$ so that it contains the same number of terms as the source of the transport equations it was intended to solve. Using this modified ansatz the author was able to find the three and nine coefficients respectively required to compute $\phi_{2}$ for the second and third 'excited' states. The source term for the the k=3 transport equation vanished when n=2 because its $\phi_{2}$ was  $17496 \pi ^2 e^{4 (\alpha+\beta_+)}$, thus allowing us to construct a closed form solution for the second 'excited' state. However when n=3 the source term did not vanish, but rather was composed of only one exponential $c_{1}e^{c_{2}\alpha+c_{3}\beta_+}$, whose coefficients the author was able to solve for which resulted in $\phi_{3}$ being $14171760 \pi ^3 e^{6 \alpha+6 \beta_+}$. We will explicitly present the aforementioned results for the n=1, 2 and 3 states after we prove that a countably infinite number of closed form 'excited' states exist which are of a particular form that we will introduce shortly. Using the three closed form solutions the author found, he was able to notice a pattern in how powers of $e^{\beta_+}$ and $e^{\alpha}$ varied between terms, and in the total number of terms present for each kth quantum correction given a value of n and deduced the following ansatz for the kth quantum correction given any positive integer value of n

\begin{equation}
\begin{aligned}
\phi_{k}= \sum _{i=0}^{n-k} \left(\sum _{j=0}^{-i-k+n} b(\{i,j,k\}) e^{ (2 \alpha (i-k+2 n)+2 \beta_+ (4 i+3 j+2 k-n))}\right).
\end{aligned}
\end{equation}
In the above expression $b(\{i,j,k\})$ is a constant determined by substituting (64) back into the 'excited' state transport equations (63). The $b(\{i,j,0\})$ coefficients can be read directly from (60). 

After we present the explicit n=1, 2, and 3 solutions it will become clear to the reader how the author was able to deduce this particular form for the solutions of (63) when $\Lambda \ne 0$. When we set n=k we obtain

\begin{equation}
\begin{aligned}
\phi_{n}= e^{2 \alpha n+2 \beta_+ n} b(\{0,0,n\}),
\end{aligned}
\end{equation}
which results in the source term of the $\phi_{n+1}$ equation vanishing, and subsequently allowing the $\phi_{n+1}$ transport equation to be satisfied by zero. This allows one to truncate the infinite sequence of transport equations and show that if (64) is indeed the general form for a family of solutions to the transport equations then one only needs to solve n transport equations to obtain the closed form nth 'excited' state of the $\Lambda \ne 0$ quantum Taub models. To prove that there exists a family composed of a countably infinite number of solutions, all with the form of (64), we will insert (64) into our source term (62) to obtain after simplification

\begin{equation}
\begin{aligned}
& f(i,j,k,n)=k (i+j+k-n-1) (5 i+3 j+k+n) \\&
-6 \sum _{i=0}^{-k+n+1} \left(\sum _{j=0}^{-i-k+n+1}  f(i,j,k,n)b(\{i,j,k-1\}) e^{ (2 \alpha (i-k+2 n+1)+2 \beta_+ (4 i+3 j+2 (k-1)-n))}\right).
\end{aligned}
\end{equation}

The coefficients of our source term vanishes when $j = 1 - i - k + n$, which is the upper limit of our sum over j. Thus we can rewrite our sum for the source term as 

\begin{equation}
\begin{aligned}
& f(i,j,k,n)=k (i+j+k-n) (5 i+3 j+k+n) \\&
-6 \sum _{i=0}^{n-k} \left(\sum _{j=0}^{-i-k+n}  f(i,j,k,n)b(\{i,j,k-1\}) e^{ (2 \alpha (i-k+2 n+1)+2 \beta_+ (4 i+3 j+2 (k-1)-n))}\right).
\end{aligned}
\end{equation}
We were able to change the range of our sum for i because (k-n-1-k+n)=-1 which is below our starting value of j=0. Our next step is to insert (64) into (63) which after much simplification results in

\begin{equation}
\begin{aligned}
&-18 \pi  \sum _{i=0}^{n-k} \left(\sum _{j=0}^{-i-k+n}  (3 i+2 j+k) b(\{i,j,k\}) e^{ (2 \alpha (i-k+2 n+1)+2 \beta_+ (4 i+3 j+2 k-n-2))}\right) \\& +36 \pi  \sum _{i=0}^{n-k} \left(\sum _{j=0}^{-i-k+n}  (i+j+k-n) b(\{i,j,k\}) e^{ (2 \alpha (i-k+2 n+1)+2 \beta_+ (4 i+3 j+2 k-n+1))}\right) \\& -6 \Lambda \sum _{i=0}^{n-k} \left(\sum _{j=0}^{-i-k+n}  (i+j+k-n) b(\{i,j,k\}) e^{ (2 \alpha (i-k+2 n+2)+2 \beta_+ (4 i+3 j+2 k-n+2))}\right) \\& -54 \pi \sum _{i=0}^{n-k} \left(\sum _{j=0}^{-i-k+n}  f(i,j,k,n)b(\{i,j,k-1\}) e^{ (2 \alpha (i-k+2 n+1)+2 \beta_+ (4 i+3 j+2k -n-2))}\right)=0.
\end{aligned}
\end{equation}
For the second and third sum their coefficients vanish when $j = -i - k + n$, thus we can rewrite our kth 'excited' state transport equation for n to be

\begin{equation}
\begin{aligned}
&-18 \pi  \sum _{i=0}^{n-k} \left(\sum _{j=0}^{-i-k+n}  (3 i+2 j+k) b(\{i,j,k\}) e^{ (2 \alpha (i-k+2 n+1)+2 \beta_+ (4 i+3 j+2 k-n-2))}\right) \\& +36 \pi  \sum _{i=0}^{n-k-1} \left(\sum _{j=0}^{-i-k+n-1}  (i+j+k-n) b(\{i,j,k\}) e^{ (2 \alpha (i-k+2 n+1)+2 \beta_+ (4 i+3 j+2 k-n+1))}\right) \\& -6 \Lambda \sum _{i=0}^{n-k-1} \left(\sum _{j=0}^{-i-k+n-1}  (i+j+k-n) b(\{i,j,k\}) e^{ (2 \alpha (i-k+2 n+2)+2 \beta_+ (4 i+3 j+2 k-n+2))}\right) \\& -54 \pi \sum _{i=0}^{n-k} \left(\sum _{j=0}^{-i-k+n}  f(i,j,k,n)b(\{i,j,k-1\}) e^{ (2 \alpha (i-k+2 n+1)+2 \beta_+ (4 i+3 j+2k -n-2))}\right)=0.
\end{aligned}
\end{equation}

If we examine the first and fourth sums we see that their summation bounds are the same, the exponentials terms $e^{ (2 \alpha (i-k+2 n+1)+2 \beta_+ (4 i+3 j+2k -n-2))}$ they contain are the same, and their coefficients do not vanish within the range which they are summed. Thus if we were to relate the variables $b(\{i,j,k\})$ and $b(\{i,j,k-1\}$ using only the first and fourth sum in (68) it is evident we would obtain a $\frac{1}{2} (1 - k + n) (2 - k + n) \cross \frac{1}{2} (1 - k + n) (2 - k + n)$ diagonal matrix with a non zero determinant \def\horzbar{\text{magic}}

\begin{align*}
 & \left[\begin{array}{ccc}
  -18 \pi  k & & \\
    & \ddots & \\
    & & -18 \pi  (3 (n-k)+k)
\end{array}\right]
\begin{bmatrix}
  b(\{0,0,k\}) \\
           b(\{0,1,k\}) \\
           \vdots \\
           b(\{n-k,0,k\})
\end{bmatrix} \\
{} &= \begin{bmatrix}
  54 \pi  k (k-n-1) (k+n)b(\{0,0,k-1\}) \\
           54 \pi  k (k-n) (k+n+3)b(\{0,1,k-1\}) \\
           \vdots \\
           -54 \pi  k (5 (n-k)+k+n)b(\{n-k,0,k-1\})
\end{bmatrix}
\end{align*}

To prove that a solution of the form (64) exists for the kth quantum correction to the nth 'excited' state we must prove that the matrix associated with (69) always has a non-zero determinant when both k and n are positive integers, and  $1 \leq k \leq n$. Luckily this isn't difficult. All the exponentials terms that show up in the second  and thirds sums in (69) are a subset of the exponentials that appear in the first and fourth sums. This is easy to see, for the second term in our sum if we were to reexpress j as (J-1) we would obtain

\begin{equation}
\begin{aligned}
36 \pi  \sum _{i=0}^{n-k} \left(\sum _{J=1}^{-i-k+n}  (i+J+k-n-1) b(\{i,J-1,k\}) e^{ (2 \alpha (i-k+2 n+1)+2 \beta_+ (4 i+3 J+2 k-n-2))}\right)
\end{aligned}
\end{equation}

which is a summation of the same expotential terms present in the first and fourth sums of (69), except that the range of the variable J which is being summed over is reduced. Hence only a subset of the exponentials which are summed in the first and fourth terms are present in the second term. The same can be easily shown for the 3rd term by substituting into i, I-1

\begin{equation}
\begin{aligned}
-6 \Lambda \sum _{\text{I}=1}^{n-k} \left(\sum _{j=0}^{-\text{I}-k+n} (\text{I}+j+k-n-1) b(\{\text{I}-1,j,k\}) e^{ (2 \alpha (\text{I}-k+2 n+1)+2 \beta_+ (4
   \text{I}+3 j+2 k-n-2))}\right).
\end{aligned}
\end{equation}

Now that we established that (69) can in principle be expressed as a system of $\frac{1}{2} (1 - k + n) (2 - k + n)$ linear equations, where the equations are grouped by the exponential term that the known $b({i,j,k-1})$ and unknown $b({i,j,k})$ coefficients are multiplied to; our next task is to deduce the characteristics of the matrix which represents this system. We aim to prove that this matrix is always invertible for the values of n and k we are exclusively considering. To accomplish this we will explicitly write out the matrix elements and show that they clearly are the elements of a triangular matrix with non vanishing diagonal elements.

If we start with the first sum in (69) we see that initially the number of terms which are summed are n-k+1 when i=0, then n-k when i=1, then n-k-1 when i=2,etc...; this pattern continues until the single i=n-k term $b(\{n-k,0,k\})$ is summed. With this information for each kth quantum correction we aim to relabel our $b(\{i,j,k\})$ using only a single index $b(\{x(i,j),k\})$ that is a function of both i and j and monotonically increases in units of one by noticing that when the subsequent i index increases by one, then one less term is summed in the j index. Thus to relabel our sum we want to find a labeling scheme such that for every summation over i starting with i=1 we add (-i-k+n+2) to our index $x(i,j)$ before summing over j again. This can be found by solving this recursion relation $\{f(i+1)=f(i)-i-k+n+1,f(0)=0\}$ which yields $-\frac{1}{2} i (i+2 k-2 n-3)$. This allows us to rewrite the first sum in (69) as

\begin{equation}
\begin{aligned}
-18 \pi  \sum _{i=0}^{n-k} \left(\sum _{j=0}^{-i-k+n}  (3 i+2 j+k) b(\{j-\frac{1}{2} i (i+2 k-2 n-3),k\}) e^{ (2 \alpha (i-k+2 n+1)+2 \beta_+ (4 i+3 j+2 k-n-2))}\right),
\end{aligned}
\end{equation}
and subsequently define the diagonal elements of our matrix as 

\begin{equation}
\begin{aligned}
M_{\left\{j-\frac{1}{2} i (i+2 k-2 n-3)+1,j-\frac{1}{2} i (i+2 k-2 n-3)+1\right\}}=-18
   \pi  (3 i+2 j+k)
\end{aligned}
\end{equation}

For the second term using our aforementioned relabeling of $b(\{i,j,k\})$ and replacing j with J-1 we obtain 

\begin{equation}
\begin{aligned}
& g(i,J,k,n)=(i+J+k-n-1) \\&
36 \pi  \sum _{i=0}^{n-k} \left(\sum _{J=1}^{-i-k+n}  g  e^{ (2 (\alpha (i-k+2 n+1)+\beta_{+} (4 i+3 J+2 k-n-2)))} b\left(\left\{J-\frac{1}{2} i (i+2
   k-2 n-3)-1,k\right\}\right)\right).
\end{aligned}
\end{equation}
Because the labels of $b(\{x(i,j),k\})$ in (72) and (74) differ by 1 in our labeling scheme, the terms which correspond to (74) are off diagonal terms to the left of the diagonal terms of (72). In other words the coefficients of $b(\{x(i,j),k\})$ which are multiplied to the same exponential $ e^{ (2 (\alpha (i-k+2 n+1)+\beta_{+} (4 i+3 J+2 k-n-2)))}$ differ between (72) and (74) by 1 in $x(i,j)$, thus the terms in (74) appears either right before the diagonal terms of (72) or not at all because not all of the exponents in (72) are summed in (74). These one space to the left non diagonal terms are

\begin{equation}
\begin{aligned}
M_{\left\{J-\frac{1}{2} i (i+2 k-2 n-3)+1,J-\frac{1}{2} i (i+2 k-2 n-3)\right\}}=36 \pi  (i+J+k-n-1).
\end{aligned}
\end{equation}

For the third and final sum we need to consider for our matrix, all of the terms are also to the left of our diagonal terms (72). This can be shown by expressing the coefficients in $b(\{x(i,j),k\})$ (71) as  $b\left(\left\{j-\frac{1}{2} (\text{I}-1) (\text{I}+2 k-2 n-4),k\right\}\right)$. In relation to the diagonal elements (72) these elements will be $-i-k+n+2$ terms to the left

\begin{equation}
\begin{aligned}
M_{\left\{j-\frac{1}{2} I (I+2 k-2 n-3)+1,j-\frac{1}{2} (\text{I}-1) (\text{I}+2 k-2 n-4)+1\right\}}=-6 \Lambda (\text{I}+j+k-n-1).
\end{aligned}
\end{equation}

All of the other elements in our matrix besides (76), (75), and (73) vanish, leaving us with a triangular matrix with non vanishing diagonal elements. Hence we have proved that the system of linear equations for the coefficients for the kth quantum correction to the nth 'excited' states (64) are always uniquely solvable. We can obtain explicit values of the  $b(\{i,j,0\}$ coefficients from our $\phi_{0}$ (60), which enable us to solve the system of linear equations for  $b(\{i,j,1\}$, whose solutions allow us to solve for the system of linear equations for $b(\{i,j,2\}$ and etc all the way up to the coefficents $b(\{i,j,n\}$ for the nth quantum correction. Afterwards the rest of the transport equations can be satisifed by zero and thus we can write out a closed form solution for the nth 'excited' state to the Taub WDW equation when $\Lambda \ne 0$. 

We will attach in the appendix a Mathematica code which has as its only input the positive integer 'n' and gives all 'n' non trivial quantum corrections to the nth 'excited' state and its corresponding closed form solution. Using similar techniques and reasoning to our proof, one can find a simpler ansatz for the quantum corrections for the case when $\Lambda=0$. 

\begin{equation}
\begin{aligned}
\phi_{k}:=e^{4 \alpha n-2 \alpha k} \sum _{j=0}^{n-k} e^{-6 \beta_+ j-2 \beta_+ k+4 \beta_+ n}
   b_{\{j,k\}}
\end{aligned}
\end{equation}

The advantage of this simpler ansatz is that it allows us to write a code which we will also include in the appendix that is able to compute exact closed form 'excited' state solutions to the Taub WDW equation far faster than the code which uses the $\Lambda \ne 0$ ansatz.

Below are the explicit quantum corrections to the first five 'excited states generated by our code.

\begin{equation}
\begin{aligned}
\phi_{1}:= 54 \pi  e^{2 (\alpha+\beta_+)}\\
\phi_{k > 1}:= 0
\end{aligned}
\end{equation}
\begin{equation}
\begin{aligned}
\phi_{2}:= 17496 \pi ^2 e^{4 (\alpha+\beta_+))}\\
\phi_{1}:= \frac{459}{2} \pi  \Lambda e^{8 \alpha+8 \beta_+}-1944 \pi ^2 e^{6 \alpha+6 \beta_+}+1458 \pi ^2 e^{6 \alpha}\\
\phi_{k > 2}:= 0
\end{aligned}
\end{equation}
\begin{equation}
\begin{aligned}
&\phi_{3}:= 14171760 \pi ^3 e^{6 (\alpha+\beta_+)}\\&
\phi_{2}:= 144342 \pi ^2 \Lambda e^{10 \alpha+10 \beta_+}+787320 \pi ^3 e^{8 \alpha+2 \beta_+}-1180980 \pi ^3 e^{8 \alpha+8 \beta_+}\\& \phi_{1}:=\frac{1053}{2} \pi  \Lambda^2 e^{14\alpha+14\beta_+}+\frac{15309}{2} \pi ^2 \Lambda e^{12 \alpha+6
   \beta_+}-\frac{18225}{2} \pi ^2 \Lambda e^{12 \alpha+12 \beta_+}\\&+26244 \pi ^3 e^{10 \alpha-2 \beta_+}-65610 \pi ^3
   e^{10 \alpha+4 \beta_+}+39366 \pi ^3 e^{10 \alpha+10 \beta_+}\\&\phi_{k > 3 }:= 0
\end{aligned}
\end{equation}
\begin{equation}
\begin{aligned}
&\phi_{4}:= 21427701120 \pi ^4 e^{8 \left(\alpha +\beta _+\right)}\\&
\phi_{3}:= 892820880 \pi ^4 e^{10 \alpha +4 \beta _+}-1428513408 \pi ^4 e^{10 \alpha +10 \beta
   _+}+178564176 \pi ^3 V e^{12 \alpha +12 \beta _+}\\& \phi_{2}:=-59521392 \pi ^4 e^{12 \alpha +6 \beta _+}+39680928 \pi ^4 e^{12 \alpha +12 \beta
   _+}+21257640 \pi ^4 e^{12 \alpha } \\& +\frac{1121931}{2} \pi ^2 \Lambda^2 e^{4 \left(\alpha +4
   \beta _+\right)+12 \alpha }+7164612 \pi ^3 \Lambda e^{14 \alpha +8 \beta _+}-9447840 \pi ^3
   \Lambda e^{14 \alpha +14 \beta _+}\\& \phi_{1}:=393660 \pi ^4 e^{14 \alpha -4 \beta _+}-1417176 \pi ^4 e^{14 \alpha +2 \beta _+}+1653372
   \pi ^4 e^{14 \alpha +8 \beta _+}-629856 \pi ^4 e^{14 \alpha +14 \beta _+}\\&+945 \pi 
   \Lambda^3 e^{20 \alpha +20 \beta _+}+21870 \pi ^2 \Lambda^2 e^{18 \alpha +12 \beta _+}-24786 \pi
   ^2 \Lambda^2 e^{18 \alpha +18 \beta _+}\\&+164025 \pi ^3 \Lambda e^{16 \alpha +4 \beta _+}-380538
   \pi ^3 \Lambda e^{16 \alpha +10 \beta _+}+216513 \pi ^3 \Lambda e^{16 \alpha +16 \beta _+}\\& \phi_{k > 4}:=0
\end{aligned}
\end{equation}
\begin{equation}
\begin{aligned}
&\phi_{5}:= 52069313721600 \pi ^5 e^{10 \left(\alpha +\beta _+\right)}\\&\phi_{4}:=1735643790720 \pi ^5 e^{6 \left(2 \alpha +\beta _+\right)}-2892739651200 \pi ^5 e^{12
   \alpha +12 \beta _+}+367332019200 \pi ^4 \Lambda e^{14 \alpha +14 \beta _+}\\&\phi_{3}:=32141551680 \pi ^5 e^{2 \left(7 \alpha +\beta _+\right)}-96424655040 \pi ^5 e^{14 \alpha
   +8 \beta _+}+68874753600 \pi ^5 e^{14 \alpha +14 \beta _+} \\& +1014166575 \pi ^3 \Lambda^2
   e^{18 \alpha +18 \beta _+}+11861763120 \pi ^4 \Lambda e^{16 \alpha +10 \beta
   _+}-16740391500 \pi ^4 \Lambda e^{16 \alpha +16 \beta _+}\\&\phi_{2}:=\frac{3072735}{2} \pi ^2 \Lambda ^3 e^{22 \alpha +22 \beta _+}+\frac{63871335}{2} \pi ^3
   \Lambda ^2 e^{20 \alpha +14 \beta _+}\\&-\frac{78830415}{2} \pi ^3 \Lambda ^2 e^{20
   \alpha +20 \beta _+}+212576400 \pi ^4 \Lambda  e^{18 \alpha +6 \beta _+}\\&-542069820
   \pi ^4 \Lambda  e^{18 \alpha +12 \beta _+}+336579300 \pi ^4 \Lambda  e^{18 \alpha +18
   \beta _+}+446410440 \pi ^5 e^{16 \alpha -2 \beta _+}\\&-1785641760 \pi ^5 e^{16 \alpha
   +4 \beta _+}+2295825120 \pi ^5 e^{16 \alpha +10 \beta _+}-956593800 \pi ^5 e^{16
   \alpha +16 \beta _+}\\& \phi_{1}:=1485 \pi  \Lambda ^4 e^{26 \alpha +26 \beta _+}+47385 \pi ^2 \Lambda ^3 e^{24 \alpha +18
   \beta _+}-52245 \pi ^2 \Lambda ^3 e^{24 \alpha +24 \beta _+}+557685 \pi ^3 \Lambda ^2
   e^{22 \alpha +10 \beta _+}\\&-1246590 \pi ^3 \Lambda ^2 e^{22 \alpha +16 \beta
   _+}+688905 \pi ^3 \Lambda ^2 e^{22 \alpha +22 \beta _+}+2854035 \pi ^4 \Lambda  e^{20
   \alpha +2 \beta _+}\\&-9743085 \pi ^4 \Lambda  e^{20 \alpha +8 \beta _+}+10924065 \pi ^4
   \Lambda  e^{20 \alpha +14 \beta _+}-4035015 \pi ^4 \Lambda  e^{20 \alpha +20 \beta
   _+}+5314410 \pi ^5 e^{18 \alpha -6 \beta _+}\\&+42515280 \pi ^5 e^{18 \alpha +6 \beta
   _+}-31886460 \pi ^5 e^{18 \alpha +12 \beta _+}+8857350 \pi ^5 e^{18 \alpha +18 \beta
   _+}-24800580 \pi ^5 e^{18 \alpha } \\&\phi_{k > 5}:=0
\end{aligned}
\end{equation}

The first three plots(figs 3a-3c) are of a superposition of the first five closed form 'excited' states $\psi_{n}$ and the 'ground state $\left(\sum^5_{n=0}e^{-n^{2}}\psi_{n}\right)^{2}$ for three different value of $\alpha$.

\begin{figure}
\centering
\begin{subfigure}{.4\textwidth}
  \centering
  \includegraphics[scale=.17]{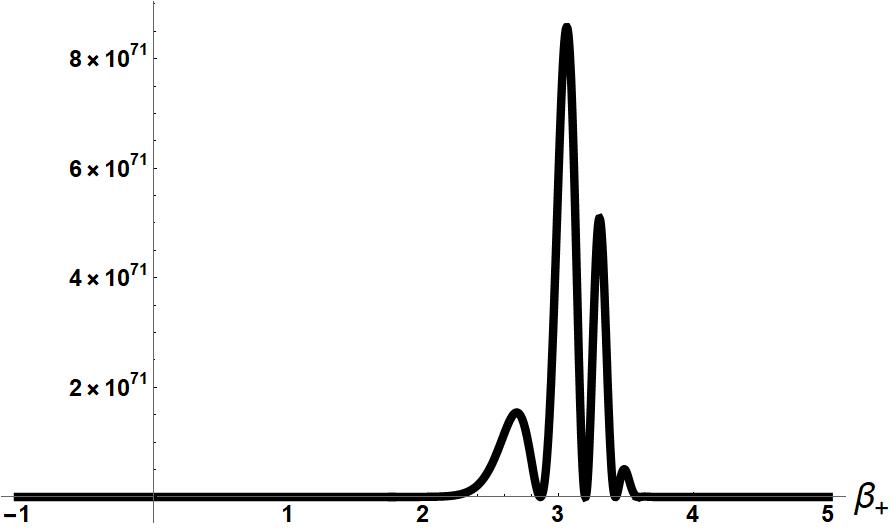}
  \caption{ $\alpha=-2$ \hspace{1 mm} $\Lambda=-1$}
  \label{fig:sub1}
\end{subfigure}%
\begin{subfigure}{.4\textwidth}
  \centering
  \includegraphics[scale=.16]{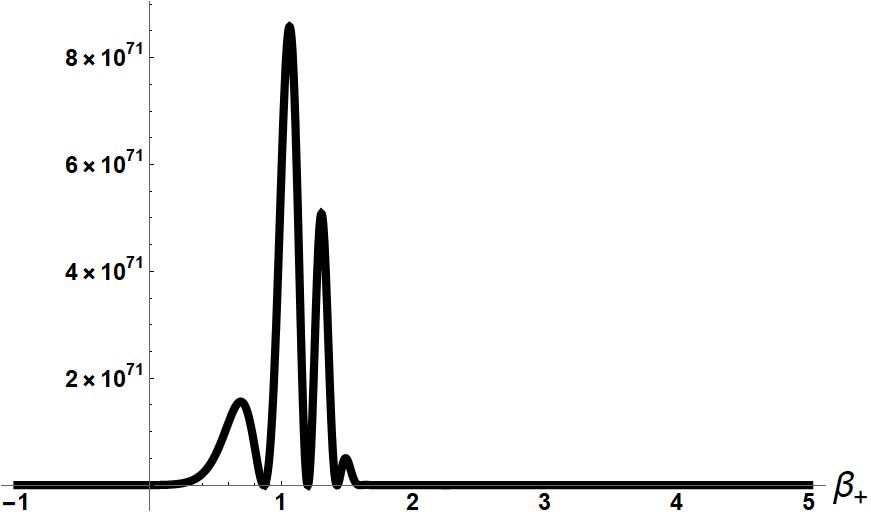}
  \caption{ $\alpha=0$ \hspace{1 mm} $\Lambda=-1$}
  \label{fig:sub2}
\end{subfigure}
\begin{subfigure}{.4\textwidth}
  \centering
  \includegraphics[scale=.19]{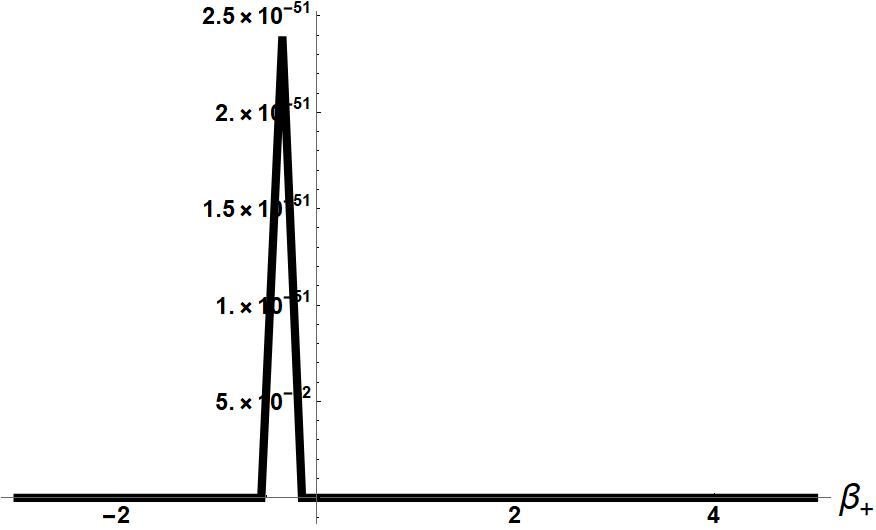}
  \caption{ $\alpha=2.5$ \hspace{1 mm} $\Lambda=-1$}
  \label{fig:sub1}
\end{subfigure}%
\caption{Superposition $\left(\sum^5_{n=0}e^{-n^{2}}\psi_{n}\right)^{2}$  of the first five closed form 'excited' states $\psi_{n}$ and the 'ground state for three different value of $\alpha$.}
\label{fig:test}
\end{figure}

The last three plots(figs 4a-4c) are semi-classical 'excited' states with both of our matter sources constructed using (59) $\left(\sum^5_{n=0}e^{-1.5\left(n+1.1\right)^{2}}\phi_{0}\Psi_{matter}\right)^{2}$ for a fixed value of $\alpha$ for three different values of $b$.
We will discuss the qualitative properties of these plots in our discussion section towards the end of this paper.

\begin{figure}
\centering
\begin{subfigure}{.4\textwidth}
  \centering
  \includegraphics[scale=.12]{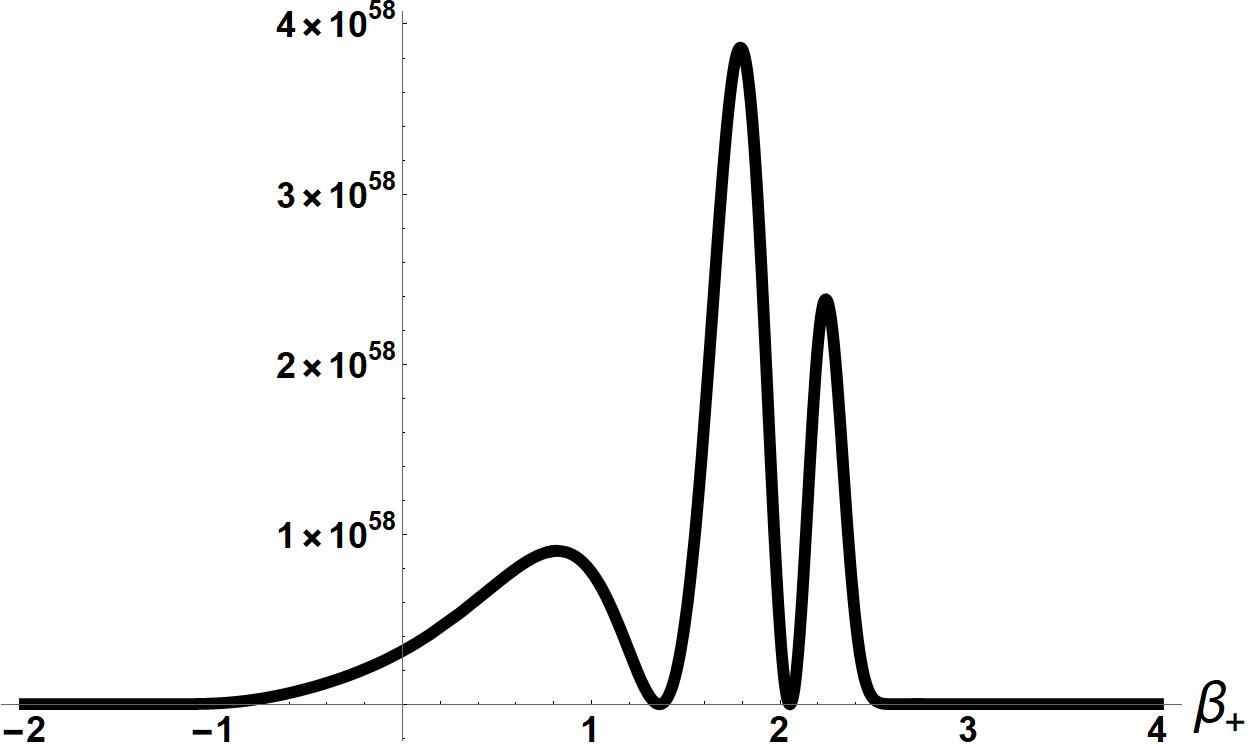}
  \caption{ $\alpha=-1$ \hspace{1 mm} $\Lambda=-1$  \hspace{1 mm} $b=0$}
  \label{fig:sub1}
\end{subfigure}%
\begin{subfigure}{.4\textwidth}
  \centering
  \includegraphics[scale=.12]{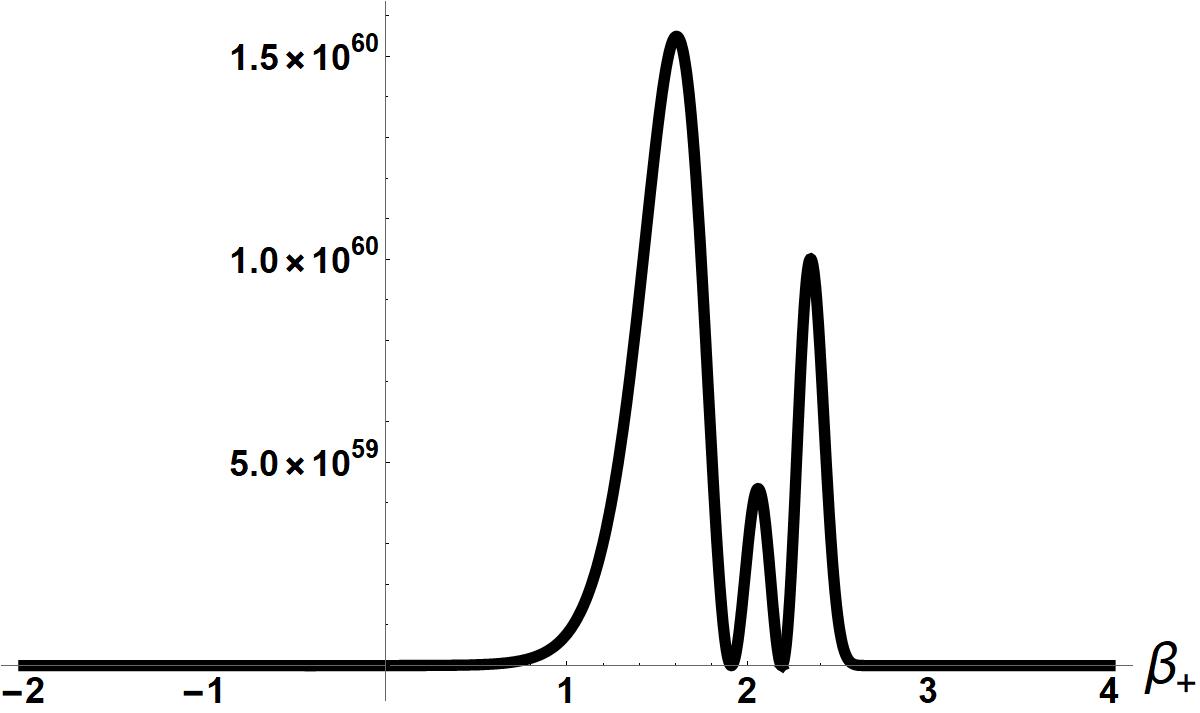}
  \caption{ $\alpha=-1$ \hspace{1 mm} $\Lambda=-1$  \hspace{1 mm} $b=1.5$}
  \label{fig:sub2}
\end{subfigure}
\begin{subfigure}{.4\textwidth}
  \centering
  \includegraphics[scale=.13]{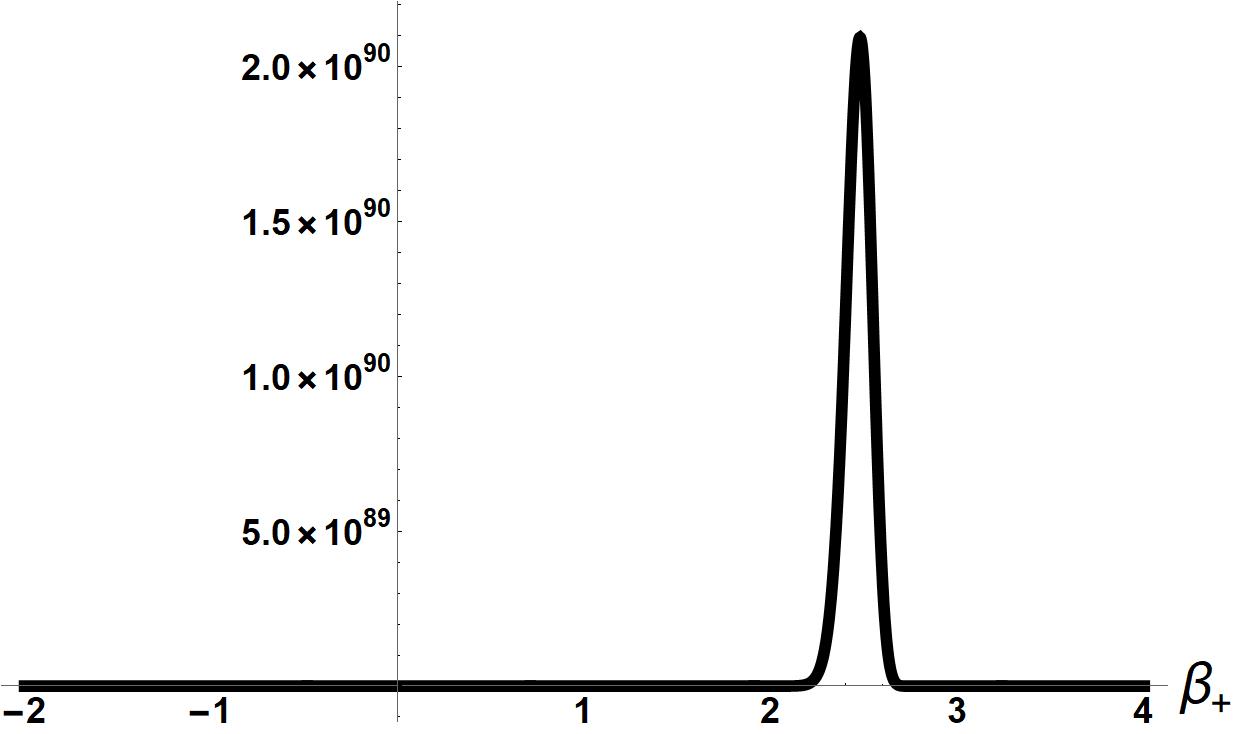}
  \caption{ $\alpha=-1$ \hspace{1 mm} $\Lambda=-1$  \hspace{1 mm} $b=5$}
  \label{fig:sub1}
\end{subfigure}%
\caption{Plots of $\left(\sum^5_{n=0}e^{-1.5\left(n+1.1\right)^{2}}\phi_{0}\Psi_{matter}\right)^{2}$ where $\Psi_{matter}$ is (50) and $\phi_{0}$ is (59) for three different values of $b$. }
\label{fig:test}
\end{figure}

\section{\label{sec:level1}Quantum Taub 'Wormhole' Model Via Euclidean-Signature Semi Classical Method} 

The goal of this section will be to prove the existence of asymptotic solutions $\stackrel{(0)}{\Psi}_{\hbar}=e^{- \mathcal{S}_{(0)}-\mathcal{S}_{(1)}-\frac{1}{2 !} \mathcal{S}_{(2)}-\cdots}$ to the Taub WDW equation for arbitrary Hartle-Hawking ordering parameter B.  Beginning with the 'wormhole' $\mathcal{S}_{(0)}$ and the following $\mathcal{S}_{(1)}$ 
\begin{equation}
\begin{aligned}
&\mathcal{S}^{wh}_{(0)} := \frac{1}{6} \left(2 e^{6 \beta_+}+1\right) e^{2\alpha-4 \beta_+},\\&
\mathcal{S}^{wh}_{(1)} := \left(-\frac{\text{B}}{2}-1\right)\alpha -\beta_+
\end{aligned}
\end{equation}
 we will integrate the transport equations along the classical Hamilton Jacobi flows which are produced by (83) and use them to show via induction that smooth and globally defined solutions exist for all of the higher order $k \geq 2$ transport equations.

With the appropriate units upon setting $c=1, G=1$, and $L=1$ if we were to apply standard variational techniques to the Einstein-Hilbert action

\begin{equation}
I_{EH}=\frac{1}{16 \pi} \int_{\Omega} \sqrt{-\operatorname{det}^{(4)} g} \left(^{4}R\left(^{(4)} g\right)\right) d^{4} x
\end{equation}
for the metric (1) we would derive the following action 

\begin{equation}
I_{\mathrm{ADM}}=\int_{I} d t\left\{p_{\alpha} \dot{\alpha}+p_{+} \dot{\beta}_{+}-N \mathcal{H}_{\perp}\right\},
\end{equation}
where

\begin{equation}
\begin{aligned} \mathcal{H}_{\perp} : &=\frac{(6 \pi)^{1 / 2} }{4  e^{3 \alpha}} \left( \left(-p_{\alpha}^{2}+p_{+}^{2}\right) +  e^{4 \alpha} \Biggl[\frac{e^{-8 \beta_{+}}}{3}-\frac{4 e^{-2 \beta_{+}}}{3}   \Biggr] \right).
\end{aligned}
\end{equation}
If we vary (85) with respect to the lapse $N$ which acts as a ‘Lagrange multiplier’ we obtain the Einstein equation known as the Hamilitonian constraint 

\begin{equation}
\begin{aligned} \mathcal{H}_{\perp}\left(\alpha,\beta_+,p_{\alpha},p_{+}\right)=0.
\end{aligned}
\end{equation}
An explicit demonstration of these standard variational techniques applied to the full Bianchi IX models which can trivially be repurposed for metric (1) as can be seen in \cite{moncrief2014euclidean}.

It can be identified that our Hamililton Jacobi equation (12) when both $\Lambda $ and $b$ vanish is related the Euclidean-signature form of our Hamililtonian constraint 

\begin{equation}
\begin{aligned} 
N_{\text {Eucl }}\mathcal{H}_{Eucl:=}N_{\text {Eucl }}\frac{(6 \pi)^{1 / 2} }{4  e^{3 \alpha}} \left( \left(p_{\alpha}^{2}-p_{+}^{2}\right) +  e^{4 \alpha} \Biggl[\frac{e^{-8 \beta_{+}}}{3}-\frac{4 e^{-2 \beta_{+}}}{3}   \Biggr] \right)
\end{aligned}
\end{equation}
 by the substitution

\begin{equation}
\begin{aligned} 
p_{\alpha}\xrightarrow{}\frac{\partial \mathcal{S}^{wh}_{(0)}}{\partial \alpha} \\ p_{+}\xrightarrow{}\frac{\partial \mathcal{S}^{wh}_{(0)}}{\partial \beta_+}.
\end{aligned}
\end{equation}
Keeping this in mind we can obtain the following differential equations for $\alpha(t)$ and  $\beta_+(t)$

\begin{equation}
\begin{aligned}
\dot{\alpha} &=\frac{(6 \pi)^{1 / 2} }{2 e^{3 \alpha}} N_{\text {Eucl }} p_{\alpha} \\
\dot{\beta}_{+} &=\frac{-(6 \pi)^{1 / 2} }{2  e^{3 \alpha}} N_{\text {Eucl }} p_{+}
\end{aligned}
\end{equation}
where $p_{\alpha}$ and $p_{+}$ are (89). We can easily obtain explicit solutions to these flow equations by exploiting our freedom to choose the lapse $N_{\text {Eucl }}$ to be any  smooth, globally defined nonvanishing function of the Misner variables. We have this freedom to choose $N_{\text {Eucl }}$ because it acts as a choice of gauge which bestows a physical meaning to our evolution parameter "t". We require our lapse to not vanish in order to ensure that no catastrophic signature changes occur in our space-time. 

Via (89) and (83) we obtain 

\begin{equation}
\begin{aligned} 
p_{\alpha}=\frac{1}{3} \left(2 e^{6 \beta_+(t)}+1\right) e^{2 \alpha(t)-4 \beta_+(t)}\\p_{+}=\frac{2}{3} \left(e^{6 \beta_+(t)}-1\right) e^{2 \alpha(t)-4 \beta_+(t)}
\end{aligned}
\end{equation}
which results in these flow equations for $\alpha(t)$ and $\beta_+(t)$ for general $N_{\text {Eucl }}$

\begin{equation}
\begin{aligned}
\dot{\alpha} &=\sqrt{\frac{\pi }{6}} \left(2 e^{6 \beta_+(t)}+1\right) N_{\text {Eucl }} e^{-\alpha(t)-4 \beta_+(t)} \\
\dot{\beta}_{+} &=-\sqrt{\frac{2 \pi }{3}} \left(e^{6 \beta_+(t)}-1\right) N_{\text {Eucl }} \left(e^{-\alpha(t)-4 \beta_+(t)}\right).
\end{aligned}
\end{equation}
If we set 

\begin{equation}
\begin{aligned}
N_{\text {Eucl }}=\sqrt{\frac{6}{\pi }} e^{\alpha(t)+4 \beta_+(t)}
\end{aligned}
\end{equation}
and insert it into our flow equations (92) we obtain 

\begin{equation}
\begin{aligned}
\dot{\alpha} &=1+2e^{6\beta_+(t)} \\
\dot{\beta}_{+} &=2-2e^{6\beta_+(t)}
\end{aligned}
\end{equation}
which can be trivially solved resulting in 

\begin{equation}
\begin{aligned}
\alpha(t) &=\alpha_{0}+\frac{1}{6} \log \left(e^{6 \beta_{+0}
} \left(e^{12 t}-1\right)+1\right)+t \\
\beta_+(t) &=-\frac{1}{6} \log \left(\left(e^{-6 \beta_{+0}
}-1\right) e^{-12 t}+1\right)
\end{aligned}
\end{equation}
where $\alpha_{0}$ and $\beta_{+0}$ are the initial values of $\alpha$ and $\beta_{+}$ when t=0. For the following allowable range of initial values $\alpha_{0} \in (-\infty,\infty)$ ,  $\beta_{+0} \in (-\infty,\infty)$ our solutions are globally defined and real when the evolution parameter t ranges from $(0,\infty)$. Classically these solutions are the flow in minisuperspace induced by (83), measured with respect to our evolution parameter whose physical meaning is derived from our choice of lapse (93). Our solutions for the evolution of $\alpha$ and $\beta_+$ (95) reveal that the flow in minisuperspace induced by our 'wormhole' solution leads to some interesting geometrical implications for the Euclidean-signature form of our Taub space-time(1)

\begin{equation}
\begin{aligned}
&d s^{2}=N_{\text {Eucl }}^{2}d t^{2}+e^{2 \alpha(t)}\left(e^{2 \beta(t)}\right)_{a b} \omega^{a} \omega^{b}\\& \left(e^{2 \beta(t)}\right)_{a b}=\operatorname{diag}\left(e^{2 \beta\left(t\right)_{+}}, e^{2 \beta\left(t\right)_{+}}, e^{-4 \beta\left(t\right)_{+}}\right).
\end{aligned}
\end{equation}
Regardless of our initial conditions $\left(\alpha_{0},\beta_{+0}\right)$ when our evolution parameter $t \to \infty$, $\alpha \xrightarrow{} \infty $ and $\beta_+ \xrightarrow{} 0 $. Geometrically this means our 'wormhole' flow induces in our metric (96) the following evolution; initially the space described by (96) has a finite radius dictated by $\alpha_{0}$ and a certain amount of anisotropy determined by $\beta_{+0}$. However as our evolution parameter grows our space-time will approach a flat isotropic state due to the aforementioned $t \xrightarrow{} \infty$ behavior of $\alpha$ and $\beta+$. It is because of this behavior that our $\mathcal{S}^{wh}_{(0)}$ is called the 'wormhole solution'. It should be noted that despite using Euclidean-signature Hamilton Jacobi flows; we are proving the existence of asymptotic solutions to the Lorentzian-signature Wheeler DeWitt equation.

Moving forward with our proof we will choose this natural looking ansatz as our form for the higher order $k \geq 2$ quantum corrections 

\begin{equation}
S_{(k)}^{\mathrm{wh}}=6 e^{-2(k-1) \alpha} \Sigma_{(k)}^{\mathrm{wh}}\left(\beta_{+}\right).
\end{equation}
In terms of this ansatz our 'wormhole' $\mathcal{S}^{wh}_{(0)}$ has the followings $\Sigma_{(0)}$

\begin{equation}
\Sigma_{(0)}^{\mathrm{wh}}=\frac{1}{6}\left(e^{-4 \beta_{+}}+2 e^{2 \beta_{+}}\right)
\end{equation}
so that 

\begin{equation}
\mathcal{S}_{(0)}^{\mathrm{wh}}=e^{2 \alpha} \Sigma_{(0)}^{\mathrm{wh}}\left(\beta_{+}\right).
\end{equation}
Using (99), (98), (97), and (83) we can rewrite our  $3\geq k \geq 2$ transport equations as 

\begin{equation}
\frac{\partial \Sigma_{(0)}^{\mathrm{wh}}}{\partial \beta_{+}} \frac{\partial \Sigma_{(2)}^{\mathrm{wh}}}{\partial \beta_{+}}+4 \Sigma_{(0)}^{\mathrm{wh}} \Sigma_{(2)}^{\mathrm{wh}}=-\frac{B^{2}}{4},
\end{equation}
\begin{equation}
\frac{\partial \Sigma_{(0)}^{\mathrm{wh}}}{\partial \beta_{+}} \frac{\partial \Sigma_{(3)}^{\mathrm{wh}}}{\partial \beta_{+}}+8 \Sigma_{(0)}^{\mathrm{wh}} \Sigma_{(3)}^{\mathrm{wh}}=\frac{3}{2}\frac{\partial^{2} \Sigma_{(2)}^{\mathrm{wh}}}{\partial \beta_{+}^{2}}+3\frac{\partial \Sigma_{(2)}^{\mathrm{wh}}}{\partial \beta_{+}},
\end{equation}
and our $ k \geq 4$ equations as 
\begin{equation}
\begin{aligned}
&\frac{\partial \Sigma_{(0)}^{\mathrm{wh}}}{\partial \beta_{+}} \frac{\partial \Sigma_{(k)}^{\mathrm{wh}}}{\partial \beta_{+}}+4 (k-1) \Sigma_{(0)}^{\mathrm{wh}} \Sigma_{(k)}^{\mathrm{wh}}=\sum _{l=2}^{k-2} \frac{k! \left(4 (l-1) (k-l-1) \Sigma_{(l)}^{\mathrm{wh}} \Sigma_{(k-l)}^{\mathrm{wh}}-\frac{\partial \Sigma_{(l)}^{\mathrm{wh}}}{\partial \beta_{+}} \frac{\partial \Sigma_{(k-l)}^{\mathrm{wh}}}{\partial \beta_{+}}\right)}{2 l!
   (k-l)!} \\& -\frac{1}{2} k \left(-2 \frac{\partial \Sigma_{(k-1)}^{\mathrm{wh}}}{\partial \beta_{+}}-\frac{\partial^{2} \Sigma_{(k-1)}^{\mathrm{wh}}}{\partial \beta_{+}^{2}}+4 (k-3) (k-2) \Sigma_{(k-1)}^{\mathrm{wh}}\right).
\end{aligned}
\end{equation}

When $\beta_+=0$, $\frac{\partial \Sigma_{(0)}^{\mathrm{wh}}}{\partial \beta_{+}}=\frac{2 e^{ 0}}{3}-\frac{2}{3} e^{0}=0$  and $\Sigma_{(0)}^{\mathrm{wh}}(0)=\frac{1}{2}$. Thus we can sequentially write out $\Sigma_{(k)}^{\mathrm{wh}}(0)$ as follows 

\begin{equation}
\Sigma_{(2)}^{\mathrm{wh}}(0)=-\frac{B^{2}}{8},
\end{equation}

\begin{equation}
\begin{aligned}
\Sigma_{(3)}^{\mathrm{wh}}(0)=\frac{1}{4}\left[\frac{3}{2}\frac{\partial^{2} \Sigma_{(2)}^{\mathrm{wh}}}{\partial \beta_{+}^{2}}+3\frac{\partial \Sigma_{(2)}^{\mathrm{wh}}}{\partial \beta_{+}}\right]&(0), \\& etc \cdots
\end{aligned}
\end{equation}

As the reader can verify using (94) and (98)

\begin{equation}
-3e^{4\beta_+}\left(\frac{\partial \Sigma_{(0)}^{\mathrm{wh}}}{\partial \beta_{+}} \frac{\partial \Sigma_{(k)}^{\mathrm{wh}}}{\partial \beta_{+}}\right)= \frac{d\Sigma_{(k)}^{\mathrm{wh}}}{dt},
\end{equation}
which allows us to convert (100), (101) and (102) into   

\begin{equation}
\frac{d\Sigma_{(k)}^{\mathrm{wh}}}{dt}+4 \left(-e^{6 \beta_+}-\frac{1}{2}\right) (k-1)\Sigma_{(k)}^{\mathrm{wh}} = \frac{d\Sigma_{(k)}^{\mathrm{wh}}}{dt}-2\frac{d\alpha}{dt} (k-1)\Sigma_{(k)}^{\mathrm{wh}}= \Lambda_{k}
\end{equation}
where $\Lambda_{k}$ denotes the right hand side of the original equations (100), (101), or (102) multiplied by $-3e^{4\beta_+}$. If we start with k=2 and integrate (106) we obtain

\begin{equation}
\Sigma_{(2)}^{\mathrm{wh}}(\beta(t)_{+})= e^{2 \alpha(t)} \left(\Sigma_{(2)}^{\mathrm{wh}}(\beta_{+0})+\int_1^t \frac{3}{4} \text{B}^2 e^{4 \beta_+(s)-2 \alpha(s)} \, ds\right).
\end{equation}
As we previously established as $t \to \infty$ so does $\alpha(t)$. Thus in order to ensure that our quantum corrections $\mathcal{S}_{(k \geq 2)}$ are smooth and globally defined we must use our freedom to pick $\Sigma_{(2)}^{\mathrm{wh}}(\beta_{+0})$ so that the term proportional to $e^{2 \alpha(t)}$ vanishes as $t \to \infty$. Because our range of t is from $(0,\infty)$, in order for our term which is proportional to $e^{2 \alpha(t)}$ to vanish we must equate 

\begin{equation}
\Sigma_{(2)}^{\mathrm{wh}}(\beta_{+0})=-\int_1^\infty \frac{3}{4} \text{B}^2 e^{4
   \beta_+(s)-2 \alpha(s)} \, ds.
\end{equation}
This choice of $\Sigma_{(2)}^{\mathrm{wh}}(\beta_{+0})$ allows us to rewrite (107) as

\begin{equation}
\Sigma_{(2)}^{\mathrm{wh}}(\beta(t)_{+})= -e^{2 \alpha(t)} \left(\int_t^\infty \frac{3}{4} \text{B}^2 e^{4
   \beta_+(s)-2 \alpha(s)} \, ds\right).
\end{equation}
which facilitates us using L'H$\hat{o}$pital's rule to show that the desired limit (103) as $t\to \infty$ is reached
\begin{equation}
\begin{aligned}
& -\frac{ \left(\int_t^\infty \frac{3}{4} \text{B}^2 e^{4
   \beta_+(s)-2 \alpha(s)} \, ds\right)}{e^{-2 \alpha(t)}}\\&  -\frac{\frac{3}{4} \text{B}^2 e^{4
   \beta_+(t)-2 \alpha(t)} }{2\dot{\alpha}e^{-2 \alpha(t)}} \\& -\frac{\frac{3}{4} \text{B}^2 e^{4
   \beta_+(t)} }{2\dot{\alpha}} \\&\lim_{t\to\infty} \Sigma_{(2)}^{\mathrm{wh}}(\beta(t)_{+})=\lim_{t\to\infty} -\frac{\frac{3}{4} \text{B}^2 e^{4
   \beta_+(t)} }{2\dot{\alpha}}=-\frac{B^2}{8},
\end{aligned}
\end{equation}
where $\lim_{t\to\infty} \beta(t)_{+}=0$. 
The reader can easily verify the $\lim_{t\to\infty} \dot{\alpha}=3$. After inserting our expressions for $\alpha(t)$ and $\beta_+(t)$ into (108) 
\begin{equation}
\Sigma_{(2)}^{\mathrm{wh}}(\beta_{+0})=-\int_1^\infty \frac{3 \text{B}^2 e^{-2 (\alpha_{0}+s)}}{4 \left(\left(e^{-6 \beta_{+0}}-1\right) e^{-12 s}+1\right)^{2/3} \sqrt[3]{e^{6 \beta_{+0}} \left(e^{12
   s}-1\right)+1}} ds
\end{equation}
we can verify that $\Sigma_{(2)}^{\mathrm{wh}}$ is smooth and globally defined by observing that as long as $\beta_{+0}$ is real that taking the derivative of (111) an arbitrary number of times with respect to  $\beta_{+0}$ does not disturb the convergence of this integral(111). The $e^{-2\alpha_{0}}$ is a constant which could have been absorbed into $\Sigma_{(2)}^{\mathrm{wh}}(\beta_{+0})$. 

Now that we have shown that $\Sigma_{(2)}^{\mathrm{wh}}$ is smooth and globally defined we can move on to computing the higher order $\Sigma_{(k)}^{\mathrm{wh}}$ terms. Assuming that $\left\{\Sigma_{(2)}^{\mathrm{Wh}}, \ldots, \Sigma_{(k-1)}^{\mathrm{Wh}}\right\}$, for $k \geq 2$ have been shown to be smooth and globally defined we can express $\Sigma_{(k)}^{\mathrm{wh}}$ as 

\begin{equation}
\Sigma_{(k)}^{\mathrm{wh}}(\beta(t)_{+})=e^{2 (k-1) \alpha(t)} \left(\Sigma_{(k)}^{\mathrm{wh}}(\beta_{+0})+\int_1^t  e^{-2 (k-1) \alpha(s)}\Lambda_{\left(k\right)}(s)  ds\right).
\end{equation}
There is only one choice for $\Sigma_{(k)}^{\mathrm{wh}}(\beta_{+0})$ which allows the kth quantum correction to be smooth and globally defined 
\begin{equation}
\Sigma_{(k)}^{\mathrm{wh}}(\beta_{+0})=-\int_1^\infty  e^{-2 (k-1) \alpha(s)}\Lambda_{\left(k\right)}(s)  ds.
\end{equation}
Via inspection of (101) and (102) it can be concluded that our $\Lambda_{\left(k\right)}$ term is solely composed of a sum of our aforementioned smooth and globally defined functions $\left\{\Sigma_{(2)}^{\mathrm{Wh}}, \ldots, \Sigma_{(k-1)}^{\mathrm{Wh}}\right\}$ and  their derivatives with respect to $\beta_+$, multiplied by the smooth and globally defined function $-3e^{4\beta_+(t)}$. Thus our source terms for our transport $\Sigma_{(k)}^{\mathrm{wh}}$ equations are always globally defined. Furthermore because of the exponential decay of $ e^{-2 (k-1) \alpha(s)}$ for $k \geq 2$ as $s \to \infty$, our integral (113) always converges and is smooth and globally defined. Using this choice of $\Sigma_{(k)}^{\mathrm{wh}}(\beta_{+0})$ we can rewrite (112) as

\begin{equation}
\Sigma_{(k)}^{\mathrm{wh}}(\beta(t)_{+})=-e^{2 (k-1) \alpha(t)} \left(\int_t^\infty  e^{-2 (k-1) \alpha(s)}\Lambda_{\left(k\right)}(s) ds\right),
\end{equation}
which allows one to easily apply  L'H$\hat{o}$pital's rule to show that the desired limit as $\beta_+ \to 0$ when $t \to \infty$ is achieved. If we were to insert our expressions for $\alpha(t)$ into (114) it would be straightforward to conclude that arbitrary differentiation with respect to $\beta_{+0}$ does not disturb its convergence and that (114) remains smooth and globally defined. 

This conclude ours proof by induction that smooth and globally defined solutions exist for arbitrary ordering parameter B for all of the 'ground' state quantum corrections of the form (97) when (98), (99) and (83) hold. In the process we have also proved the existence of a full asymptotic solution of the form $\stackrel{(0)}{\Psi}_{\hbar}=e^{- \mathcal{S}_{(0)}-\mathcal{S}_{(1)}-\frac{1}{2 !} \mathcal{S}_{(2)}-\cdots}$ for any arbitrary ordering parameter. Hopefully the nature of the convergence of this asymptotic solution can be explored in a future work. For information on how to prove that the quantum corrections associated with the 'excited' state transport equations are smooth and globally defined we refer the reader to section 5 of \cite{moncrief2014euclidean} and to the references contained within it. 

Even though the vacuum quantum Taub models are solvable\cite{martinez1983exact} using separation of variables for any ordering parameter, it isn't trivial to obtain solutions which possess the forms of (11) and (17). Via ordinary separation of variables the Taub WDW equation admits two Bessel functions as its solutions which in principle can be used to construct all of the solutions of the quantum Taub models via superposition. However only a limited number of integrals involving Bessel function are known in closed form. In addition even if one can evaluate those integrals numerically there is still the issue of using the correct kernel to obtain wave functions which possess non-trivial characteristics. 

These non-trivial characteristics include the wave function's behavior being dependent upon $\alpha$ which as was previously mentioned is our internal clock and also dictates the size of our Taub universes. Another non-trivial feature is the manifestation of discreteness in our wave functions which is showcased in our 'excited states that we have computed thus far. The Euclidean-signature semi classical method is able to bypass the aforementioned difficulties and obtain closed form wave functions which possess these non-trivial characteristics.

We went through the trouble of integrating the transport equations along the flow generated by $\mathcal{S}^{wh}_{(0)}$ in order to prove that smooth and globally defined solutions exist for all of them. However due to the mathematical simplicity present in the quantum Taub models we can directly solve for an arbitrary kth level quantum correction $\mathcal{S}^{wh}_{(k)}(\alpha,\beta_+)$ by solving elementary differential equations. The explicit form of the  $\mathcal{S}^{wh}_{(2)}(\alpha,\beta_+)$ transport equation when our ansatz (96) is employed is 

\begin{equation}
8 e^{-4 \beta_+} \left(
\frac{\partial \Sigma_{(2)}^{\mathrm{wh}}}{\partial \beta_+}-e^{6 \beta_+} \left(\frac{\partial \Sigma_{(2)}^{\mathrm{wh}}}{\partial \beta_+}+2 \Sigma_{(2)}^{\mathrm{wh}}\right)-\Sigma_{(2)}^{\mathrm{wh}}\right)-3 \text{B}^2=0;
\end{equation}
which can be easily solved, yielding 
\begin{equation}
\mathcal{S}^{wh}_{(2)}(\alpha,\beta_+):= e^{-2 \alpha} \left(\frac{e^{\beta_+} \text{B}^2 \sin ^{-1}\left(e^{3 \beta_+}\right)}{8 \sqrt{1-e^{6
   \beta_+}}}+\frac{e^{\beta_+} c_1}{\sqrt{1-e^{6 \beta_+}}}\right).
\end{equation}
This expression is only smooth and globally defined when a specific value of $c_{1}$ is chosen. As $\beta_+ \to 0$ the first term of (116) which is proportional to $\frac{1}{\sqrt{1-e^{6 \beta_+}}}$ approaches $\frac{B^{2}\pi}{16}$. In order for (116) to not blow up we must ensure that $\frac{1}{8} e^{\beta_+} \text{B}^2 \sin ^{-1}\left(e^{3 \beta_{+}}\right)+e^{\beta_+} c_1$ vanishes when  $\beta_+ \to 0$, which is accomplished if we set $c_{1}=-\frac{B^{2}\pi}{16}$. This results in the following $k=2$ quantum correction 

\begin{equation}
\mathcal{S}^{wh}_{(2)}(\alpha,\beta_+):= -e^{-2\alpha}\left(\frac{e^{\beta_+} \text{B}^2 \cos ^{-1}\left(e^{3 \beta_{+}}\right)}{8 \sqrt{1-e^{6 \beta_+}}}\right)
\end{equation}
which has the established limit (103) when $\beta_+ \to 0$ as the reader can verify. 

Thanks to the existence and uniqueness theorem for ordinary differential equations, each one of our solutions to the 'ground' state transport equations we found by integrating along the flow of $\mathcal{S}^{wh}_{(0)}$ are equivalent to the explicit solutions of those same transport equations as long as they are smooth and globally defined. Because we were able to show that picking the right integration constant for a particular $\mathcal{S}^{wh}_{(k)}$ results in it being a smooth and globally defined function, we know that we can obtain smooth and globally defined explicit solutions such as (116) by picking a unique constant of integration. Thus all of our results regarding smooth and globally defined solutions to the 'ground' state transport equations we previously obtained also apply to our explicit solutions. 

Our $\mathcal{S}^{wh}_{(2)}$ quantum correction has the interesting property that it is undefined in the real plane when $\beta_+ >0$ due to the domain of $\cos ^{-1}\left(e^{3 \beta_{+}}\right)$. However $\cos ^{-1}\left(e^{3 \beta_{+}}\right)$ can be analytically continued into the complex plane by expressing it as an integral 
\begin{equation}
\cos ^{-1}\left(e^{3 \beta_+}\right):= \int_{-\infty }^{\beta_+} \frac{3 e^{3 x}}{\sqrt{1-e^{6 x}}} \, dx-\frac{\pi}{2}.
\end{equation}
Despite analytically continuing $\cos ^{-1}$ into the complex plane the  quantum corrections we computed thus far remain real valued functions because every instance of $\cos ^{-1}\left(e^{3 \beta_+}\right)$ is multiplied by a term which becomes complex valued when $\beta_{+} >0$, which ultimately results in a real expression.

We can go further and explicitly compute the 'ground' state $k=3$ quantum correction which is 

\begin{equation}
\begin{aligned} 
 \mathcal{S}^{wh}_{(3)}(\alpha,\beta_+):= \frac{\text{B}^2 e^{-4\alpha +2\beta_{+}}}{32 \left(1-e^{6 \beta_{+}}\right)^{5/2}}\Biggl(2& e^{6 \beta_{+}} \sqrt{1-e^{6 \beta_{+}}} \text{B} \cos ^{-1}\left(e^{3 \beta_{+}}\right)^2+2
   \sqrt{1-e^{6 \beta_{+}}} \text{B} \cosh ^{-1}\left(e^{3 \beta_{+}}\right)^2\\&+3
   \sqrt{1-e^{6 \beta_{+}}} \left(e^{6 \beta_{+}}-4\right)-3 e^{3 \beta_{+}} \left(2 e^{6
   \beta_{+}}-5\right) \cos ^{-1}\left(e^{3 \beta_{+}}\right)\Biggr)
\end{aligned}
\end{equation}

and in principle compute further more complicated explicit expressions for the higher order 'ground' state quantum corrections. 

We will now move on and compute the $\phi^{wh}_{(2)}$ quantum correction for the first and second 'excited' states of the wormhole quantum Taub models. We will seek 'excited' state quantum corrections which possess the form 

\begin{equation}
\phi^{wh}_{(k)}=e^{(4n-2 k) \alpha} \chi_{(k)}\left(\beta_{+}\right).
\end{equation}
For our $\phi^{wh}_{(0)}$ we will use $\left(\left(e^{6 \beta_+}-1\right) e^{4 \alpha-2 \beta_+}\right)^{n}$ , and when n=1 our $\phi_{(1)}$ is $-6 e^{2 (\alpha+\beta_+)}$. Thus using (117) we can write out the $\phi^{wh}_{(2)}$ transport equation for the first 'excited' wormhole state  

\begin{equation}
\begin{aligned}
9 e^{3 \beta_+} \text{B}^2 \left(\sqrt{1-e^{6 \beta_+}} \left(e^{3 \beta_+}+2 e^{9 \beta_+}\right)+\left(1-4 e^{6 \beta_+}\right) \cos ^{-1}\left(e^{3 \beta_+}\right)\right)-8
   \left(1-e^{6 \beta_+}\right)^{5/2} 
\frac{\partial \chi_{(2)}}{\partial \beta_+}=0,
\end{aligned}
\end{equation}
whose solution is 
\begin{equation}
\begin{aligned}
\phi^{wh}_{2}=\chi_{(2)}=-\frac{3 \text{B}^2 \left(e^{6 \beta_+}+e^{3 \beta_+} \sqrt{1-e^{6 \beta_+}} \left(2 e^{6 \beta_+}-1\right) \cos ^{-1}\left(e^{3 \beta_+}\right)-1\right)}{8 \left(e^{6
   \beta_+}-1\right)^2}.
\end{aligned}
\end{equation}

For $n=2$, $\phi^{wh}_{1}=-6 e^{6 \alpha} \left(4 e^{6 \beta_+}-3\right)$, which allows us to write down the following $\phi^{wh}_{(2)}$ equation for the second 'excited' state 

\begin{equation}
\begin{aligned}
\\& 9 e^{\beta_{+}} \left(e^{3 \beta_+} \sqrt{1-e^{6 \beta_+}} \left(\left(2 e^{6  \beta_+}+1\right) \text{B}^2-576\right)-\left(4 e^{6  \beta_+}-1\right) \text{B}^2 \cos ^{-1}\left(e^{3
    \beta_+}\right)\right) \\& +4 \sqrt{1-e^{6  \beta_+}} \left(e^{6  \beta_+} \left(4 \chi_{(2)}(\beta_+)-\frac{\partial \chi_{(2)}}{\partial \beta_+}\right)+\frac{\partial \chi_{(2)}}{\partial \beta_+}+2 \chi_{(2)}(\beta_+)\right)=0,
\end{aligned}
\end{equation}
whose solution allows us to write the following as our $\phi^{wh}_{(2)}$ quantum correction for the second 'excited' state for arbitrary Hartle-Hawking ordering parameter B
\begin{equation}
\begin{aligned}
\phi^{wh}_{(2)}=e^{4\alpha}\frac{3 e^{-2 \beta_+} \left(\left(e^{6 \beta_+}-1\right) \left(288-\text{B}^2\right)-e^{3 \beta_+} \sqrt{1-e^{6 \beta_+}} \left(2 e^{6 \beta_+}-1\right) \text{B}^2 \cos
   ^{-1}\left(e^{3 \beta_+}\right)\right)}{4 \left(e^{6 \beta_+}-1\right)}.
\end{aligned}
\end{equation}
In principle with the help of a computer algebraic system we can continue to find closed form quantum corrections for the 'ground' state equations and then use those quantum correction to find additional quantum corrections for the 'excited' state transport equations. 

Below we will plot 'ground' and 'excited' states for our 'wormhole' wave functions which include our analytically continued quantum corrections that we calculated above. The first three set of plots(figs 5a-5c) are of the 'wormhole' 'ground' states when we include the $\mathcal{S}^{wh}_{(2)}$ and $\mathcal{S}^{wh}_{(3)}$ quantum corrections when $B=-1$. The second set of three plots(figs 6a-6c) are of the second 'wormhole' 'excited' state which include the $\mathcal{S}^{wh}_{(2)}$, $\mathcal{S}^{wh}_{(3)}$, and $\phi^{wh}_{2}$ quantum corrections when $B=-1$.

\begin{figure}
\centering
\begin{subfigure}{.4\textwidth}
  \centering
  \includegraphics[scale=.12]{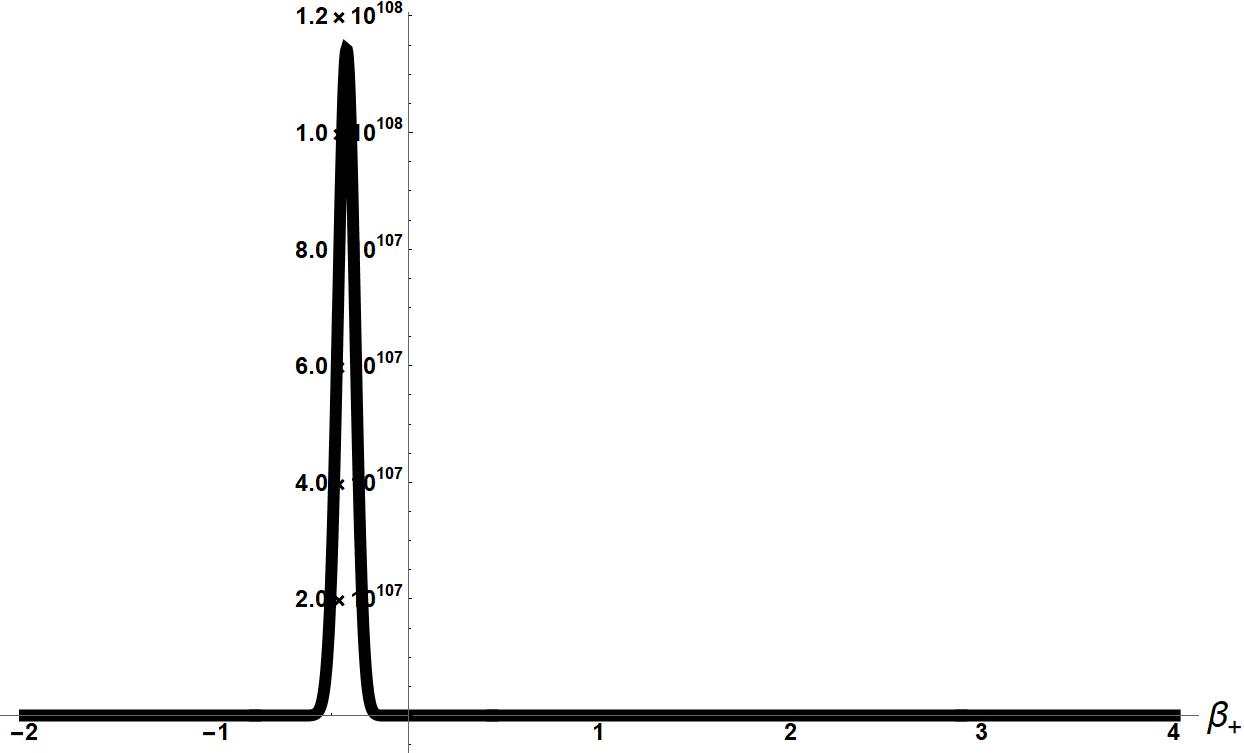}
  \caption{ $\alpha=-2$ \hspace{1mm} $B=-1$ }
  \label{fig:sub1}
\end{subfigure}%
\begin{subfigure}{.4\textwidth}
  \centering
  \includegraphics[scale=.12]{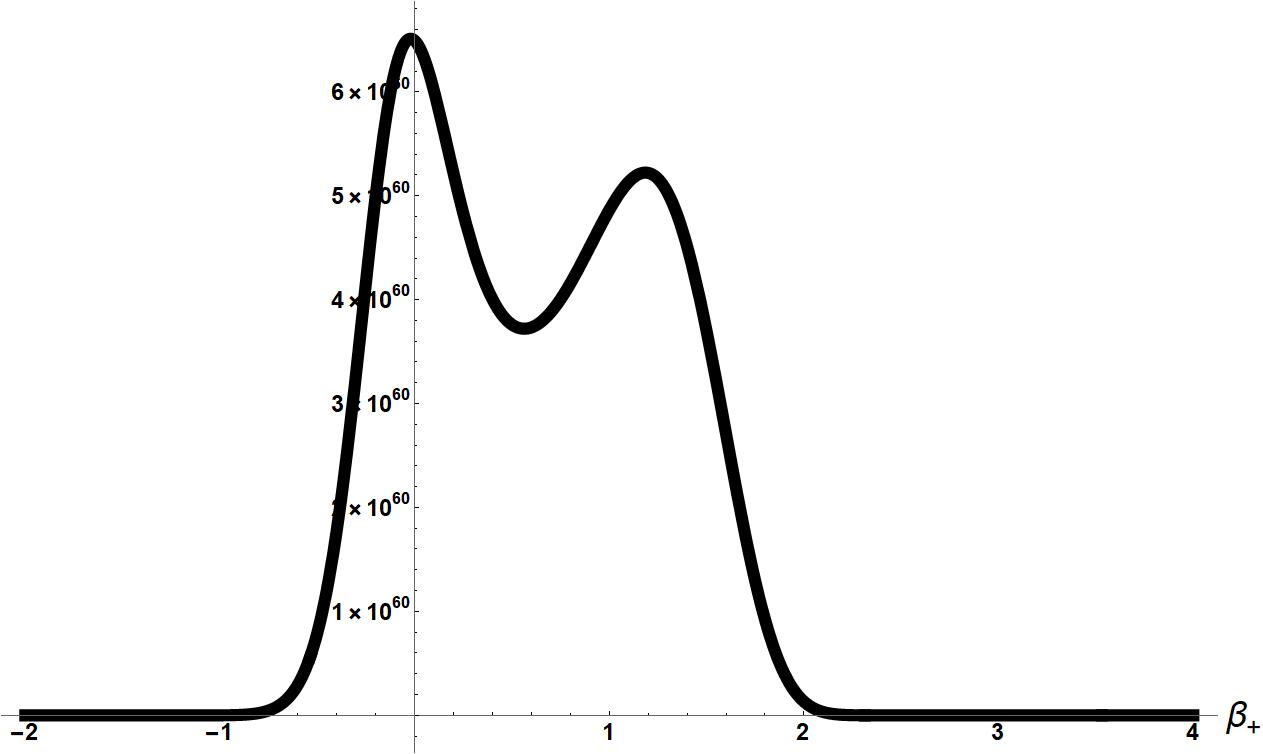}
  \caption{ $\alpha=-1.1$ \hspace{1mm} $B=-1$  }
  \label{fig:sub2}
\end{subfigure}
\begin{subfigure}{.4\textwidth}
  \centering
  \includegraphics[scale=.12]{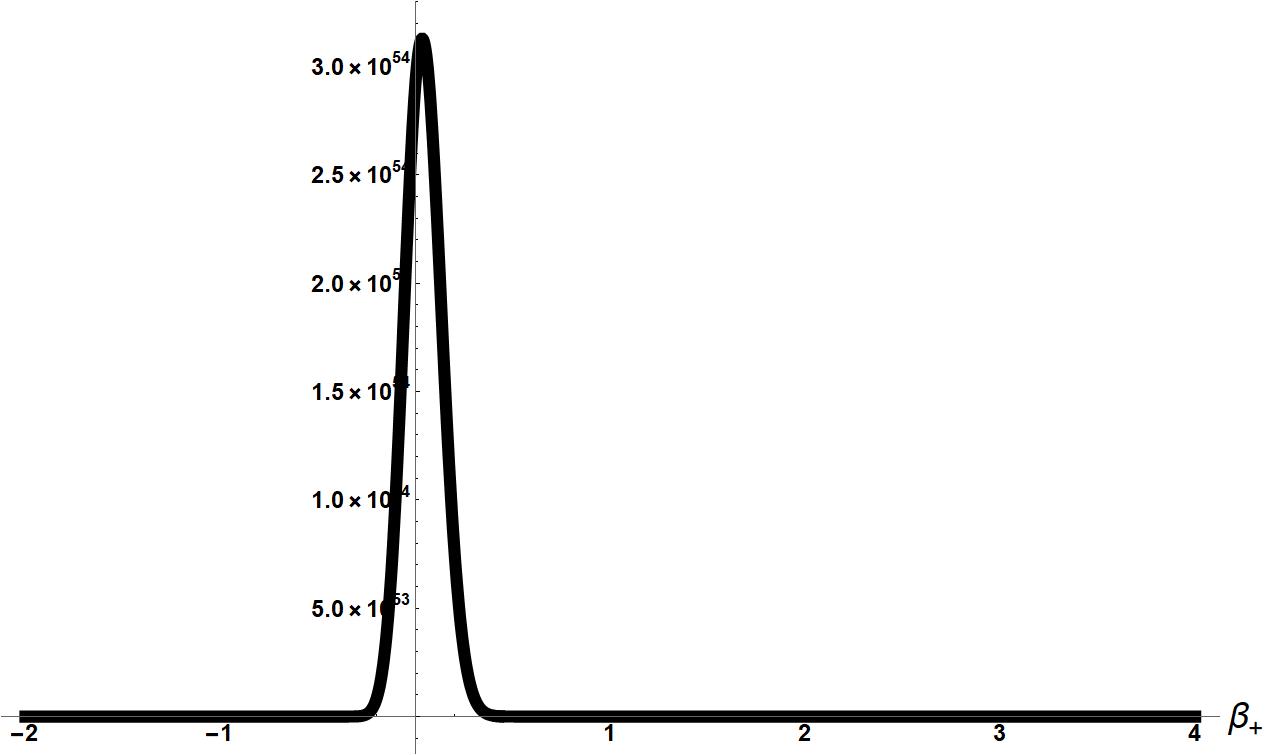}
  \caption{ $\alpha=1$ \hspace{1mm} $B=-1$  }
  \label{fig:sub1}
\end{subfigure}%
\caption{Plots of $\abs{\psi^{wh}}^{2}$ for the 'wormhole' vacuum quantum Taub 'ground' state constructed out of $\mathcal{S}^{wh}_{(0)}$, $\mathcal{S}^{wh}_{(1)}$,$\mathcal{S}^{wh}_{(2)}$, and $\mathcal{S}^{wh}_{(3)}$.}
\label{fig:test}
\end{figure}

\begin{figure}
\centering
\begin{subfigure}{.4\textwidth}
  \centering
  \includegraphics[scale=.13]{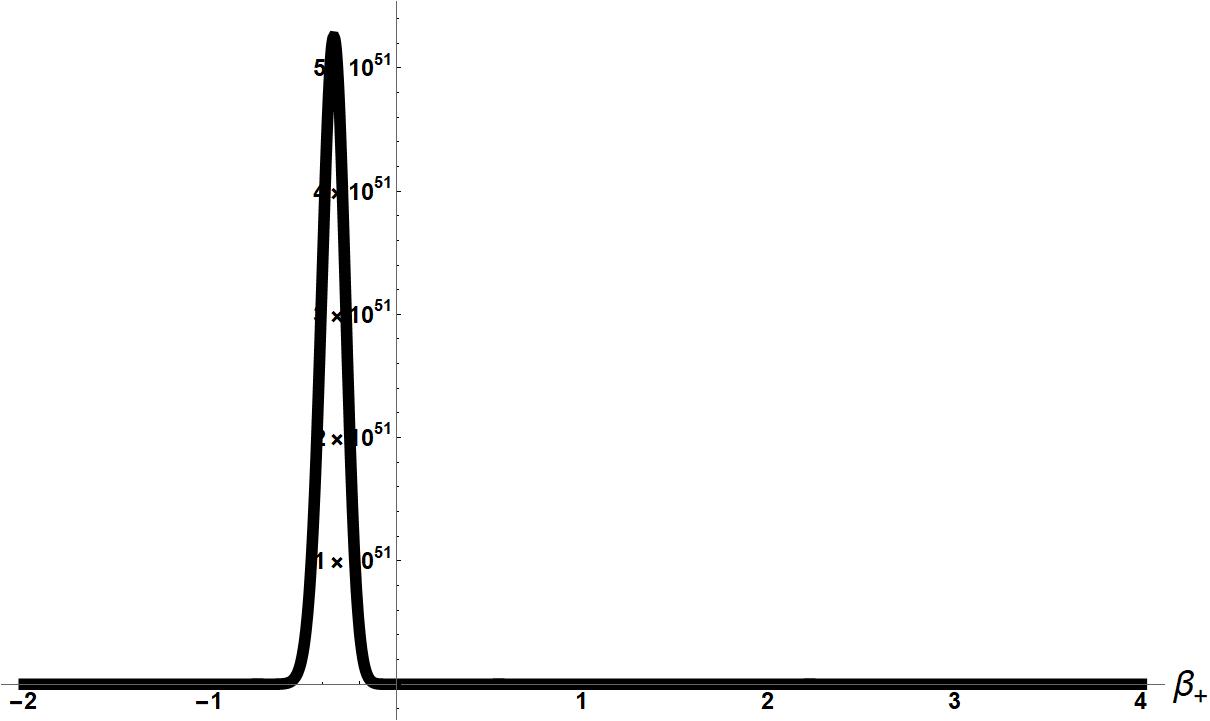}
  \caption{ $\alpha=-2$ $B=-1$  }
  \label{fig:sub1}
\end{subfigure}%
\begin{subfigure}{.4\textwidth}
  \centering
  \includegraphics[scale=.13]{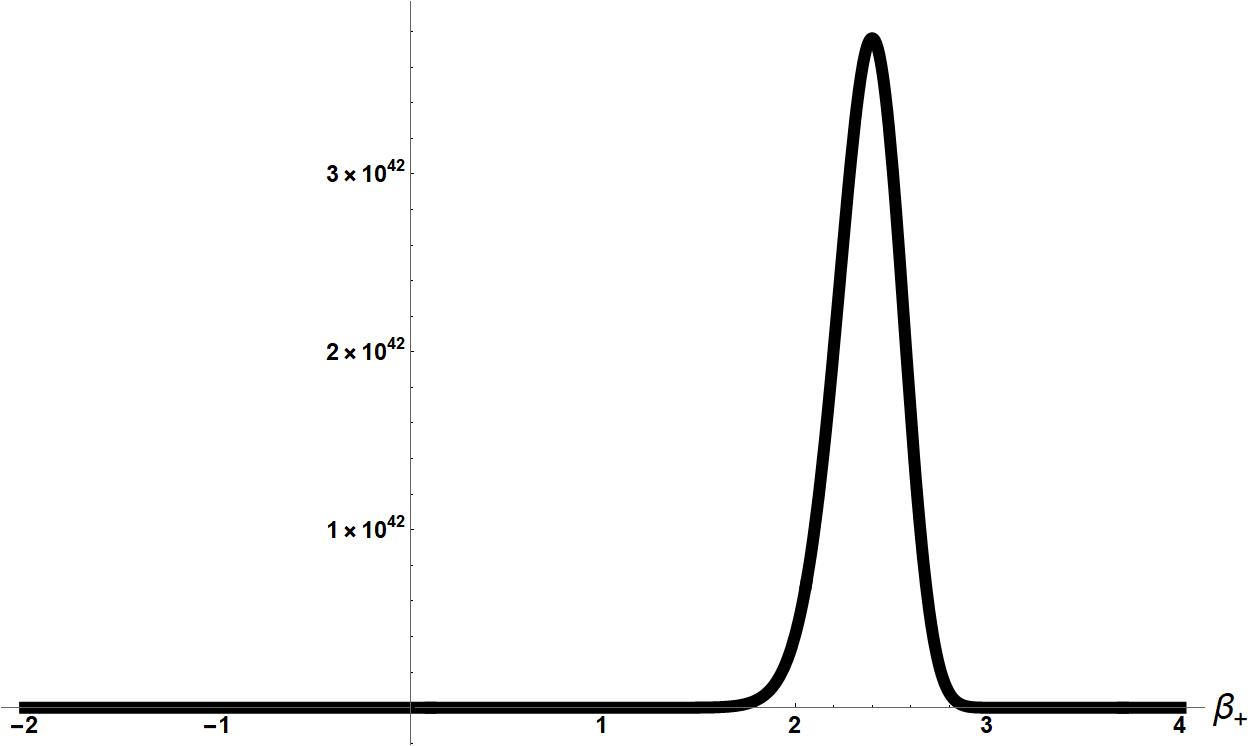}
  \caption{ $\alpha=-1.1$ $B=-1$  }
  \label{fig:sub2}
\end{subfigure}
\begin{subfigure}{.4\textwidth}
  \centering
  \includegraphics[scale=.13]{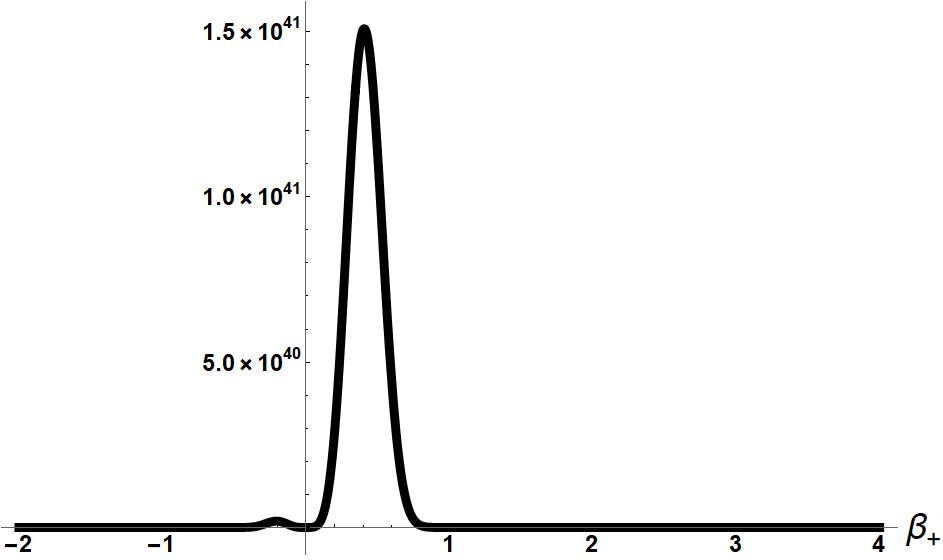}
  \caption{ $\alpha=1$ $B=-1$  }
  \label{fig:sub1}
\end{subfigure}%
\caption{Plots of $\abs{\psi_{n=2}^{wh}}^{2}$ for the second 'wormhole' vacuum quantum Taub 'excited' state constructed out of both 'ground' and 'excited' state quantum corrections up to $k=2$.}
\label{fig:test}
\end{figure}

We will qualitatively discuss the interesting features of these plots in our discussion section later. For now we will move on and derive the 'no boundary' asymptotic solutions for the quantum Taub models, which have different properties than our wormhole solutions and whose existence have implications for the Bianchi IX models.

\section{\label{sec:level1}Quantum Taub 'No Boundary' Wave Functions } 

The 'no boundary' $\mathcal{S}_{(0)}^{\mathrm{nb}}$ solution for the quantum Taub models can easily be found by setting the $\beta_{-}$ term in the 'no boundary' $\mathcal{S}_{(0)}$ \cite{graham1993anisotropic} for the full Bianchi IX models to zero, which results in 
\begin{equation}
\mathcal{S}_{(0)}^{\mathrm{nb}}:=\frac{1}{6} \left(1-4 e^{3 \beta_+}\right) e^{2 \alpha-4 \beta_+}.
\end{equation}

The reader can easily verify that the following $\mathcal{S}_{(1)}^{\mathrm{nb}}$ term satisfies equation (13) when $\mathcal{S}_{(0)}^{\mathrm{nb}}$ is inserted into it 
\begin{equation}
\begin{aligned}
\mathcal{S}_{(1)}^{\mathrm{nb}}:= \frac{1}{2} \alpha (4-\text{B})-\frac{5 \beta_+}{2}.
\end{aligned}
\end{equation}
If we seek a $\mathcal{S}_{(2)}^{\mathrm{nb}}$ which possesses the form of our ansatz (97) which we used earlier for the 'wormhole case', the resultant $\mathcal{S}_{(2)}$ differential equation is satisfied by a remarkably simple function compared to the 'wormhole' case
\begin{equation}
\begin{aligned}
\mathcal{S}_{(2)}^{\mathrm{nb}}:=\frac{1}{8} \left(\text{B}^2+9\right) e^{\beta_+-2 \alpha}.
\end{aligned}
\end{equation}
Unlike the 'wormhole' case whose $\Sigma_{(2)}$ was the rather exotic $-\text{B}^2e^{-2\alpha}\left(\frac{e^{\beta_+}  \cos ^{-1}\left(e^{3 \beta_{+}}\right)}{8 \sqrt{1-e^{6 \beta_+}}}\right)$ with its $ \cos ^{-1}\left(e^{3 \beta_{+}}\right)$ which needs to be analytically continued when $\beta_+ >0$, 'the no boundary' $\Sigma_{(2)}$ is simply $\frac{1}{8} \left(\text{B}^2+9\right) e^{\beta_+}$. This points to a fundamental difference between the 'wormhole' and 'no boundary' quantum Taub models. In addition this could suggest that the 'wormhole' and the 'no boundary' Bianchi IX models \cite{moncrief2014euclidean} possess significant differences as well. As a result of the simple nature of the $\mathcal{S}_{(2)}^{\mathrm{nb}}$ quantum correction we will be able to prove that an asymptotic solution exists for arbitrary ordering parameter for the 'no boundary' case using a more straightforward method than the one we applied to the 'wormhole' case. The existence of this 'no bounadry' asymptotic solution for the Taub models is possible evidence that an asymptotic 'no boundary' solution exists for the full quantum Bianchi IX models as well.   

Before we proceed it should be pointed out that if we allow the Hartle-Hawking ordering parameter to take on complex values, then we can form two closed form 'ground' state solutions to the Wheeler DeWitt equation using just (125), (126) when $B=\pm 3i$. A potential complication though of allowing the Hartle Hawking parameter to assume complex values is that it may fundamentally change the nature of the Wheeler DeWitt equation in such a way that it yields non-physical solutions. This issue of a complex Hartle-Hawking ordering parameter should be further investigated. For now though we will move forward assuming that the ordering parameter B is an arbitrary real number.  

After continuing the process of inserting the explicit forms of the $\mathcal{S}_{(k-1)}^{\mathrm{nb}}$ quantum corrections into the kth 'ground' state transport equations; the author found that each kth order quantum correction can be expressed as 
\begin{equation}
\begin{aligned}
\mathcal{S}_{(k \geq 2)}^{\mathrm{nb}}:=g(k) e^{\beta_+ (k-1)-2 \alpha (k-1)}.
\end{aligned}
\end{equation}
This is very straightforward to show. First we'll rewrite equation (14) as

\begin{equation}
\begin{aligned}
& 2\left[\frac{\partial \mathcal{S}_{(0)}}{\partial \alpha} \frac{\partial \mathcal{S}_{(k)}}{\partial \alpha}-\frac{\partial \mathcal{S}_{(0)}}{\partial \beta_{+}} \frac{\partial \mathcal{S}_{(k)}}{\partial \beta_{+}}\right]  {+k\left[B \frac{\partial \mathcal{S}_{(k-1)}}{\partial \alpha}-\frac{\partial^{2} \mathcal{S}_{(k-1)}}{\partial \alpha^{2}}+\frac{\partial^{2} \mathcal{S}_{(k-1)}}{\partial \beta_{+}^{2}}\right]} \\ & + \sum_{\ell=2}^{k-2} \frac{k !}{\ell !(k-\ell) !}\Biggr(\frac{\partial \mathcal{S}_{(\ell)}}{\partial \alpha} \frac{\partial \mathcal{S}_{(k-\ell)}}{\partial \alpha}-\frac{\partial \mathcal{S}_{(\ell)}}{\partial \beta_{+}} \frac{\partial \mathcal{S}_{(k-\ell)}}{\partial \beta_{+}}\Biggl)+2k\Biggr(\frac{\partial \mathcal{S}_{(1)}}{\partial \alpha} \frac{\partial \mathcal{S}_{(k-1)}}{\partial \alpha}-\frac{\partial \mathcal{S}_{(1)}}{\partial \beta_{+}} \frac{\partial \mathcal{S}_{(k-1)}}{\partial \beta_{+}}\Biggl) \\& =0.
\end{aligned}
\end{equation}
Next we will insert into (129) the following equations (128), (126) and (125) which will yield 

\begin{equation}
\begin{aligned}
\sum _{\ell=2}^{k-2} \frac{3 (\ell-1) k! g(\ell) (k-\ell-1) g(k-\ell)}{\ell! (k-\ell)!}+(k-1) (4 g(k)-3 (k-2) k g(k-1)) =0,
\end{aligned}
\end{equation}
where all of the dependency on $\alpha$ and $\beta_+$ has dropped out which has turned the problem of finding higher order quantum corrections into simply an algebraic one. We can easily solve for g(k) 
\begin{equation}
\begin{aligned}
g(k)=\frac{\sum _{\ell=2}^{k-2} \frac{3 (\ell-1) k! g(\ell) (k-\ell-1) g(k-\ell)}{\ell! (k-\ell)!}}{4-4
   k}+\frac{3}{4} (k-2) k g(k-1),
\end{aligned}
\end{equation}
and it is evident that g(k) is always a well defined number assuming that $\left(g(2), \ldots, g(k-1)\right)$ are also well defined numbers. All that we need to compute sequentially an arbitrary number of coefficients g(k) is g(2), which can clearly be read off from (127) to be $g(2)=\frac{1}{8} \left(\text{B}^2+9\right)$. With g(2) known we can write out the following asymptotic 'ground' state solution to the Taub WDW equation 

\begin{equation}
\begin{aligned}
\psi^{nb}:=e^{\left(-\sum _{k=2}^{\infty } \frac{g(k) e^{\beta_+ (k-1)-2 \alpha (k-1)}}{k!}-\frac{1}{6}
   \left(1-4 e^{3 \beta_+}\right) e^{2 \alpha-4 \beta_+}-\frac{1}{2} a (4-\text{B})+\frac{5 \beta_+}{2}\right)}.
\end{aligned}
\end{equation}
Attached in the appendix section will be a Mathematica code which explicitly computes this asymptotic solution using (131) up to any order k. 

We will leave the proof for the existence of asymptotic 'excited' state solutions for another time. For now though we will be content to find the semi classical $\phi_{0}$ term and the $\phi_{1}$ quantum correction for the 'no boundary' case. The author found the following family of conserved quantities along the flow of $\mathcal{S}_{(0)}^{\mathrm{nb}}$ which satisfies (19) as the reader can easily verify

\begin{equation}
\begin{aligned}
\phi_{0}^{\mathrm{nb}}:=\left(\left(e^{3 \beta_+}-1\right) e^{\beta_+-2 \alpha}\right)^n.
\end{aligned}
\end{equation}
Because our $\phi_{0}^{\mathrm{nb}}$ possesses zeros within the domain of real Misner variables we must restrict n to be a positive integer so our wave functions are smooth and globally defined. Using a computer algebraic system and the ansatz (120) the ordinary differential equation for $\phi_{1}$ can be solved which yields the following quantum correction 
\begin{equation}
\begin{aligned}
\phi_{1}^{\mathrm{nb}}:=-\frac{3 n e^{-2 \alpha n-2 \alpha} \left(e^{\beta_+} \left(e^{3 \beta_+}-1\right)\right)^{n+1} \left(4 e^{6  \beta_+} (n+1)-e^{3  \beta_+} (4 n+7)+n+2\right)}{4 \left(e^{3  \beta_+}-1\right)^3}.
\end{aligned}
\end{equation}

The general form of $\phi_{2}^{\mathrm{nb}}$ for any n is far too cumbersome to include in this paper, however below we will display this quantum correction for arbitrary ordering parameter for the first and second 'excited' states.
\begin{equation}
\begin{aligned}
\\& \phi_{2 \hspace{1 mm} n=1 }^{\mathrm{nb}}:=w\Biggl(e^{6 \beta_+} \left(\text{B}^2+432 e^3-27\right)+\left(e^3-3\right) e^{3 \beta_+} \left(\text{B}^2+432 e^3-27\right) \\& +e^6 \text{B}^2-3 e^3 \text{B}^2+3
   \text{B}^2-27 e^6+81 e^3+351\Biggr) \\& w=-e^{-6\alpha}\frac{e^{3 \beta_+} \left(e^{3 \beta_+}-e^3\right)}{16 \left(e^3-1\right)^3}
\end{aligned}
\end{equation}

\begin{equation}
\begin{aligned}
\\& \phi_{2 \hspace{1 mm} n=2 }^{\mathrm{nb}}:=w\biggl(\left(e^3-1\right) \left(e^{3 \beta_+}-1\right) \left(e^{6 \beta_+}+\left(e^3-3\right) e^{3 \beta_+}+3-3 e^3+e^6\right) \text{B}^2 \\& +27 \biggl(\left(8-65 e^3+88
   e^6\right) e^{9 \beta_+}+\left(e^3-4\right) \left(8-65 e^3+88 e^6\right) e^{6 \beta_+} \\& -\left(40+70 e^3-268 e^6+65 e^9\right) e^{3 \beta_+}+33+8 e^3 \left(-5-4
   e^3+e^6\right)\biggr)\biggr) \\& w=-e^{-8\alpha}\frac{e^{4 \beta_+} \left(e^{3 \beta_+}-e^3\right)}{8 \left(e^3-1\right)^4}
\end{aligned}
\end{equation}

We will now plot some interesting Taub no boundary wave functions composed of (136), (134), (133), and (132) which are in the form of (11) and (17). The first three plots(figs 7a-7c) are of the 'no boundary' 'ground' states (132) computed up to the $\mathcal{S}_{(2)}^{\mathrm{nb}}$ quantum correction for different values of $\alpha$ when B=0.

\begin{figure}
\centering
\begin{subfigure}{.4\textwidth}
  \centering
  \includegraphics[scale=.13]{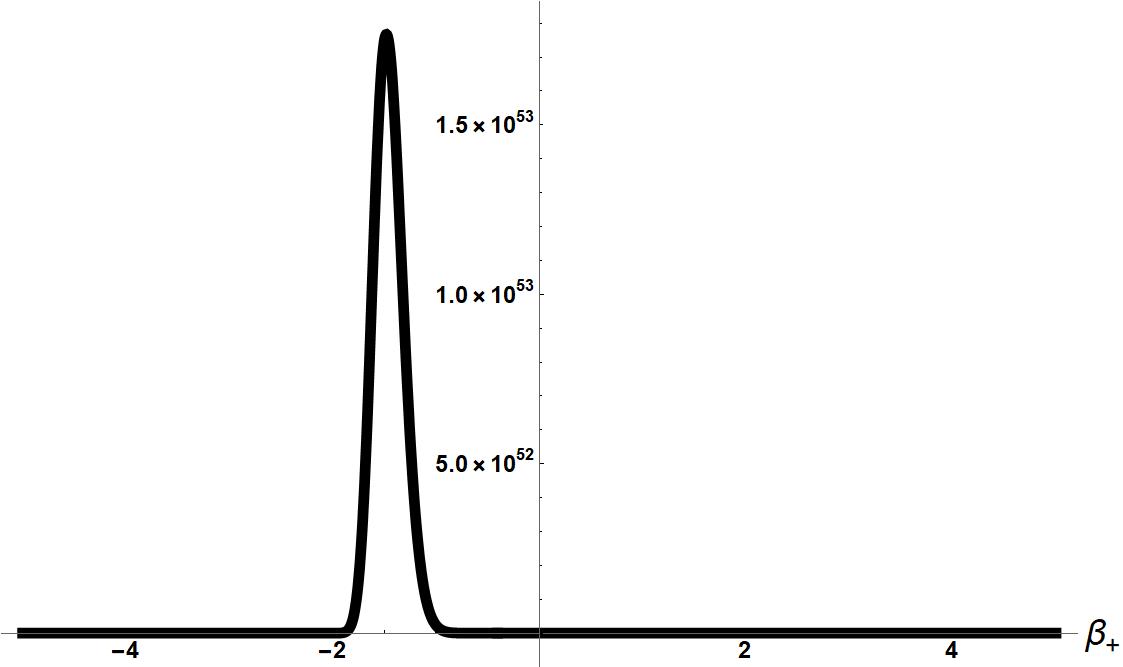}
  \caption{ $\alpha=-2$ }
  \label{fig:sub1}
\end{subfigure}%
\begin{subfigure}{.4\textwidth}
  \centering
  \includegraphics[scale=.13]{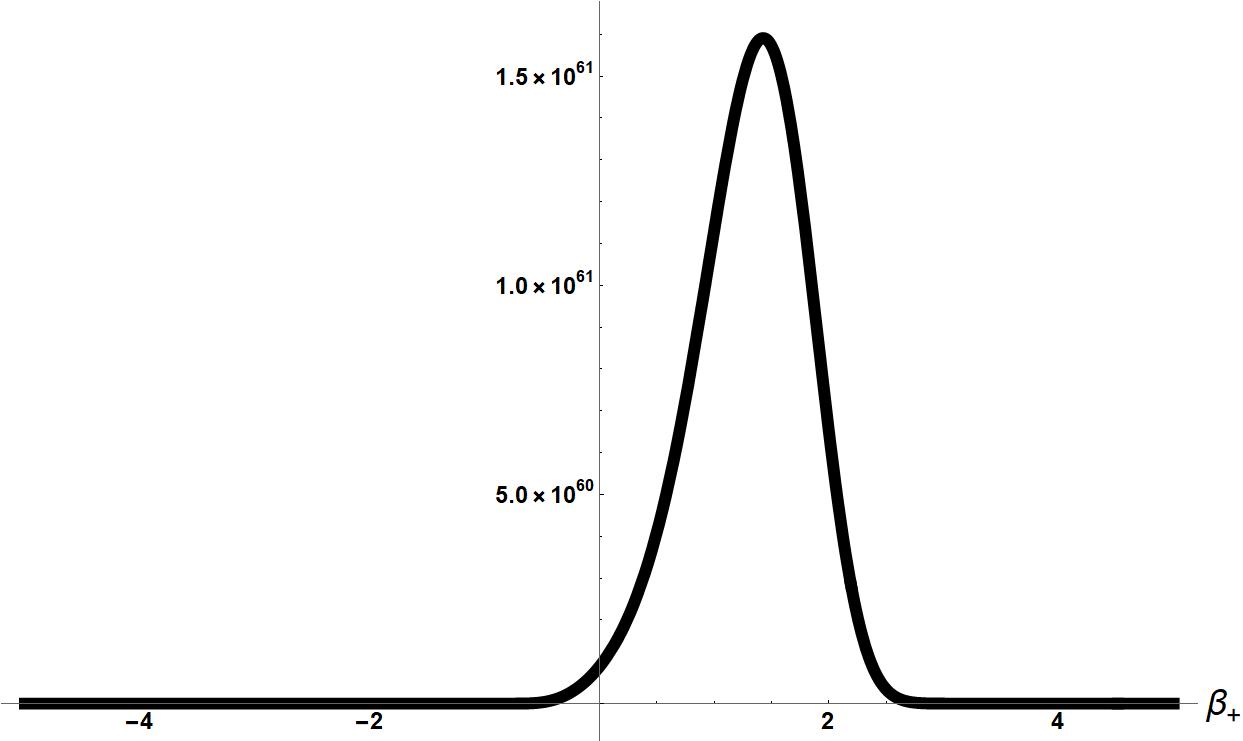}
  \caption{ $\alpha=0$ }
  \label{fig:sub2}
\end{subfigure}
\begin{subfigure}{.4\textwidth}
  \centering
  \includegraphics[scale=.13]{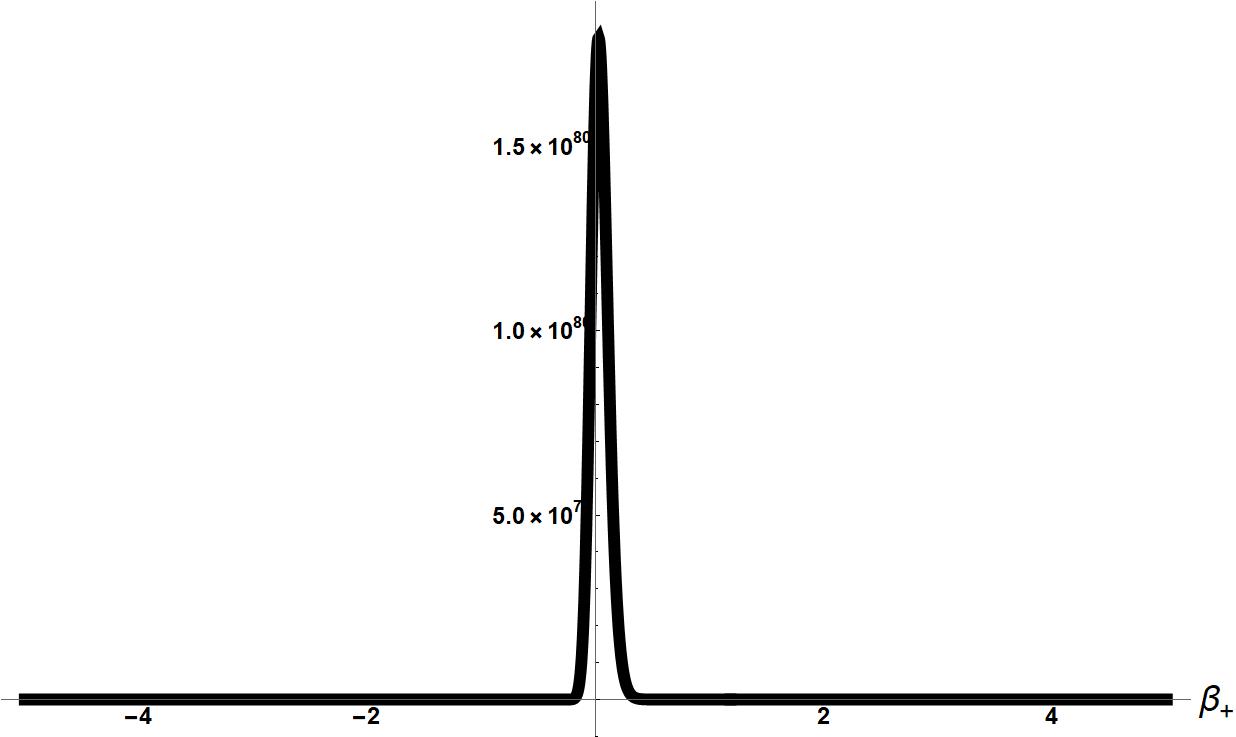}
  \caption{ $\alpha=2$ }
  \label{fig:sub1}
\end{subfigure}%
\caption{Plots of $\abs{\psi^{nb}}^{2}$ for the 'no boundary' vacuum Taub 'ground' state (132) which includes quantum corrections up to $k=2$.}
\label{fig:test}
\end{figure}

The next three plots(figs 8a-8c) are of the second 'excited' state where we include quantum corrections up to $k=2$ for different values of $\alpha$.

\begin{figure}
\centering
\begin{subfigure}{.4\textwidth}
  \centering
  \includegraphics[scale=.13]{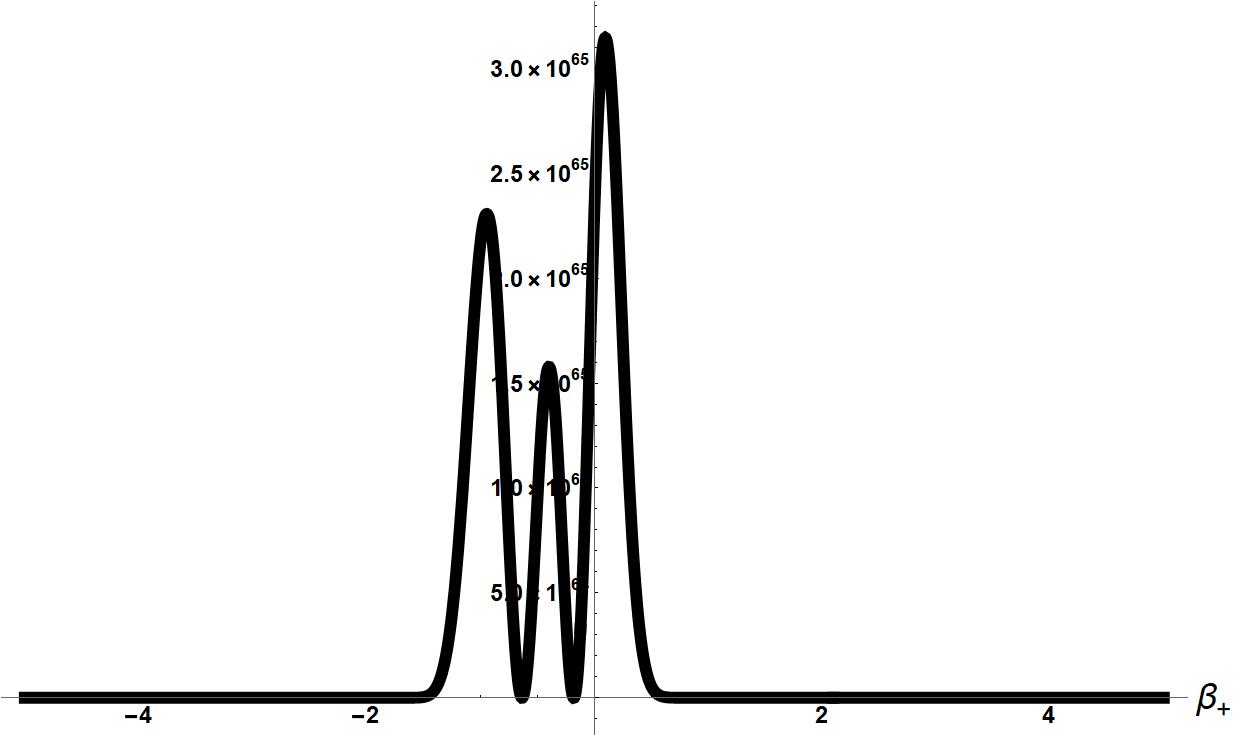}
  \caption{ $\alpha=-1.6$ }
  \label{fig:sub1}
\end{subfigure}%
\begin{subfigure}{.4\textwidth}
  \centering
  \includegraphics[scale=.13]{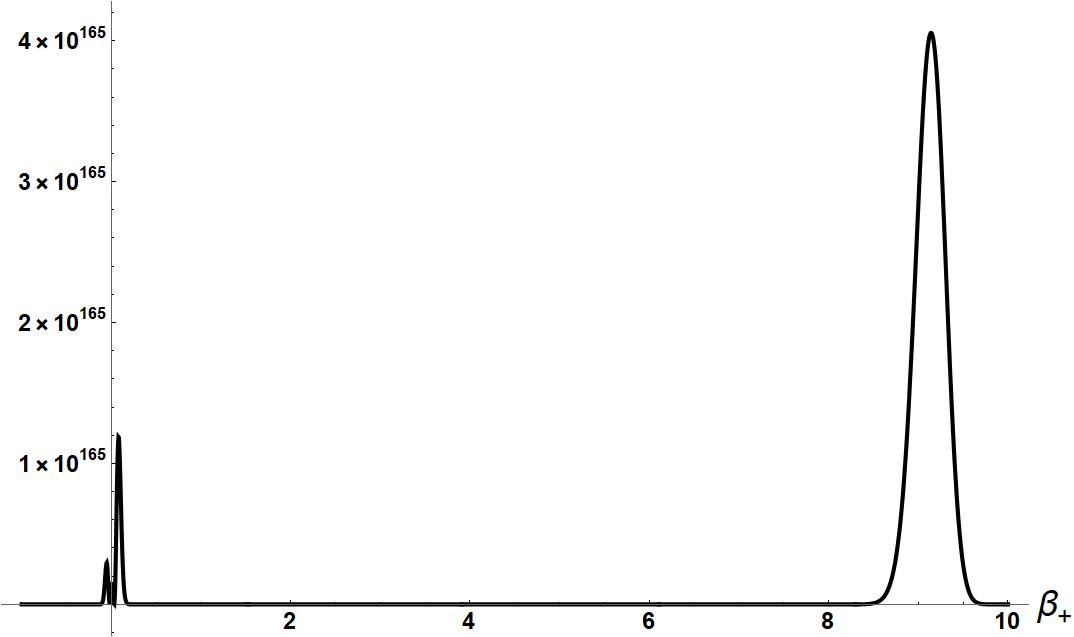}
  \caption{ $\alpha=2.825$ }
  \label{fig:sub2}
\end{subfigure}
\begin{subfigure}{.4\textwidth}
  \centering
  \includegraphics[scale=.13]{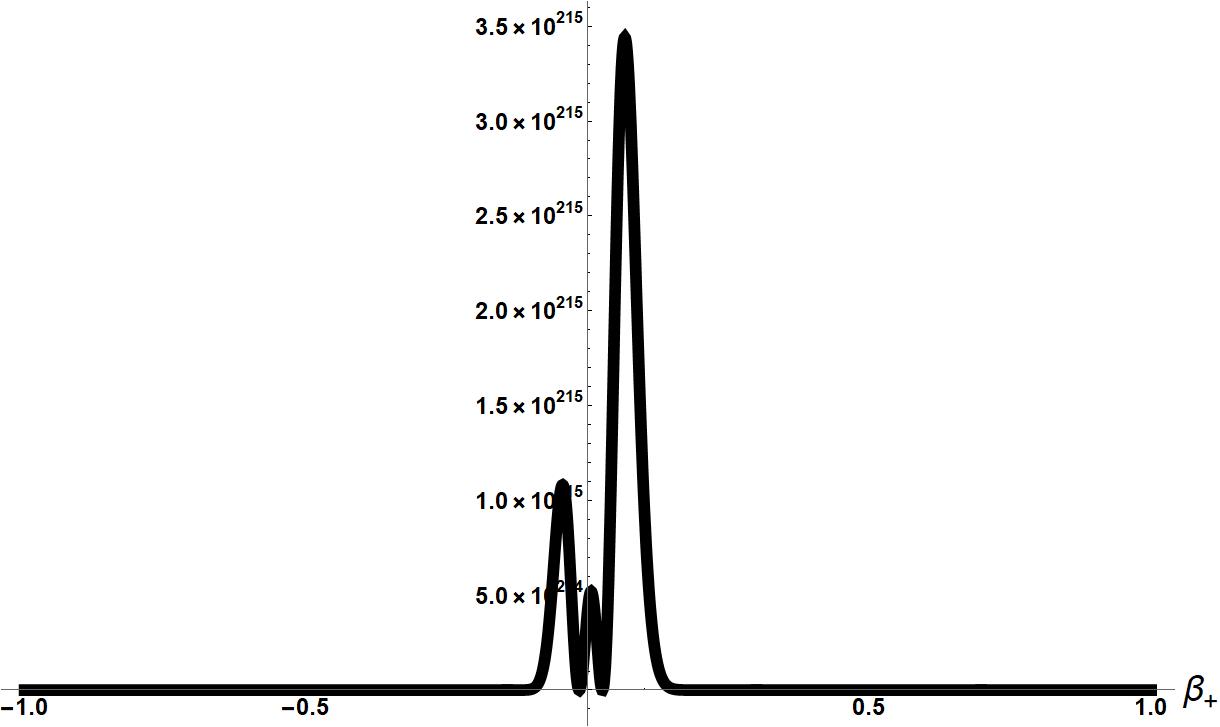}
  \caption{ $\alpha=3$ }
  \label{fig:sub1}
\end{subfigure}%
\caption{Plots of $\abs{\psi_{n=2}^{nb}}^{2}$ for the second 'no boundary' vacuum quantum Taub 'excited' state including terms up to $k=2$.}
\label{fig:test}
\end{figure}

In the next section we will discuss qualitatively the behavior of our wave functions and the physical implications that can be extrapolated from them. 

\section{\label{sec:level1}Discussion} 

To begin analyzing our results we first need to adopt an interpretation for the wave functions we computed. Two interpretations of quantum mechanics which in the past have been used to extrapolate physics from Wheeler Dewitt wave functions within the context of quantum cosmology are the consistent histories approach \cite{griffiths1984consistent} and pilot wave theory \cite{bohm1952suggested}. However, for our purposes, we will use the following admittingly naive interpretation which we will briefly outline. Even though we cannot interpret $|\psi|^{2}$ as a probability density due to the lack of a known dynamical unitary operator for the symmetry reduced Wheeler Dewitt equation, if we fix $\alpha$ and only consider wave functions which do not grow without bound when $\beta_{+}\to{\pm \infty}$ then our wave functions are reminiscent of normalizable probability densities as can be seen from our plots. Each point in those plots at a fixed $\alpha$ represents a potential geometric configuration that a quantum universe described by those wave functions can possess. Associated with each of those points in $\beta_{+}$ space at a fixed $\alpha$ is a value of $|\psi|^{2}$; it is not unreasonable to conjecture that the greater the value of $|\psi|^{2}$ is, the more likely a Taub universe will possess the geometry dictated by $\beta_+$. For example if $|\psi\left(\alpha,\beta1_+\right)|^{2} > |\psi\left(\alpha,\beta2_+\right)|^{2}$  we would interpret this to mean that a Taub universe described by $\psi$ when it reaches a size dictated by $\alpha$ is more likely to have a spatial geometry which possesses a level of anisotropy described by the value of $\beta1_+$ as opposed to  $\beta2_+$.

A shortcoming of our interpretation is that it cannot assign numerical values of probability to a micro ensemble of Taub universes with different values of $\alpha$, because $\int^{\infty}_{\infty}\abs{\psi}^{2} d\beta_+$ is not conserved in $\alpha$. Nonetheless we are picking this interpretation because it is intuitive for the solutions we are dealing with and facilitates the elucidation of the points the author wishes to make. In essence the author was inspired to pick this interpretation because he would like to let the bare solutions speak for themselves. The author strongly encourages future work to be done in extrapolating physics for the results presented in this paper using both the Bohmian and consistent histories approaches, in conjunction to other quantitative methods. What follows will be an attempt to extrapolate the physical implications of a cosmological constant and an electromagnetic field in the development of a quantum Taub universe from our wave functions by examining their aesthetic characteristics. 

Our first set of figures(1a-1d) emphatically shows how our matter sources can effect the quantum cosmological evolution of a Taub universe. If we compare figures 1(a) and 1(b) we see that the electromagnetic field $b^{2}$ causes the wave function originally in figure 1(a) to be peaked at a higher value of anisotropy and be more narrow. Physically this indicates a quantum mechanism by which an electromagnetic field can increase the level of anistropy in the early universe. Additionally it can also cause the geometry of a universe to be more defined in the sense that its wave function of the universe becomes sharply peaked at a specific value of $\beta_{+}$ as opposed to being spread out in $\beta_{+}$ space as figure 1(a) is. 

The above two effects that our electromagnetic field has on our wave functions in figures(1a-2b) could have had a profound impact on nucleogenesis and the formation of pockets of anistropy in our early universe if similar behavior was present in its wave function of the universe. Furthermore because the electromagnetic field causes the geometry of our quantum Taub universe to be more sharply defined it could have also played a role in the transition of a universe from one which could only be adequately described using quantum mechanics to one which can be described adequately using classical mechanics. Our results encourage studying further other anistropic quantum cosmologies such as the Bianchi I models using the Euclidean-signature semi classical method when an electromagnetic field and a cosmological constant are present in order to establish if the plethora of effects we detailed in this paper are generic to anistropic quantum cosmologies. 

Figures(1c-1d) shed interesting light on other effects that our two matters sources can have on the evolution of a quantum universe. For the vacuum quantum Taub 'wormhole' when $\alpha=1.5$ our wave function is sharply peaked at isotropy$\left(\beta_{+}=0\right)$. As $\alpha$ continues to grow that peak at isotropy rapidly sharpens. However when $\Lambda < 0$ no matter how large $\alpha$ becomes the wave function peaks at some non-zero value of $\beta_{+}$. In other words no matter how large $\alpha$ becomes there is always some residual anisotropy present as is shown in figure 1(c). However in figure 1(d) we see that an electromagnetic field can actually decrease the level of anistropy present in the early universe at certain stages of its development in terms of our clock $\alpha$ as can be seen by how the wave function is centered at isotropy. This points to how the electromagnetic field can have a myriad of different effects on the quantum evolution of a universe, thus justifying further investigations into primordial electromagnetic fields within the context of quantum cosmology. 

Moving on to our 'excited' states(figs 2a-2f) when $\Lambda=0$ we see even more spectacular effects from our primordial electromagnetic field. When $b^{2}=0$ our 'excited' states have two peaks which represent two potential geometric configurations our quantum Taub universe can tunnel in and out of. For the vacuum 'wormhole' case as $\alpha$ grows the multiple peaks merge into one central peak located at isotropy$\left(\beta_{+}=0\right)$. This behavior makes sense because as we previously mentioned $e^{\alpha}$ classically represents the scale factor of our cosmology and we expect that quantum effects, such as tunneling, would be most prominent when a universe is small and highly energetic and diminish as it grows in size. Of course no matter how large a universe becomes it is fundamentally quantum mechanical, but the probabilities of those quantum mechanical effects manifesting on large macroscopic scales rapidly diminishes as it grows in size. In other words our 'excited' state quantum Taub universes experience a phase transition over a certain range of $\alpha$ where they transition from a universe where tunneling between different geometric configurations is common to one where it is exceedingly rare. 

When our electromagnetic field is turned on we see some profound effects. For $b=1.5$ an additional peak emerges which represents another likely geometric configuration our quantum Taub universe can tunnel into as can be seen by comparing figures 2(d) and 2(e). This additional state was created quantum mechanically by our electromagnetic field. As our field grows in strength all of the peaks eventually merge into one central peak as can be seen in figure 2(f). This effect of an electromagnetic field creating an additional state which a quantum universe can tunnel in and out of can be seen also in figure 4(b). Thus it appears for 'excited' states that the effects which a  primordial electromagnetic field has on the evolution of a quantum Taub universe are highly dependent upon its strength. When the field is somewhat weak it can create additional states that our quantum universe can tunnel in and out of and when it is strong it can destroy those states, leaving only one sharply defined state left. 

Using our aforementioned analysis of the first 10 plots the reader can handily interpret in the manner which we have done thus far the wave functions for our superposition of 'excited' states(3a-3c) and our semi-classical 'excited' states(figs 4a-4c) when both matter sources are present.

The vacuum 'wormhole' case for arbitrary ordering parameter has some interesting mathematical properties which we already mentioned. It should be stressed that these properties are only present when $B\ne0$ and that they manifest themselves differently within our wave functions for different values of the ordering parameter. Nonetheless for $B=-1$ our vacuum 'wormhole' wave functions behave peculiarly in comparison to the ones we computed earlier in this paper. For large negative $\alpha <<0$ our vacuum 'wormhole' asymptotic wave function which possesses terms up to $\mathcal{S}^{wh}_{(3)}$ behaves as a 'ground' state as we would expect it to as indicated in figure 5(a). As $\alpha$ continues to grow the wave function behaves as a pseudo-Gaussian traveling to the right in the positive $\beta_{+}$ direction. 

However around $\alpha=-1.1$ something drastic happens, a second peak forms which is illustrated in figure 5(b) and thus makes our 'ground' state aesthetically resemble an 'excited' state. This is why we constantly employ '    ' to denote 'ground' and 'excited' states in our work. Unlike in ordinary quantum mechanics the lines between 'ground' and 'excited' states are not as sharply defined. Using our 'ground' state equations we can obtain terms which do not manifest the discretization in their quantum numbers that we showed to be present in our $\phi_{0}$ terms, and yet the states constructed from them behave as 'excited' states. This further suggests that more theoretical work is necessary to rigorously delineate between the 'ground' and 'excited' states of a theory which possesses a vanishing Hamiltonian. Another interesting feature of our wave functions is that these two peaks which our Taub universe can tunnel in between emerges as our $\alpha$ reaches a threshold, as opposed to being continuously present for small $\alpha <0$'s as is the case of our closed form 'excited' states(figs 3a-3c). Our 'excited' states(figs 6a-6c) which include terms up to  $\mathcal{S}^{wh}_{(3)}$ and $\phi^{wh}_{2}$ don't overtly manifest the properties one typically expects from 'excited' states as is exemplified in figures (3a-4c). At most we see a small second peak form when $\alpha=1$ in figure 6(c).

Moving on to our 'no boundary' asymptotic solutions, if we start with our 'ground' state figures $(7a-7c)$ we see that for small values of $\alpha$ our wave function forms a peak around a negative value of $\beta_{+}$ and is moving to the right from the negative $\beta_{+}$ axis. For larger values of $\alpha$ our wave function forms a peak around a positive value of $\beta_{+}$ as it continues to travel to the right. However at a certain value of $\alpha$ our travelling Gaussian changes direction and eventually resides at the origin as its magnitude continues to grow as $\alpha$ grows. This behavior of our wave function not being bounded from above for real values of the Misner variables poses a complication in our efforts to obtain a physically meaningful picture of what is going on using our method. If we naively interpreted our 'no boundary' states as such using our purely qualitative method we would conclude that the universe described by our wave function can only exist when $\alpha = \infty$ because that state is infinitely more likely to occur then any other geometric configuration because our wave function approaches $\infty$ as $\alpha \to \infty$. Such an interpretation makes no sense. A potential way to remedy this problem can be by choosing a different way to construct an inner product between $\psi_{nb}s$. 

Our 'no boundary' excited states can be interpreted similarly to our other 'excited' states with the aforementioned caveat. One note worthy feature of our plots is that in 8(b) our wave function is peaked at an extremely anistropic value of $\beta{+}$, but then quickly snaps back towards isotropy at $\alpha=3$. It will require more work to determine if this "snapping" behavior is a generic feature belonging to all quantum Taub 'no boundary' 'excited' states.

\section{\label{sec:level1}Concluding Remarks} 
Despite the vast number of results detailed in this work it now seems that there are as many unanswered questions as results. In the author's own estimation this is a positive sign that the research we are conducting is worthwhile and that continued pursuit of it will lead to a further understanding on how matter sources such as an electromagnetic field and a cosmological constant impacted the evolution of the early universe. 

Moving forward the author would like to carry out the analysis he did in this paper for the case when a scalar field is present. Additionally as mentioned he would like to prove that a countably infinite number of 'excited' states exist for the case when both an electromagnetic field and cosmological constant are present. Due to the non-trivial nature of the solutions that the Euclidean-signature semi classical method has been able to generate we have learned much about how matter sources could have effected the evolution of the early universe. In order for the results in this paper to be potentially more relevant to our own universe in its early infancy it is crucial that the analysis that we carried out here be duplicated for other anisotropic quantum cosmologies  models\cite{ryan1997hamiltonian,christodoulakis1996new} to establish how typical these effects are for these types of solutions. The author looks forward to what future works will have to say about the role that matter sources played in the evolution of our early universe.

\section{\label{sec:level1}ACKNOWLEDGMENTS}
 
I am grateful to Professor Vincent Moncrief for valuable discussions at every stage of this work. I would also like to thank George Fleming for facilitating my ongoing research in quantum cosmology. Daniel Berkowitz acknowledges support from the United States Department of Energy through grant number DE-SC0019061. I also must thank my aforementioned parents.

\section{\label{sec:level1}Appendix} 

\begin{verbatim}
Table[{Subscript[S, 
      0] = ((-9 E^(
           4 \[Alpha] - 2 \[Beta]) (-1 + E^(6 \[Beta])) \[Pi] + 
          E^(6 (\[Alpha] + \[Beta])) \[CapitalLambda]))^(n);, 
    Subscript[S, k] = 
      Sum[Sum[b[{i, j}]*
         E^(6*n*\[Alpha] - 2*(n - i)*\[Alpha] - 
            2*k*\[Alpha] + -2*n*\[Beta] + 6*j*\[Beta] + 8*\[Beta]*i + 
            4*\[Beta]*k), {j, 0, n - k - i}], {i, 0, n - k}];, 
    asd = ((-D[Subscript[S, k], \[Alpha]]*
           D[1/6 E^(2 \[Alpha] - 4 \[Beta]) (1 + 2 E^(6 \[Beta])) - (
             E^(4 (\[Alpha] + \[Beta])) \[CapitalLambda])/(
             36 \[Pi]), \[Alpha]] + 
          D[Subscript[S, k], \[Beta]]*
           D[1/6 E^(2 \[Alpha] - 4 \[Beta]) (1 + 2 E^(6 \[Beta])) - (
             E^(4 (\[Alpha] + \[Beta])) \[CapitalLambda])/(
             36 \[Pi]), \[Beta]]) + 
        k*(D[Subscript[S, -1 + k], \[Alpha]] - 
           D[Subscript[S, -1 + k], \[Beta]]) + (k/
           2)*(D[D[Subscript[S, -1 + k], \[Alpha]], \[Alpha]] - 
           D[D[Subscript[S, -1 + k], \[Beta]], \[Beta]]));, \[Alpha] =
       Log[x];, \[Beta] = Log[y];, 
    Subscript[S, 
      k] = ((Sum[
          Sum[b[{i, j}]*
            E^(6*n*\[Alpha] - 2*(n - i)*\[Alpha] - 
               2*k*\[Alpha] + -2*n*\[Beta] + 6*j*\[Beta] + 
               8*\[Beta]*i + 4*\[Beta]*k), {j, 0, n - k - i}], {i, 0, 
           n - k}]) /. 
        Solve[Simplify[
           MonomialList[Simplify[asd], {x, y, 1/x, 1/y}]] == 
          Table[0, {i, 1, 
            Length[Flatten[
              Table[Table[b[{i, j}], {j, 0, n - k - i}], {i, 0, 
                n - k}], 1]]}], 
         Flatten[Table[
           Table[b[{i, j}], {j, 0, n - k - i}], {i, 0, n - k}], 1]]);,
     ClearAll[\[Alpha]];, ClearAll[\[Beta]];, 
    Subscript[S, k] = 
     Expand[Simplify[
       First[Subscript[S, 
         k] /. {x -> E^(\[Alpha]), y -> E^(\[Beta])}]]], ClearAll[x];,
     ClearAll[y];};, {k, 1, n}];
\end{verbatim}

\begin{verbatim}
Table[Subscript[S, g], {g, 0, n}]
\end{verbatim}

These two first lines of code produce the n non-trivial quantum corrections to the nth 'excited' state of the $\Lambda \ne 0$ Taub models which can be used to construct a closed form solution when $B=0$. The only required input from the user is setting n equal to a positive integer.

\begin{verbatim}
Table[{Subscript[S, 
      0] = ((-9 E^(
          4 \[Alpha] - 2 \[Beta]) (-1 + E^(6 \[Beta])) \[Pi]))^(n);, 
    Subscript[S, k] = 
      E^(4*\[Alpha]*n - 2*\[Alpha]*k)*
       Sum[E^(-6*\[Beta]*j - 2*\[Beta]*k + 4*\[Beta]*n)*b[j, k], {j, 
         0, n - k}];, 
    asd = ((-D[Subscript[S, k], \[Alpha]]*
           D[1/6 E^(
             2 \[Alpha] - 
              4 \[Beta]) (1 + 2 E^(6 \[Beta])), \[Alpha]] + 
          D[Subscript[S, k], \[Beta]]*
           D[1/6 E^(
             2 \[Alpha] - 
              4 \[Beta]) (1 + 2 E^(6 \[Beta])), \[Beta]]) + 
        k*(D[Subscript[S, -1 + k], \[Alpha]] - 
           D[Subscript[S, -1 + k], \[Beta]]) + (k/
           2)*(D[D[Subscript[S, -1 + k], \[Alpha]], \[Alpha]] - 
           D[D[Subscript[S, -1 + k], \[Beta]], \[Beta]]));, \[Alpha] =
       Log[x];, \[Beta] = Log[y];, 
    Subscript[S, 
      k] = ((E^(4*\[Alpha]*n - 2*\[Alpha]*k)*
          Sum[E^(-6*\[Beta]*j - 2*\[Beta]*k + 4*\[Beta]*n)*
            b[j, k], {j, 0, n - k}]) /. 
        Solve[Simplify[
           MonomialList[Simplify[asd], {x, y, 1/x, 1/y}]] == 
          Table[0, {i, 1, 
            Length[Flatten[Table[b[j, k], {j, 0, n - k}], 1]]}], 
         Flatten[Table[b[j, k], {j, 0, n - k}], 1]]);, 
    ClearAll[\[Alpha]];, ClearAll[\[Beta]];, 
    Subscript[S, k] = 
     Expand[Simplify[
       First[Subscript[S, 
         k] /. {x -> E^(\[Alpha]), y -> E^(\[Beta])}]]], ClearAll[x];,
     ClearAll[y];};, {k, 1, n}];
\end{verbatim}

\begin{verbatim}
Table[Subscript[S, g], {g, 0, n}]
\end{verbatim}

The Mathematica code above computes the n 'excited' state quantum corrections for the 'wormhole' quantum Taub models which can be used to construct a closed form solution to the corresponding Wheeler DeWitt when $B=0$.

\begin{verbatim}
Total[{g[2] = 1/8 (9 + BB^2), 
   Table[{g[k] = 3/4 (-2 + k) k g[-1 + k] + \!\(
\*UnderoverscriptBox[\(\[Sum]\), \(L = 2\), \(\(-2\) + k\)]
\*FractionBox[\(3\ \((\(-1\) + k - L)\)\ \((\(-1\) + L)\)\ \(k!\)\ g[
            k - L]\ g[L]\), \(\(\((k - L)\)!\)\ \(L!\)\)]\)/(
        4 - 4 k), (1/(k!)) g[k]*
       E^(-2 \[Alpha] (-1 + k) + \[Beta] (-1 + k))}[[2]], {k, 3, 
     n}]}[[2]]]
\end{verbatim}

This last piece of codes computes the higher order  $-\mathcal{S}^{nb}_{(k \geq 3)}$ terms of the asymptotic series present in the exponent of our 'ground' state 'no boundary' wave functions $\psi^{nb}:=e^{\left(-\sum _{k=2}^{\infty } \frac{g(k) e^{\beta_+ (k-1)-2 \alpha (k-1)}}{k!}-\frac{1}{6}
   \left(1-4 e^{3 \beta_+}\right) e^{2 \alpha-4 \beta_+}-\frac{1}{2} a (4-\text{B})+\frac{5 \beta_+}{2}\right)}$.

\bibliography{Taub}

\end{document}